\documentclass[a4paper,11pt]{article}

\usepackage{amsmath}
\usepackage{graphicx}
\usepackage{amsfonts}
\usepackage{amssymb}
\usepackage{epsfig}
\usepackage{color}
\usepackage{psfrag}

 \newtheorem{theorem}{Theorem}[section]
    \newtheorem{rema}{Remark}[section]
    \newtheorem{propo}[rema]{Proposition}
    \newtheorem{defi}[rema]{Definition}
    \newtheorem{lemma}[rema]{Lemma}
    \newtheorem{corol}[rema]{Corollary}
    \newtheorem{proper}[rema]{Property}

\newcommand{\EQ}{\begin{equation}}
\newcommand{\EN}{\end{equation}}

\newcommand{\sect}[1]{\setcounter{equation}{0}\section{#1}}

\newcommand{\bc}{\begin{center}}
\newcommand{\ec}{\end{center}}
\def\ba#1{\begin{array}{#1}\displaystyle}
\newcommand{\ea}{\end{array}}
\newcommand{\z}{\\[2mm] \displaystyle}

\newcommand{\beq}{\begin{equation}}
\newcommand{\eeq}{\end{equation}}
\newcommand{\beqa}{\begin{eqnarray}}
\newcommand{\eeqa}{\end{eqnarray}}
\newcommand{\no}{\nonumber}
\newcommand{\n}{\nonumber\\}
\newcommand{\bi}{\begin{itemize}}
\newcommand{\ei}{\end{itemize}}

\def\lt#1{\left#1}
\def\rt#1{\right#1}
\def\t#1{\tilde{#1}}

\def\b#1{\bar{#1}}
\def\frc#1#2{\frac{#1}{#2}}

\newcommand{\p}{\partial}

\newcommand{\bra}{\langle}
\newcommand{\ket}{\rangle}
\newcommand{\Z}{{\mathbb{Z}}}
\newcommand{\N}{{\mathbb{N}}}
\newcommand{\R}{{\mathbb{R}}}
\newcommand{\C}{{\mathbb{C}}}
\newcommand{\hC}{{\hat{\mathbb{C}}}}

\newcommand{\Or}{{\cal O}}

\newcommand{\ep}{\epsilon}

\newcommand{\id}{{\rm id}}

\newcommand{\End}{{\rm End}}

\newcommand{\halmos}{\rule{1ex}{1.4ex}}
\newcommand{\eproof}{\hspace*{\fill}\mbox{$\halmos$}}
\newcommand{\proof}{{\em Proof.\ }}

\newcommand{\spa}{{\tt H}}
\newcommand{\spag}{{\tt F}}

\def\cl#1{\overline{#1}}

\renewcommand{\:}{\mbox{\tiny ${\circ\atop\circ}$}}

\begin{document}

\begin{titlepage}

\begin{center}
{\Large {\bf
Higher conformal variations and the Virasoro vertex operator algebra
}

\vspace{1cm}

Benjamin Doyon}

Department of Mathematics, King's College London\\
Strand, London WC2R 2LS, U.K.\\
email: benjamin.doyon@kcl.ac.uk

\end{center}

\vspace{1cm}

\noindent We develop a calculus of variations for functionals on certain spaces of conformal maps. Such a space $\Omega$ is composed of all maps that are conformal on domains containing a fix compact annular set of the Riemann sphere, and that are ``sense-preserving''. The calculus of variations is based on describing infinitesimal variations of such maps using vector fields. We show that derivatives of all orders with respect to such conformal maps, upon conjugation by an appropriate functional, give rise to the structure of the Virasoro vertex operator algebra. Our construction proceeds in three steps. We first put a natural topology on $\Omega$ and define smooth paths and an operation of differentiation to all orders (``conformal derivatives''). We then study certain second-order variational equations and their solutions. We finally show that such solutions give rise to representations of the Virasoro algebra in terms of conformal derivatives, from which follows constructions for the Virasoro vertex operator algebra. It turns out that series expansions of multiple covariant conformal derivatives are equal to products of multiple vertex operators. This paper extends a recent work by the author, where first-order conformal derivatives were associated with the stress-energy tensor of conformal field theory.
\vfill

{\ }\hfill 7 October 2011

\end{titlepage}

\tableofcontents

\sect{Introduction}

Vertex operator algebras \cite{FLM88,FHL93,LL04}, a variant of Borcherds' notion of vertex algebras \cite{Bo86}, provide the algebraic underpinning for conformal field theory (CFT) \cite{BPZ,Gins,DFMS97}. The fields of CFT corresponding to conserved currents, like the identity field, the stress-energy tensor, and its descendants, are seen as basis elements for a graded quasi-finite vector space $V$. This space is endowed with an algebraic structure provided by the vertex operator map $Y(\cdot,x): V\to \End(V)[[x,x^{-1}]]$ with formal variable $x$ -- this can be interpreted as giving an infinite number of multiplication laws in $V$. Correlation functions in CFT involving conserved currents at different points can then be obtained by evaluating appropriate ``matrix elements'' of products of vertex operator maps at different formal variables. General correlation functions, involving other fields of CFT, are obtained through modules and intertwiners, as well as a pairing of ``right-movers'' with ``left-movers'', giving rise to the general structure of a full field algebra \cite{Hu07}. Vertex operator algebras encode all the properties of correlation functions that are expected to follow from general principles of CFT (for instance: commutativity, associativity). In particular, the algebraic structure formed by the identity field, the stress-energy tensor and its descendants is called the Virasoro vertex operator algebra, and this is a subalgebra of any vertex operator algebra. The Virasoro vertex operator algebra structure can be given to any identity module of the Virasoro algebra (see for instance \cite{LL04}), and it contains, as a subalgebra, the Virasoro algebra itself.

In the present paper, we study the Virasoro vertex operator algebra structure of certain spaces of smooth functionals of conformal maps, where the endomorphisms determining $Y(\cdot,x)$ are expressed in terms of differential operators representing infinitesimal variations of such conformal maps. This is a general construction, independent of CFT. In its application to CFT, this essentially gives rise to the opposite route to that described above: instead of constructing correlation functions from a vertex operator algebra, we give vertex operator algebra structures to spaces of correlation functions. Indeed, correlation functions can be seen as appropriate functionals of conformal maps. Every (genus-zero) CFT correlation function is then associated with an ``identity functional", and the application of products of vertex operators at different formal variables is identified with the insertion of the stress-energy tensor and its descendants at different points. This identification is the expression of the so-called conformal Ward identities in CFT.

The present paper substantially extends a study by the author of conformal derivatives of first order \cite{DCFT}, where the relation was made between a single application of such a conformal derivative and a single insertion of the stress-energy tensor in correlation functions of boundary CFT. A consequence of the results of the present paper is the extension of the construction of \cite{DCFT} to all descendants of the stress-energy tensor, and to multiple insertions in correlation functions. The present paper is mostly self-contained, so that it is not based in any essential way on \cite{DCFT}.

The main goal of our construction is to offer the possibility of identifying such structures as those of CFT correlation functions in other contexts, and hence of using powerful algebraic arguments in these contexts. In particular, a potential application is to conformal loop ensembles (CLE) \cite{Sh06,ShW07a,ShW07b}, which provide measures for random loops argued to describe the same scaling limits as those conjecturally described by CFT. In this context, expectation values are seen as functionals of conformal maps, and first-order conformal differential operators were shown (up to certain assumptions) to give rise to the insertion of a particular ``limiting'' random variable \cite{DCLE} (extending the work \cite{DRC}) -- whence this random variable could be identified with the stress-energy tensor thanks to \cite{DCFT}. In its application to CLE, the present construction provides an opportunity to identify all descendants of the stress-energy tensor with random variables, and to give a vertex operator algebra structure to such random variables; it is this potential application which guides the present construction. 

Identifying the vertex operator algebra structure of CLE is a crucial step in re-constructing CFT from CLE, an endeavor which can be seen as constructive quantum field theory. This gives an understanding of CFT as a theory for emergent behaviors rather than, more traditionally in the mathematics community, as a geometric construction. This is very important especially in light of the recent progress in establishing the existence and conformal invariance of certain scaling limits \cite{Sm}.

Our construction proceeds in three steps. We first define a Hausdorff topological space $\Omega$ of conformal maps all of whose domain contain an annular compact set, and all of which are sense-preserving (to be defined below). On $\Omega$, we put a natural structure of smooth paths which guide the definition of differential operators. We show, in particular, that $\Omega$ is connected by piecewise smooth paths. Then, on this space, we study certain second-order conformal differential equations which have a close connection with the Virasoro algebra; these equations essentially ask for an unknown $\log(Z)$ to have constant second derivatives, and solutions $Z$ are referred to as (normal) conformal Gaussians. We show that for spaces connected by piecewise smooth paths, such equations are equivalent to certain transformation equations for $Z$. Finally, assuming the existence of a normal conformal Gaussian, we provide a new (to our knowledge) construction of the Virasoro algebra in terms of conformal differential operators, and we study the associated Virasoro vertex operator algebra structure on quasi-finite subspaces of the space of smooth functionals. Our main result is Theorem \ref{theovoa}. It says that multiple applications of covariant conformal derivatives on the identity functional are reproduced by multiple applications of vertex operators, where the identification of covariant conformal derivatives with vertex operators is simple.

From the viewpoint of the Virasoro vertex operator algebra, our construction provides an interpretation of normal-ordered products of vertex operators in terms of multiple conformal derivatives. The normal-ordering operation is used in the construction of the Virasoro, and other, vertex operator algebras, and appears naturally through the representation theory for vertex operator algebras \cite{Li96,LL04}. Our results give a clear analytic meaning to this operation, which points to an analytic interpretation of the representation theory.

In its application to CFT, our construction gives a quite transparent re-writing of the conformal Ward identities associated with descendants of the stress-energy tensor in terms of multiple conformal derivatives; further, an intermediate result provides a generating function for the transformation laws of these descendants under conformal maps. In particular, the transformation laws of products of covariant conformal derivatives has a tensor product structure, in agreement with multi-linearity of CFT correlation functions.

Our construction also gives a rather complete mathematical framework whereby the ideas of metric variations (whose role is played, here, by conformal variations) are connected to those of the algebraic structure of correlation functions (vertex operator algebras).

\subsection{Relations with previous works}

There are (at least) four groups of previous works in which related ideas are developed. One attraction of the present paper is that it may offer a framework whereby some of these ideas are naturally connected. We describe them briefly here, more or less in historical order.

{\bf Analytic geometry of CFT}. At the initial stages of the development of CFT in the physics community, there was a strong interest in understanding the geometro-analytic meaning of various aspects of CFT; in particular of the identification of the stress-energy tensor with metric variations. In the important work \cite{FS87}, these metric variations are naturally associated with variations of conformal structures of compact Riemann surfaces with punctures; such variations are described by Beltrami differentials $\mu$, which form the tangent space in this formulation. Equation (6) of \cite{FS87} states that the CFT partition function ${\rm Z}$, which depends on the Riemann surface and on the choice of metric on it, satisfies
\beq\label{formFS}
	\frc{i}{2\pi} \int d z \,d \b{z} \,T(\b{m},m,z) \mu(z,\b{z}) =
	- {\rm Z}^{-1}\, \delta_\mu {\rm Z},
\eeq
where $\delta_\mu {\rm Z}$ is the infinitesimal variation of ${\rm Z}$ in the direction $\mu$, $T(\b{m},m,z)$ is the one-point function of the stress-energy tensor $T(z)$ at the point $z$ on the Riemann surface characterized by the moduli $\b{m},m$, and on both sides a particular metric has been chosen whereby the scalar curvature is constant and the volume is 1 (since $z\mapsto T(\b{m},m,z)$ is not a quadratic differential because of the conformal anomaly, this is important for the left-hand side to be well defined).

By contrast, here and in \cite{DCFT}, conformal maps, rather than conformal structures, are the objects that are affected by infinitesimal variations, and such variations are described by holomorphic vector fields $h$ rather than Beltrami differentials. Equation (\ref{formFS}) is to be compared with a formula that follows from \cite[Sect 4.4]{DCFT}, and from Equation (\ref{T}) in the present paper:
\beq\label{formDCFT}
	\frc1{2\pi i} \int (dz\, h(z) + d\b{z} \,\b{h}(\b{z}))\,
	\bra T(z) \Or({\bf w})\ket = - Z^{-1} \nabla_h (Z \bra \Or({\bf w})\ket),
\eeq
where the correlation function $\bra \cdot\ket$ is evaluated on a simply connected domain $C$, $\Or({\bf w})$ represents a product of fields at non-coincident points $w_1, w_2,\ldots\in C$ away from $z$, the contour of integration on the left-hand side is counter-clockwise in $C$ and does not surround any point $w_i$, $h$ is a holomorphic function on a simply connected domain whose complement is surrounded by the contour of integration, $\nabla_h$ is the associated conformal derivative, and $Z$ is the {\em relative partition function}, a particular ratio of partition functions that is M\"obius invariant \cite{DCFT} (and an example of a normal conformal Gaussian).

In principle, the result (\ref{formDCFT}) can be deduced from an adaptation of (\ref{formFS}) to the context of boundary CFT on simply connected domains with punctures (based on the formal relation $\mu \propto \b\p h$; but there are nontrivial subtleties), where the punctures are the points $w_1, w_2,\ldots$. Further, it is likely that both formulations -- variations of conformal structures and variations of conformal maps -- can be connected, through the theory of quasiconformal maps, in the general context described in the present paper, which is independent of the notions of correlation functions and partition functions, and of other particularities of CFT. Developing this connection, however, is beyond the scope of this paper; we hope to come back to this in a later work \cite{DKprog}.

The present work, besides giving a CFT-independent mathematical formulation of the calculus of infinitesimal variations of conformal maps on annular domains, develops rather different aspects than those studied in \cite{FS87}: the full vertex operator algebra is described in terms of conformal derivatives, not just the stress-energy tensor; but the present work is restricted to the Riemann sphere. Note that the use of infinitesimal variations of conformal maps, rather than conformal structures, is better adapted to the potential application to CLE described above, as is clear from the construction in \cite{DCLE}.

{\bf Geometry of the Virasoro group and Teichm\"uller theory.} The Virasoro group ${\rm Diff}(S^1)$ of orientation-preserving diffeomorphisms of the unit circle is an infinite-dimensional Fr\'echet Lie group whose Lie algebra is the Virasoro algebra. Its geometry has been studied extensively both in the physics and mathematics literature \cite{diffS1}. It turns out that it has a natural complex-analytic embedding into the universal Teichm\"uller space \cite{NV90} (for Teichm\"uller theory, see for instance \cite{Lehto}). The universal Teichm\"uller space can be seen as an open ball in the complex Banach space of $L^\infty$ functions on the unit disk space (the space of Beltrami differentials), up to a natural equivalence. Further, by conformal welding \cite{Lehto}, diffeomorphisms in ${\rm Diff}(S^1)$ give rise to conformal maps of the unit disk that extend to quasiconformal maps of the Riemann sphere. Hence, there are natural relations between variations of Beltrami differentials, variations of certain conformal maps, and variations of elements in ${\rm Diff}(S^1)$, and all these variations have interpretations through infinite-dimensional manifolds. In particular, there is a nice theory for the tangent space that relates the Beltrami differential to these other descriptions \cite{N93}.

Although it would be natural to start with these constructions in order to develop the Virasoro vertex operator algebra, the approach that we take is rather guided by the potential application to loop models (cf. \cite{DCLE}). It turns out that the natural spaces are formed by conformal maps on annular domains; hence in our construction, we study differentiability on such spaces. The study of these spaces and the associated differential operators relate to that of ${\rm Diff}(S^1)$ in a way similar to that by which the problem of factorization of conformal maps on annular (or higher-connectivity) domains \cite{Hub66,Kuh70} (see also \cite{Dconf}) relates to that of conformal welding. Conformal welding is a problem of factorization of a diffeomorphism of $S^1$ into a composition of conformal maps on simply connected domains; this is essentially a ``limit case'' of the factorization problem for conformal maps on annular domains, whereby the image and pre-image domains both collapse to $S^1$. In this sense, then, a ``limit case'' of our construction should recover ${\rm Diff}(S^1)$. We do not have the full geometric structure of our spaces (here we develop only what is necessary for our present purposes), but a possible option to develop it would be to use the theorem of factorization \cite{Hub66,Kuh70}, and the known geometric structure on conformal maps of simply connected domains for the individual factors.

{\bf Vertex operator algebras of certain topological spaces.} The idea of representing the Virasoro algebra or vertex operator algebras on spaces of correlation functions is very natural, as the structure can be induced from that on the vertex operators forming the correlation functions. The vertex operator algebra of a completion of the space of rational correlation functions was studied in \cite{Hu99,Hu03} in a quite general setting; a related general construction from correlation functions (or ``amplitudes") was given in \cite{GG00}. Our construction, restricted to the Virasoro vertex operator algebra, deals however with more general spaces of smooth functionals of conformal maps, without {\em a priori} a vertex operator algebra structure or a notion of correlation functions. Our take is also from a different viewpoint, concentrating on differential operators. However, the vertex operator algebra structure that we construct is based on quasi-finite vector subspaces of the space of smooth functionals; it would be interesting to develop the vertex operator algebra of a completion like that in \cite{Hu99}, or of the full space of smooth functionals.

{\bf Virasoro algebra of differential operators on polynomials.} A close connection should exist between the present paper and works \cite{BB03,BB04,Kyt07}, which construct the Virasoro algebra and (in a slightly different language) parts of the Virasoro vertex operator algebra using certain differential operators on polynomials in infinitely-many unknown. These polynomials can be interpreted as certain functionals of germs of holomorphic functions, and these ideas were applied to the study of  Schramm-Loewner evolution (closely connected to CLE) in the works mentioned. The differential operators constructed are found from studying correlation functions of objects with a Virasoro (vertex operator) algebra structure, but are then studied independently from correlation functions. Our construction, again, differs in its origin, in that we construct differential operators from basic functional-analytic principles; as a consequence, we obtain seemingly more natural expressions for the Virasoro algebra elements. Further, we do describe the full vertex operator algebra, and we use the more general context of smooth functionals of conformal maps instead of restricting to certain polynomials. Yet, it would be extremely interesting to understand the full relation between our formulation and that of \cite{BB03,BB04,Kyt07}, especially in order to fully understand the meaning of our formulation in the SLE context.

\subsection{Organization of the paper}

In Section 2, we define the topological space $\Omega$ of conformal maps, and describe some of its differential structures, based on, but largely extending, those of \cite{DCFT}. In Section 3, we study certain conformal differential equations defining what we call conformal Gaussians. Conformal Gaussians will play the role of, but are much more general than, the relative partition function constructed in CFT and CLE, using which the stress-energy tensor was defined in the previous works \cite{DCLE,DCFT}. We obtain results about conformal differential equations that are interesting on their own. In Section 4, we suggest constructions for highest-weight and identity representations of the Virasoro algebra in terms of conformal derivatives, and we construct the Virasoro vertex operator algebra structure. In Appendix \ref{appvoa}, we recall the definition of a vertex operator algebra.

Throughout, conformal maps are bijective conformal functions from one domain to another, and a domain is a nonempty open proper subset of the Riemann sphere $\hC$ (on which we will use the global coordinates given by $\C\cup\{\infty\}$). We will use the notation $\C_0:= \C\setminus \{0\}$. We will also make use of the round metric on $\hC$ normalized so that the distance between $0$ and $\infty$ is 1, denoting the distance function by $d(\cdot,\cdot)$; we will use this symbol to represent both the distance between points in $\hC$ and the distance between subsets of $\hC$ (or a mix thereof).

{\bf Acknowledgements:} I am grateful to T. Koeppe and D. Bernard for fruitful discussions on the subject. I acknowledge support from EPSRC, Grant EP/H051619/1 ``From conformal loop ensembles to conformal field theory", First Grant scheme.

\sect{The topological space $\Omega$ and conformal derivatives} \label{sectOmega}

In this section, we construct an explicit space $\Omega$ of conformal maps on which complex functions (i.e.~functionals) and their derivatives will be considered throughout the rest of the paper. This space is characterized by an annular closed set that separates $0$ from $\infty$: it is the space of maps that are conformal on this set (that is, each map is conformal on some (open) neighbourhood of this set), such that the image has the same annular structure, still separating $0$ from $\infty$. The topology is that induced by the local topology on conformal maps near to the idendity discussed in \cite{DCFT}, and is Hausdorff. We define piecewise smooth paths as well as multiple conformal derivatives and holomorphic derivatives on this space.

\subsection{The space $\Omega$}

Let $E\subset \C_0$ be a continuum whose complement has exactly two simply connected components (an annular continuum): the interior, containing $0$, and the exterior, containing $\infty$. A conformal map $g$ on a neighbourhood of such a continuum $E$ will be said to be sense-preserving if 1) one component of $\hC\setminus g(E)$ contains 0 and the other contains $\infty$, and 2) the cluster values \cite{Pom75} of $g$ in the interior (resp.~exterior) of $E$ lie in the interior (resp.~exterior) of $g(E)$. Given such an $E$, the space $\Omega$ is the space of all sense-preserving conformal maps on neighbourhoods of $E$, up to the equivalence given by equality on $E$. Throughout we will fix $E$ and $\Omega$ as in this paragraph.

Given $\Sigma\in\Omega$, we will denote by $\alpha(\Sigma)$ (resp. $\beta(\Sigma)$) the closed set formed by the union of $\Sigma(E)$ with the interior (resp. the exterior) of $\Sigma(E)$. For uniformity of notation we will denote by $\gamma(\Sigma):=\Sigma(E)=\alpha(\Sigma)\cap \beta(\Sigma)$.

This space carries a natural action of the groupoid of conformal maps on annular domains: let $\Sigma\in\Omega$ and let $g$ be a conformal map on a domain that contains $\gamma(\Sigma)$. If $g$ is sense-preserving, then we say that it has an action on $\Sigma$, and we define $g\cdot\Sigma:=g\circ \Sigma$ (otherwise, the action is not defined). Clearly, this action agrees with the groupoid structure, and $\gamma(g\cdot\Sigma) = g(\gamma(\Sigma))$. Further, given any $\Sigma,\Sigma'\in\Omega$, there exists a unique (up to conformal continuation) conformal map $g$ such that $\Sigma'=g\cdot \Sigma$ (indeed, the map $g=\Sigma'\circ \Sigma^{-1}$). A conformal map on a simply connected domain containing $\alpha(\Sigma)$ (resp. $\beta(\Sigma)$) which has an action on $\Sigma$ will be referred to as an $\alpha$-map (resp. $\beta$-map) at $\Sigma$.

\begin{lemma}\label{lemfactor}
Let $\Sigma\in\Omega$ and $g$ be a conformal map with an action on $\Sigma$. Then there exist an $\alpha$-map $g_1$ at $\Sigma$ and a $\beta$-map $g_2$ at $g_1\cdot \Sigma$, such that $g=g_2\circ g_1$ on a neighbourhood of $\gamma(\Sigma)$.
\end{lemma}
\proof Let $g$ be a conformal map with an action on $\Sigma$. Then, by \cite{Hub66,Kuh70} (see also \cite{Dconf}), it is, on an annular neighbourhood of $\gamma(\Sigma)$, the composition $g_2\circ g_1$ of a conformal map $g_1$ on a simply connected neighbourhood of $\alpha(\Sigma)$ followed by a conformal map $g_2$ on a simply connected neighbourhood of $g_1(\gamma(\Sigma))$ whose complement lies in $g_1(\alpha(\Sigma))$. There exists a M\"obius map $m$ such that $m\circ g_1$ is sense-preserving on $\gamma(\Sigma)$. Indeed, $g_1(\gamma(\Sigma))$ is an annular continuum. Let $a$ be in the component of $\hC\setminus g_1(\gamma(\Sigma))$ where lie the cluster values of $g_1$ in the interior of $E$, and let $b$ be in the other component. We may choose $m(z)=(z-a)/(z-b)$: this sends $a$ to 0 and $b$ to $\infty$. Then, clearly $g_2\circ m^{-1}$ is sense-preserving on $(m\circ g_1)(\gamma(\Sigma))$, since $g=g_2\circ g_1$ is sense-preserving on $\gamma(\Sigma)$ by supposition. Hence, $m\circ g_1$ is an $\alpha$-map and $g_2\circ m^{-1}$ is a $\beta$-map. \eproof

\subsection{The topology on $\Omega$}

The topology on $\Omega$ is that induced, through the action of conformal maps on points in $\Omega$, by the local topology around the identity discussed in \cite{DCFT} (recalled below). We will see that this defines a topology that is strictly stronger than that of uniform convergence on $E$, and strictly weaker than the topology of compact convergence on any fixed domain that contains $E$.

Let $A\subset \hC$ be a domain. Let $D_n:n\in\N$ be a sequence of increasing compact sets $D_1\subset D_2 \subset\cdots$, such that $\cup_{n\in\N} D_n = A$. Note that if $K$ is a compact subset of $A$, then there exists a $n$ such that $K\subseteq D_n$. Given any conformal map $g$ on some domain $B$, we associate to it the positive real number
\[
	\delta_A(g):=\sum_{n=1}^\infty 2^{-n}\lt\{\ba{ll}
	{\rm max}\{d(g(z),z):z\in D_n\}& (D_n\subset B) \\
	1 & (\mbox{otherwise}). \ea\rt.
\]
The maximum value that this quantity can take is 1, and this value is attained if $g$ is a conformal map on a domain $B$ such that $B\cap A=\emptyset$. A $A$-neighbourhood of the identity is any set $\{g:\delta_A(g)<r\}$ for some $0<r<1$; all maps $g$ in such a set are conformal on a domain that intersect $A$. This can be used to define convergence towards the identity, hence it defines a local topology ($A$-topology) around the identity. This local topology is such that if $g_n:n\in\N$ is a sequence of conformal maps defined on $A$, then $g_n\to \id$ in the $A$-topology if and only if it does so in the compact (i.e.~local uniform) convergence topology on $A$. The $A$-topology is a compact convergence local topology that does not require all $g_n$ to be defined on $A$. Note that if $B\subseteq A$, then every $A$-neighbourhood is included inside a $B$-neighbourhood, and every $B$-neighbourhood contains a $A$-neighbourhood.

The following results are independent of the choices of sequences $D_n$ for every domain $A$.
\begin{lemma}\label{lem1}
Let $\Sigma\in\Omega$. For every domain $A$ such that $A\supset \gamma(\Sigma)$, there exists a $A$-neighbourhood $U$ of the identity such that $g$ has an action on $\Sigma$ for all $g\in U$.
\end{lemma}
\proof Let us consider the sequence $D_n$ as above, and choose $n$ such that $D_n\supseteq \gamma(\Sigma)$. Let $d_{min}:={\rm min}\{d(\gamma(\Sigma),0),d(\gamma(\Sigma),\infty)\}$, and $r:=\sum_{m=n}^\infty 2^{-m}d_{min}$. Then, the set $\{g:\delta_A(g)<r\}$ is a set of conformal maps $g$ defined on $\gamma(\Sigma)$, such that $d(g(z),z)<d_{min}$ for all $z\in\gamma(\Sigma)$. Hence, all such maps $g$ are sense-preserving, so that $g\cdot\Sigma\in\Omega$. \eproof
\begin{lemma}\label{lem2}
Let $A, B, C$ be domains such that $A\cup B\subseteq C$. If $U$ is a $A$-neighbourhood (of the identity) and $V$ is a $B$-neighbourhood, then there exists a $C$-neighbourhood $W$ such that $W\subseteq U\cap V$.
\end{lemma}
\proof Let $r$ and $s$ be such that $U=\{g:\delta_A(g)<r\}$ and $V=\{g:\delta_B(g)<s\}$, and let us assume without loss of generality that $r\leq s$. Let us consider the sequences $D_m,D_n',D_p''$, $m,n,p \in\N$ of increasing compact sets associated to $A,B,C$ respectively, and choose integers $n,p$ such that $D_p''\supseteq D_n\cup D_n'$. For any $n$, such a $p$ exists. Let $d_{min}$ be such that $r=\sum_{m=1}^n 2^{-m}d_{min} + \sum_{m=n+1}^\infty 2^{-m}$; let us choose $n$ large enough such that $d_{min}>0$. All maps $g$ defined on $D_n$ such that ${\rm max}\{d(g(z),z):z\in D_n\}<d_{min}$ lie in $U$, and all maps $g$ defined on $D_n'$ such that ${\rm max}\{d(g(z),z):z\in D_n'\}<d_{min}$ lie in $V$. Let $t=\sum_{m=p}^\infty 2^{-m} d_{min}$, and let $W=\{g:\delta_C(g)<t\}$. All maps $g$ that lie in $W$ are defined (and conformal) on $D_p''$, and are such that ${\rm max}\{d(g(z),z):z\in D_p''\}<d_{min}$. Hence, $W\subseteq U\cap V$.
\eproof

We define open sets in $\Omega$ as follows.
\begin{defi}\label{defiopen}
A subset ${\cal N}\subset \Omega$ is open if for every $\Sigma\in{\cal N}$ and for every domain $A\supset\gamma(\Sigma)$, there exists a $A$-neighbourhood $U$ of the identity such that $g\cdot \Sigma\subset{\cal N}$ for all $g\in U$.
\end{defi}
Clearly $\emptyset$ is open, and by Lemma \ref{lem1}, so is $\Omega$. It is immediate that any union (finite or infinite) of open sets is an open set. Thanks to Lemma \ref{lem2}, it also follows that any finite intersection of open sets is open. Hence, these open sets define a topology, which we will refer to as the conformal topology. Unless otherwise stated, topological concepts on $\Omega$ will refer to this topology.

\begin{rema}\label{remaann}{\em 
Since every $A$-neighbourhood is a subset of a $A'$-neighbourhood for $A\supseteq A'$, openness can be defined by requiring that for every simply connected domains $A\supset\alpha(\Sigma)$ and $B\supset\beta(\Sigma)$, there exists a $A\cap B$-neighbourhood $U$ of the identity such that $g\cdot \Sigma\subset{\cal N}$ for all $g\in U$.}
\end{rema}

\begin{lemma}\label{lemtopo}
The conformal topology on $\Omega$ is stronger than the topology given by uniform convergence of the maps $\Sigma\in\Omega$ on the closed set $E$.
\end{lemma}
\proof We want to show that every open set in the uniform convergence topology is open in the conformal topology. Let $\Sigma\in\Omega$. Let $0<d_{min}<1$, and let ${\cal B}:=\{\Sigma'\in\Omega:{\rm max}\{d(\Sigma'(z),\Sigma(z)):z\in E\}<d_{min}\}$ be an open ball of radius $d_{min}$ centered at $\Sigma\in\Omega$ in the uniform convergence topology. Let $A\supset \gamma(\Sigma)$ be a domain, and let $D_n:n\in\N$ be the increasing sequence of compact sets associated to $A$. Let us choose $n$ such that $D_n\supseteq \gamma(\Sigma)$, and let $r>0$ be such that $r<\sum_{m=n}^\infty 2^{-m} d_{min}$. Then, by Lemma \ref{lem1}, we may choose $r$ small enough such that all maps in the set $U:=\{g:\delta_{A}(g)<r\}$ have an action on $\Sigma$; and since $r<\sum_{m=n}^\infty 2^{-m} d_{min}$, we have that $g\cdot \Sigma$ lies in ${\cal B}$ for all $g\in U$. Since every point of an open set in the uniform convergence topology has an open-ball neighbourhood contained in that set, it follows that every open set in the uniform convergence topology is open in the conformal topology.
\eproof

\begin{corol}
The space $\Omega$ is Hausdorff.
\end{corol}
This holds because the uniform convergence topology is Hausdorff.

Lemma \ref{lemtopo} gives us many interesting open sets in $\Omega$; we will denote some of them by $\Omega(F)$ for $F$ closed subsets of $\hC$.
\begin{lemma}\label{lemOmegar}
Let $F$ be a closed subset of $\hC$. Then the subsets $\Omega(F):=\{\Sigma\in\Omega:\gamma(\Sigma)\cap F = \emptyset\}$, $\Omega(F)_\alpha:=\{\Sigma\in\Omega:\alpha(\Sigma)\cap F = \emptyset\}$ and $\Omega(F)_\beta:=\{\Sigma\in\Omega:\beta(\Sigma)\cap F = \emptyset\}$ are open.
\end{lemma}
\proof We will show that ${\cal N}=\Omega(F),\;\Omega(F)_\alpha,\;\Omega(F)_\beta$ are open in the uniform convergence topology; Lemma \ref{lemtopo} completes the proof. If $F=\emptyset$, or if $\hC\setminus F$ does not contain $\gamma(\Sigma)$ (resp.~$\alpha(\Sigma)$, $\beta(\Sigma)$) for any $\Sigma\in\Omega$, then the lemma is trivial; hence let us assume otherwise. We only need to show that every $\Sigma\in\cal N$ has an open-ball neighbourhood included inside $\cal N$. Let $\Sigma\in \Omega$ be such that $\gamma(\Sigma)\cap F$ (resp.~$\alpha(\Sigma)\cap F$, $\beta(\Sigma)\cap F$) $=\emptyset$, and let $d_{min}$ be the distance between $F\cup \{0,\infty\}$ (resp.~$F\cup \{\infty\}$, $F\cup \{0\}$) and $\gamma(\Sigma)$ (resp.~$\alpha(\Sigma)$, $\beta(\Sigma)$); this distance exists and is greater than 0 because both sets are closed and don't intersect. Then $\{\Sigma'\in\Omega:{\rm max}\{d(\Sigma'(z),\Sigma(z)):z\in E\}<d_{min}\}\subset \cal N$.
\eproof

The following lemma clarifies the meaning of the conformal topology by describing continuity at a point $\Sigma\in\Omega$ via the $A$-topology discussed above.
\begin{lemma}\label{lemcont}
Let $M$ be a topological space, let $p\in M$, and assume that $M$ has a local base at p. Then, the map $M\to\Omega : t\mapsto \Sigma_t$ is continuous at $p$ if and only if there exits a domain $A\supset \gamma(\Sigma_{p})$ such that $g_t:=\Sigma_t\circ \Sigma_{p}^{-1}$ converges to $\id$ in the $A$-topology as $t\to p$.
\end{lemma}
\proof Part 1: $\Leftarrow$. Assume that there exists a domain $A\supset\gamma(\Sigma_{p})$ such that $g_t\to\id$ in the $A$-topology. Then for every $A$-neighbourhood $U$ of $\id$ there exists a neighbourhood $T$ of $p$ such that $g_t\in U$ for all $t\in T$. Since for every open subset ${\cal N}\ni\Sigma_{p}$, there exists a $A$-neighbourhood $U$ of $\id$ such that $g\cdot\Sigma_{p}\subset {\cal N}$ for all $g\in U$, then for every such open subset ${\cal N}$, there exists a neighbourhood $T$ of $p$ such that $\Sigma_t\in {\cal N}$ for all $t\in T$. Hence, $t\mapsto \Sigma_t$ is continuous at $t=p$.

Part 2: $\Rightarrow$. Assume that there does not exist a domain $A\supset\gamma(\Sigma_{p})$ such that $g_t\to\id$ in the $A$-topology. That is, for every domain $A\supset\gamma(\Sigma_p)$, there exists a $A$-neighbourhood $U_A$ such that $\{g_t:t\in T\}\not\subseteq U_A$ for every neighbourhood $T$ of p. We will construct a neighbourhood ${\cal N}_p$ of $\Sigma_{p}$ with the property that $\{\Sigma_t:t\in T\}\not\subseteq {\cal N}_p$ for every neighbourhood $T$ of $p$.

If $g_t$ does not tend to $\id$ in the uniform convergence topology on $\gamma(\Sigma_p)$ as $t\to p$, then by Lemma \ref{lemtopo}, we may choose ${\cal N}_p$ to be a uniform-convergence open ball around $\Sigma_p$. Hence, let us assume on the contrary that $\lim_{t\to p}g_t=\id$ uniformly on $\gamma(\Sigma_{p})$.

We start by constructing a set of maps $V_{p}$ all of which have an action on $\Sigma_{p}$, with the properties that (a) for every domain $A\supset\gamma(\Sigma_{p})$, there exists a $A$-neighbourhood $U$ of $\id$ such that $U\subset V_{p}$, and (b) there exists no neighbourhood $T$ of {p} such that $\{g_t:t\in T\}\subseteq V_{p}$.

We recall the following fact noted above: (*) for every domains $A$ and $B$ such that $A\supseteq B$, and for every $B$-neighbourhood $U$ of $\id$, there exists a $A$-neighbourhood $U'$ of $\id$ such that $U'\subseteq U$.

Consider a decreasing sequence of domains $A_n\supset\gamma(\Sigma_{p}):n\in\N$ such that $\lim_{n\to\infty} A_n = \gamma(\Sigma_{{p}})$ (say, $A_n = \{z\in\hC:d(\gamma(\Sigma_p),z)<2^{-n}\}$). Let $r_n\in(0,1),\;n\in\N$ and define the $A_n$-neighbourhoods $U_n := \{g:\delta_{A_n}(g)<r_n\},\;n\in\N$. Let us construct $V:=\cup_{n\in\N} U_n$. If there does not exist a neighbourhood $T$ of ${p}$ such that $\{g_t:t\in T\}\subseteq V$, then we may choose $V_{p}=V$ which has property (b), and which by construction and thanks to the fact (*) also has property (a). Suppose on the contrary that there exists a neighbourhood $T$ of ${p}$ such that $\{g_t:t\in T\}\subseteq V$. Hence, for every $t\in T$, there exists a $n\in\N$ such that $g_t\in U_n$. For every $t\in T$, let $n_t$ be the smallest positive integer such that $g_t \in U_{n_t}$. Further, let us denote by $T_k:k\in\N$ a local base at ${p}$, with $T_{k+1}\subset T_k$ for all $k$.

At least one of the following two statements is true: (1) $\lim_{k\to\infty} \sup\{\delta_{A_{n_{t}}}(g_{t}):t\in T_k\}>0$, or (2) $\sup\{n_t:t\in T'\}=\infty$ for every neighbourhood $T'\subseteq T$ of ${p}$. Indeed: if (2) does not hold, then there exists a $T'\subseteq T$ and a domain $B\supset\gamma(\Sigma_{p})$ such that $A_{n_t}\supseteq B$ for all $t\in T'$, and there exists a $N<\infty$ such that $n_t\in\{1,\ldots,N\}$ for all $t\in T'$. If additionally (1) does not hold, then for every $A_n$-neighbourhoods $W_n$, $n=1,\ldots,N$, there exists a neighbourhood $T''\subset T'$ of $p$ such that $\{g_t:t\in T''\}\subset \cup_{n=1}^N W_n$. By the fact (*), we may choose $W_n$ small enough such that $\cup_{n=1}^n W_n\subset U_B$, which leads to a contradiction with the assumption of Part 2.

If (1) holds, then set $r:=\lim_{k\to\infty} \sup\{\delta_{A_{n_{t}}}(g_{t}):t\in  T_k\}$ as well as $U_n' := \{g:\delta_{A_n}(g)<r/2\},\;n\in\N$. Then in every neighbourhood $T'\subset T$ of $p$, there is a $t$ such that such $g_t\not\in U_n'$ for all $n\in \N$. Let $V_{p}=\cup_{n\in\N} U_n'$. Then the last statement implies that there is no neighbourhood $T'$ of $p$ such that $\{g_t:t\in T'\}\subseteq V_{p}$; this is property (b) above, and property (a) holds by construction and thanks to the fact (*).

On the other hand, suppose that (2) holds. For every $k,n\in \N$, we construct a $A_n$-neighbourhood $U_n^{(k)}$, and for every $k\in\N$ and $t\in T$, an integer $n^{(k)}_t$, as follows. We set $U_n^{(1)} = U_n,\;n\in\N$, and we define $n_t^{(k)}:=\min\{n\in\N:g_t\in U_n^{(k)}\}$ (whence $n_t^{(1)}=n_t$). We define $U_n^{(k+1)}$ in terms of $U_n^{(k)}$ for all $k\geq 1$ as follows. Assume that statement (2) holds for $n_t^{(k)}$ in place of $n_t$. This holds for $k=1$, and we will show by induction that it is true for all $k$. Let us consider the smallest integer $m_1$ such that $m_1>k$ and that $n_{t'}^{(k)}=m_1$ for some $t'\in T_k\cap T$ (and choose such a $t'$). Since statement (2) holds for $n_t^{(k)}$ in place of $n_t$, then $m_1$ exists. Clearly, $g_{t'}\neq \id$ since $g_{t'}\not\in U_n^{(k)}$ for all $n<m_1$; whence $\delta_{A_{n}}(g_{t'})>0$ for all $n$. We set $U_{m_1}^{(k+1)}:=\{g:\delta_{A_{m_1}}(g)<\delta_{A_{m_1}}(g_{t'})\}$. Let us consider all the integers $m_2<m_3<\ldots$ such that $g_{t'}\in U_{m_\ell}^{(k)}$, $\ell=2,3,\ldots$. For every $\ell=2,3,\ldots$, we set $U_{m_\ell}^{(k+1)}:=\{g:\delta_{A_{m_\ell}}(g)<\delta_{A_{m_\ell}}(g_{t'})\}$. Finally, we set $U^{(k+1)}_{n} = U_n^{(k)}$ for all $n\in\N\setminus\{m_1,m_2,\ldots\}$. Then by construction, $\{g_t:t\in T_k\cap T\}\not\subseteq \cup_{n\in\N}U_n^{(k+1)}$. In particular, since the $A_n$-neighbourhoods have been reduced or kept the same, $n_t^{(k+1)}\geq n_t^{(k)}$ for all $t\in T$, whence statement (2) holds for $n_t^{(k+1)}$ in place of $n_t$. This completes the construction of $U_n^{(k)}$ for all $n,k\in\N$. By construction, for every $n\in\N$, the limit $\lim_{k\to\infty} U_n^{(k)}$ exists, and is given by $U_n^{(n)}$. We set $V_{p}=\cup_{n\in\N}U_n^{(n)}$. This has the property that $\{g_t:t\in T_k\cap T\}\not\subseteq V_{p}$ for all $k\in \N$, hence it has property (b). Again by construction, and thanks to the fact (*), it also has property (a).

Hence we have constructed $V_{p}$ with properties (a) and (b) above. The neighbourhood ${\cal N}_{p}$ of $\Sigma_{p}$ is built by taking ${\cal K}:=\{g\cdot \Sigma_{p}:g\in V_{p}\}$ and attaching to every $\Sigma\in {\cal K},\;\Sigma\neq\Sigma_{p}$ a uniform-convergence open ball ${\cal M}_\Sigma$ centered at $\Sigma$. Certainly, ${\cal K}$ is such that for every domain $A\supset\gamma(\Sigma_{p})$, there exists a $A$-neighbourhood $U$ of the identity such that $g\cdot\Sigma\subset {\cal K}$ for all $g\in U$, thanks to property (a) of the set $V_{p}$. Further, by Lemma \ref{lemtopo}, the set ${\cal M}:=\cup_{\Sigma\in{\cal K}\setminus\{\Sigma_{p}\}}{\cal M}_{\Sigma}$ is open. Hence, the set ${\cal N}_{p}:=\{\Sigma_{p}\}\cup {\cal M}$ is also open. The open balls ${\cal M}_\Sigma$ are chosen as follows. Clearly, $\{\Sigma_t:t\in T\}\not\subseteq {\cal K}$ for every neighbourhood $T$ of $p$, thanks to property (b) of the set $V_{p}$. Then, let $t_k\in T_k\cap T:k\in \N$ be such that $\Sigma_{t_k}\not\in {\cal K}$ for every $k\in\N$. Further, for every $\Sigma\in {\cal K}$, and for every $k\in\N$, let $d_k(\Sigma)$ be the uniform-convergence distance between $\Sigma_{t_k}$ and $\Sigma$. We have $d_k(\Sigma)>0$, otherwise we would have $\Sigma_{t_k}=\Sigma\in{\cal K}$ (a contradiction) by the fact that the uniform-convergence topology is Hausdorff. For every $\Sigma\in{\cal K}$, let $d_\infty(\Sigma)$ be the uniform-convergence distance between $\Sigma$ and $\Sigma_{p}$. Again, $d_\infty(\Sigma)>0$. Since $\Sigma_t\to\Sigma_{p}$ in the uniform-convergence topology as $t\to{p}$ by assumption, then $d_k(\Sigma)\to d_\infty(\Sigma)$ as $k\to\infty$. Hence, $d_{\rm inf}(\Sigma):=\inf\{d_k(\Sigma):k\in\N\}>0$. The open ball ${\cal M}_\Sigma$ is the uniform-convergence ball of radius $d_{\rm inf}(\Sigma)/2$ centered at $\Sigma$. By construction, we then have $\Sigma_{t_k}\not\in{\cal M}$, as well as $\Sigma_{t_k}\neq\Sigma_{p}$, for every $k\in\N$, whence $\{\Sigma_t:t\in T\}\not\subseteq {\cal N}_{p}$ for every neighbourhood $T$ of {p}.
\eproof

Let us display an example of a set of conformal maps $g_t$ such that $g_t\to\id$ uniformly on a compact set but not compactly on any domain containing this set. Let the compact set be the closed annulus $D:=\{z\in\C:1\leq |z|\leq r\}$ (for some $0<r<1$). Consider $g_t(z):=z+t^2/(z-1-|t|-t^2)$ for $t\in(-1,1)$. For every $t$, $g_t$ is conformal on neighbourhoods of $D$; but if $t\neq 0$, it is not conformal on those that contain the point $t^2+1$ or $t^2+2|t|+1$. Since both these points approach $D$ as $t\to0$, then $g_t$ does not converge to $\id$ in the $A$-topology for any neighbourhood $A$ of $D$. However, for $z\in D$, we have $d(g_t(z),z)\leq |g_t(z)-z|= |t|^2/|z-1-|t|-t^2|\leq|t|^2/(|t|+t^2) \leq |t|$, whence $g_t$ tends to $\id$ uniformly on $D$.

\begin{rema}\em The example given above guarantees that in the proof of Lemma \ref{lemcont} (Part 2), we have constructed an open neighbourhood ${\cal N}_p$ which is not open in the uniform-convergence topology. In particular, it is constructed by appending the single point $\Sigma_p$ at the ``boundary'' of the uniform-convergence open set ${\cal M}$. This shows that the conformal topology is indeed strictly stronger than the uniform-convergence topology on $E$.
\end{rema}

\subsection{Smooth paths}

We introduce the notion of smooth paths, which will be closely connected with that of differentiability for complex functions on $\Omega$ defined in the next subsection.

The standard notion of smooth paths, and of differentiability, is naturally obtained from a manifold structure. However, the development of a manifold structure on $\Omega$ is beyond the scope of the present paper. Our notion of smooth paths is simply induced from the differentiable structure of the underlying Riemann surfaces mapped by the conformal maps in $\Omega$. Let $M$ be a neighbourhood of 0 in $\R$; note that $M$ is a topological space satisfying the condition of Lemma \ref{lemcont}. Let $M\to \Omega: t\mapsto \Sigma_t$ be continuous at $0\in M$. Then, thanks to Lemma \ref{lemcont}, there exists a domain $A\supset \gamma(\Sigma_0)$ such that $\Sigma_t\circ\Sigma_0^{-1}(z)$ is continuous at $t=0$ for every $z\in A$; and this continuity is uniform on any compact subset of $A$. The function $t\mapsto \Sigma_t\circ\Sigma_0^{-1}(z)$, for fixed $z\in A$, is continuous at $t=0$ in the topology of the (one complex dimensional) manifold $A$, hence it is natural to ask for it to be differentiable at 0. Further, in view of Lemma \ref{lemcont}, it is also natural to ask for a stronger condition: that differentiability hold at 0 uniformly on any compact subset of $A$. This leads to our notion of a differentiable path at 0. A smooth path will be such that the derivative thus obtained is continuous.

Note that the derivative of $\Sigma_t\circ\Sigma_0^{-1}(z)$ at $t=0$ is naturally a holomorphic vector field on $A$ (that is, a $(-1,0)$-differential). This can be seen both as a vector space over $\R$ or over $\C$; for convenience in treating the derivative, we will consider it over $\R$. In global coordinates, the space of holomorphic vector fields is the space of all holomorphic functions on $A$, plus $\C \{z\mapsto z\}\oplus \C\{z\mapsto z^2\}$ (with the canonical identification $\C\simeq \R\oplus i\R$) if $\infty\in A$. We will denote the space holomorphic vector fields on $A$ by $\spa^>(A)$, with the topology of (local-coordinate) compact convergence on $A$ (this topology is coordinate independent).
\begin{defi}\label{defismoothpath}
Let $M$ be a neighbourhood of $p\in\R$, and let $\sigma:M\to \Omega:t\mapsto\Sigma_t$. If $\sigma$ is continuous at $p$, then let $A$ be as in Lemma \ref{lemcont}. Note that from Lemma \ref{lemcont}, there exists a neighbourhood $T$ of p and a domain $B\supset\gamma(\Sigma_p)$, $B\subseteq A$ such that $\Sigma_t\circ\Sigma_p^{-1}$ is a conformal map on $B$ for all $t\in T$. 
\begin{itemize}
\item We say that $\sigma$ is differentiable at $p$ if it is continuous at $p$ and if $\Sigma_t\circ\Sigma_p^{-1}$ is differentiable at $t=p$ uniformly on every compact subset of $A$. We will denote
\beq
	{\cal D}\sigma(p):=\lt.\frc{d}{dt}\Sigma_t\circ\Sigma_p^{-1}\rt|_{t=p},
\eeq
and we remark that ${\cal D}\sigma(p)\in\spa^>(A)$. This will be refereed to as the direction of $\sigma$ at $p$. We say that $\sigma$ is continuously differentiable at $p$ if for every $t'\in T$, $\Sigma_t\circ\Sigma_{t'}^{-1}$ is differentiable at $t=t'$ compactly on $B$, and ${\cal D}\sigma$ is continuous at $p$ compactly on $B$.
\item We say that $\sigma$ restricted to a compact subset $T\subset M$ is a path if it is continuous at every point in $T$; and that it is a smooth path if it is continuously differentiable at every point in $T$.
\item We say that $\sigma$ restricted to a compact subset $T\subset M$ is a (smooth) $\alpha$-path (resp.~$\beta$-path) if for every $p\in T$, the domain $A$ can be chosen such that $A\supset  \alpha(\Sigma_p)$ (resp.~$A\supset\beta(\Sigma_p)$).
\item We say that $\sigma$ restricted to a compact subset $T\subset M$ is a piecewise-smooth path (resp.~$\alpha$-path, $\beta$-path) if it is a concatenation of a finite number of smooth paths (resp.~$\alpha$-path, $\beta$-path).
\end{itemize}
\end{defi}
Let $\sigma|_T$ be a path as above, $p\in M$, and $g_t:=\Sigma_t\circ\Sigma_p^{-1}$. Then there exists a $B\supset \gamma(\Sigma_p)$ such that $g_t$ is continuous at every $t\in T$, compactly on $B$. If the path is smooth, then
\beq\label{gtsigma}
	\frc{d}{dt} g_t = {\cal D}\sigma(t)\circ g_t.
\eeq
Further, in this case, there exists a $B\supset\gamma(\Sigma_p)$ such that $g_t$ is differentiable at every $t\in T$, compactly on $B$; and for every $t$ there exists a $B_t\supset\gamma(\Sigma_p),\;B_t\subseteq B$ such that the derivative of $g_{t'}$ is continuous at $t'=t$ compactly on $B_t$. Similar statements hold for $\alpha$-paths and $\beta$-paths, were $\gamma(\Sigma_p)$ is replaced by $\alpha(\Sigma_p)$ and $\beta(\Sigma_p)$ respectively.

If we define $h_t$ for $t\in T$ via
\beq\label{gt}
	\frc{d}{dt} g_t = h_t \p g_t\quad (g_p=\id),
\eeq
then clearly $h_t$ is related to the direction of $\sigma$ at $t$ by $h_t = ({\cal D}\sigma(t)\circ g_t)/\p g_t$. By this relation and the comments above, $h_t$ is a holomorphic vector field on some domain $B\supset\gamma(\Sigma_p)$ for all $t\in T$, and for every $t\in T$, $h_{t'}$ is continuous at $t'=t$ compactly on some domain $B_t\supset\gamma(\Sigma_p),\;B_t\subseteq B$. A certain inverse of this will be quite useful, because it will allow us to apply Loewner's Theorem: given $h_t$ a holomorphic vector field on some domain $B\supset\gamma(\Sigma_p)$ for all $t\in T$, that is continuous for all $t\in T$ compactly on $B$, if the unique solution $g_t$ to (\ref{gt}) has an action on $\Sigma_p$ for all $t\in T$, then $t\mapsto g_t\cdot \Sigma_t$ is a smooth path on $T$. If the unique solution gives rise to $\alpha$-maps (resp.~$\beta$-maps) $g_t$ at $\Sigma_t$, then $t\mapsto g_t\cdot \Sigma_t$ is a smooth $\alpha$-path (resp.~$\beta$-path).

Note that given any holomorphic vector field $h$ on a domain $A\not\ni\infty$, the function $g_t(z) = z + t h(z)$ ($z\in\C$) is such that $\lim_{t=0}g_t=\id$ in the $A$-topology, and that there exists a $B\subset A$ such that the derivative with respect to $t$ exists and is continuous compactly on $B$ for every $t$ in a neighbourhood of 0, and gives $h$ at $t=0$. If $\infty\in A$, one can use conformal transport by a M\"obius map; in global coordinates,
\beq\label{pathinfty}
	g_t(z) = a + \frc{z-a}{1-\frc{t}{z-a}h(z)}=:G_{t;a}^{h}(z)
\eeq
for some $a\in\hC\setminus A$. Note that in the limit $a\to\infty$, this indeed reproduces $g_t(z) = z + th(z)$; we will use the symbol $G_{t;a}^{h}(z)$ in general for describing the function (\ref{pathinfty}) if $a\neq \infty$ and the function $z+th(z)$ if $a=\infty$. The maps $G_{t;a}^h$ then allow us to construct, for every $\Sigma\in\Omega$ and every $h\in\spa^>(A),\;A\supset\gamma(\Sigma)$, a smooth path $\sigma$ passing by $\Sigma$ at $0$ in the direction ${\cal D}\sigma(0)=h$:
\[
	\sigma: t\mapsto G_{t;a}^h\cdot\Sigma,\quad
	a\in\hC\setminus A.
\]

The existence of piecewise smooth paths connecting any two points of a subset is important for going from differential equations to global properties of functions. Such subsets will be referred to as piecewise smoothly connected (with the $\alpha$- and $\beta$- versions):
\begin{defi}\label{deficonnect}
A subset ${\cal N}\subseteq\Omega$ is piecewise $\alpha$-smoothly ($\beta$-smoothly) connected if every $\Sigma\in{\cal N}$ and $\Sigma'=g\cdot\Sigma\in{\cal N}$, where $g$ is an $\alpha$-map (resp.~$\beta$-map) at $\Sigma$, are connected by a piecewise smooth $\alpha$-path (resp.~$\beta$-path) in ${\cal N}$. Further, ${\cal N}$ is piecewise smoothly connected if every $\Sigma,\Sigma'\in{\cal N}$ are connected by a piecewise smooth path in $\cal N$.
\end{defi}

\begin{lemma}\label{lemconnect}
Let $K\subset\C$, $0\in K$ be a compact set such that $\max\{x\in [0,\infty): xz\in K\}=\min\{x\in [0,\infty):xz\in K\}$ for all $z\in \C_0$ (for instance, $K$ is a closed disk centered at 0). The open set $\Omega(K)$ is piecewise $\alpha$- and $\beta$-smoothly connected, as well as piecewise smoothly connected.
\end{lemma}
\proof Part 1. The case $K=\{0\}$, that is, the space $\Omega$.

(a) Piecewise $\alpha$-smoothness.

Let $g$ be an $\alpha$-map at $\Sigma\in\Omega$. Let $T$ be a translation by $-g(0)$, $R$ a rotation about the origin by an angle $-{\rm arg}\,(\p g(0))$, and $S$ a scaling transformation about the origin such that $(S\circ R\circ T\circ g)(\alpha(\Sigma))\supset \alpha(\Sigma)$. The map $S$ exists because $\infty\not\in\alpha(\Sigma)$ and $0\in\alpha(\Sigma)$, the same holds for the set $g(\alpha(\Sigma))$, and $T$ and $R$ preserve $\infty$ and $0$. Then, the map $\t{g}:=S\circ R\circ T\circ g$ maps a neighbourhood $A$ of $\alpha(\Sigma)$ to a domain that contains $A$, and has the properties $\t{g}(0)=0$ and $\p\t{g}(0)>0$. Further, since scaling transformations preserve $\infty$ and 0, we have $\infty\not\in\t{g}(\alpha(\Sigma))$ and $0\in\t{g}(\alpha(\Sigma))$. Hence, in particular, $\t{g}$ is an $\alpha$-map.

By Loewner Theorem \cite[Thm 6.2]{Pom75}, there exists a continuous (in the compact convergence topology) $t\mapsto h_t$, where $h_t$ is holomorphic on $A$ for all $t\in[0,1]$, such that (\ref{gt}) holds with $g_1=\t{g}$, with the property that $g_t(0)=0$ and $\p g_t(0)>0$ for all $t\in[0,1]$. The maps $g_t$ are conformal on $A$ and map to domains $g_t(A)$ that are continuously increasing (in the Carath\'eodory topology). Hence, $g_t$ is an $\alpha$-map at $\Sigma$ for all $t$, so that we may define $\Sigma_t := g_t\cdot\Sigma$, and $t\mapsto\Sigma_t$ for $t\in[0,1]$ is a smooth $\alpha$-path connecting $\Sigma_0=\Sigma$ to $\Sigma_1=\t{g}\cdot\Sigma$.

Further, the map $g_t':=(R^{-1}\circ S^{-1})^{t}$ is such that $t\mapsto g_t'\cdot\Sigma'$ is smooth for any $\Sigma'\in\Omega$, hence in particular for $\Sigma' :=\Sigma_1$. Then, we can define $\Sigma_{t+1} := g_t'\cdot\Sigma_1$ for $t\in[0,1]$, and $t\mapsto \Sigma_t$ for $t\in [0,2]$Êis a piecewise smooth $\alpha$-path connecting $\Sigma$ to $\Sigma_2=(R^{-1}\circ S^{-1}\circ \t{g})\cdot\Sigma=(T\circ g)\cdot\Sigma$.

Finally, given two points in a domain, there exists a segmental arc, lying in that domain, connecting them (because there is a simple polygon separating the two points from the complement of the domain \cite[Thm 3.3]{N61}). Hence, in the domain $\hC\setminus \beta(g\cdot \Sigma)$, there exists a path $t\mapsto z_t$ connecting $z_0=0$ to $z_1=g(0)$, that is piecewise linear. Let $T_z$ be a translation by $z\in \C$. Then the map $g_t:=T_{z_t}$ is such that $\hC\setminus \beta(g_t\cdot \Sigma_2)$ contains $0$ for all $t\in[0,1]$. Since translations preserve $\infty$, then $g_t$ is an $\alpha$-map for all $t$. Equation (\ref{gt}) holds for $h_t=const.$ on every segment of the arc formed by $t\mapsto z_t$, whence we may define $\Sigma_{t+2}:=g_t\cdot\Sigma_2$ for $t\in[0,1]$. Hence, $t\mapsto \Sigma_t$ for $t\in[0,3]$ is a piecewise smooth $\alpha$-path on $[0,3]$ connecting $\Sigma$ to $\Sigma_3 = g\cdot \Sigma$.

(b) Piecewise $\beta$-smoothness.

Let $g$ be a $\beta$-map at $\Sigma$ such that $\Sigma'=g\cdot\Sigma$. Then, with $m(z):=1/z$, the map $\t{g}:=m\circ g\circ m$ is an $\alpha$-map at $m\cdot\Sigma\in\Omega$ such that $m\cdot\Sigma'=\t{g}\cdot m\cdot\Sigma$. By (a), there exists a piecewise smooth $\alpha$-path from $m\cdot\Sigma$ to $m\cdot\Sigma'$; let $t\mapsto \t{g}_t\cdot (m\cdot \Sigma)$ be this path. Then with $g_t:=m\circ\t{g}_t\circ m$, $t\mapsto g_t\cdot \Sigma$ is a piecewise smooth $\beta$-path from $\Sigma$ to $\Sigma'$.

(c) Piecewise smoothness.

Given any $\Sigma,\Sigma'\in\Omega$, there is a conformal map $g$ acting on $\Sigma$ such that $\Sigma'=g\cdot\Sigma$. By Lemma \ref{lemfactor}, we can write $\Sigma' = g_2\cdot g_1\cdot\Sigma$ where $g_1$ is an $\alpha$-map and $g_2$ is a $\beta$-map. By (a) and (b), there exists a piecewise smooth path connecting $\Sigma$ to $g_1\cdot\Sigma$, and one connecting $g_1\cdot\Sigma$ to $g_2\cdot g_1\cdot\Sigma$.

Part 2. The case $K\neq \{0\}$.

Let $\Sigma,\Sigma'\in\Omega(K)$ and let $r=\max\{d(z,0):z\in K\}$. Let $[0,1]\to \Omega : t\mapsto \Sigma_t$ be a piecewise smooth path (resp.~$\alpha$-path, $\beta$-path) from $\Sigma_0=\Sigma$ to $\Sigma_1=\Sigma'$. This path exists by (c) (resp.~(a), (b)). Let $d_{min}:={\rm min}\{d(\gamma(\Sigma_t),0):t\in[0,1]\}$, and let $\t{d}_{min}$ and $\t{r}$ be the flat (i.e.~plane)-metric radii of discs centered at 0 of round-metric radii $d_{min}$ and $r$ respectively. Since $\Sigma_t\in \Omega$ for all $t\in[0,1]$, since $[0,1]$ is closed, since $t\mapsto\Sigma_t$ is continuous, and since $d(\gamma(\cdot),0):\Omega\to[0,1]$ is continuous by Lemma \ref{lemtopo}, $d_{min}$ exists and is greater than 0, hence so does $\t{d}_{min}$. Let $S$ be the scaling transformation $S(z) = 2\t{r}z/\t{d}_{min}$. Then, the scaled path $t\mapsto S\cdot\Sigma_t$ is a piecewise smooth path (resp.~$\alpha$-path, $\beta$-path) from $S\cdot \Sigma$ to $S\cdot\Sigma'$, such that each piece of the original path that is smooth, also is in the scaled path. The scaled path has the property that  $d(\gamma(S\cdot\Sigma_t),0)>r$ for all $t\in[0,1]$. Hence, $S\cdot\Sigma_t\in\Omega(K)$ for all $t\in[0,1]$. Further, if $z\not\in K$ then clearly $|z|>\min\{x\in[0,\infty):xz/|z|\in K\}$, whence by the property stated in the lemma, $|z|>\max\{x\in[0,\infty):xz/|z|\in K\}$, whence $S^t(\gamma(\Sigma))$ does not intersect $K$ for all $t\in[0,1]$. Hence, the path $[0,1]\to \Omega:t\mapsto S^t\cdot\Sigma$, which is both a smooth $\alpha$- and $\beta$- path from $\Sigma$ to $S\cdot\Sigma$, entirely lies in $\Omega(K)$; likewise for the path $[0,1]\to\Omega: t\mapsto S^t\cdot \Sigma'$. Whence there exists a piecewise smooth path (resp.~$\alpha$-path, $\beta$-path) lying in $\Omega(K)$ that connects $\Sigma$ and $\Sigma'$.
\eproof

\subsection{Conformal derivatives}

Let us now consider complex functions on $\Omega$ and their derivatives. Our notion of differentiability of complex functions on $\Omega$ will essentially be based on that of differentiability of functions $\sigma:M\to\Omega:t\mapsto\Sigma_t$ for $M$ a neighbourhood of $0\in\R$, and on a generalization of this concept to $\R^k$ for $k\in\N$. This definition of differentiability together with our notion of smooth paths above is sufficient to make it a powerful enough concept from which interesting results can be obtained. The definition of differentiability from differentiable functions $\sigma$ is in a sense a ``minimal'' one, in such a way that it is likely that the results of this paper can be used in further investigations of the manifold structure of $\Omega$.

\subsubsection{Single differentiability}

Differentiability on $\Omega$ is induced from the concept of conformal differentiability defined in \cite{DCFT}, which we express here for general (not necessarily simply connected) domains. Let ${\cal N}\subseteq\Omega$ be an open subset, let $\Sigma\in{\cal N}$, and let $A\supset\gamma(\Sigma)$ be a domain. A function $f$ is $A$-differentiable at $\Sigma$ if there exists a continuous $\R$-linear functional $\nabla^A f(\Sigma)$ (the differential of $f$ at $\Sigma$) on the (real-linear) space $\spa^>(A)$ of holomorphic vector fields on $A$ such that, for every $\sigma:M\to\Omega: t\mapsto \Sigma_t$ ($M$ some neighbourhood of 0 in $\R$) differentiable at $t=0$, with $\Sigma_0=\Sigma$ and ${\cal D}\sigma(0)\in \spa^>(A)$, the following derivative exists and gives
\beq\label{defder}
	\lt.\frc{df(\Sigma_t)}{dt}\rt|_{t=0}
	= \nabla^A f(\Sigma)\, h
\eeq
where $h:={\cal D}\sigma(0)$ is, as above, the derivative of $\Sigma_t\circ\Sigma^{-1}$ with respect to $t$ at $t=0$\footnote{In \cite{DCFT}, only the condition of right-differentiability was considered. However, it is simple to see that the definitions are equivalent.}, which exists compactly on $A$.

In general, it will be convenient to use the notation
\[
	 \nabla_h f(\Sigma):=\nabla^A f(\Sigma) h.
\]
In this notation, the superscript $A$ is not necessary, as enough information about the domain $A$ is encoded into the vector field $h$ (see \cite[Prop~3.7]{DCFT}). In particular, we note that if $h$ is holomorphic on $A\supset B$ and $f$ is $B$-differentiable at $\Sigma$, then $f$ is also $A$-differentiable at $\Sigma$ and $\nabla^Af(\Sigma)h = \nabla^Bf(\Sigma)h$. For convenience, we will also use the notation
\[
	\nabla_{h\,|\,g} f(g\cdot\Sigma):= \nabla_h f(\Sigma);
\]
here, $\nabla_{h\,|\,g}$ should be understood as a conformal derivative with respect to variations of the map $g$, evaluated at $g=\id$.

\begin{defi}\label{defidiff}
A function $f$ on an open subset ${\cal N}\subseteq\Omega$ is differentiable at $\Sigma\in{\cal N}$ if it is $A$-differentiable at $\Sigma$ for every domain $A\supset\gamma(\Sigma)$. It is $\alpha$-differentiable (resp.~$\beta$-differentiable) at $\Sigma$ if it is $A$-differentiable at $\Sigma$ for every domain $A\supset\alpha(\Sigma)$ (resp.~$A\supset\beta(\Sigma)$).
\end{defi}

\begin{rema}\label{remaanndiff}{\em 
In line with Remark \ref{remaann}, the requirement in the definition of differentiability can be replaced by requiring that for every simply connected domains $A\supset\alpha(\Sigma)$ and $B\supset\beta(\Sigma)$, the function $f$ is $A\cap B$-differentiable.}
\end{rema}
The following lemma is immediate.
\begin{lemma}\label{lemchainrule}
Let $\sigma:t\mapsto \Sigma_t,\;t\in[0,1]$ be a smooth path lying in an open subset ${\cal N}\subseteq\Omega$, and let $f$ be a function on ${\cal N}$. If $f$ is differentiable at $\Sigma_t$ for all $t\in[0,1]$, then for all such $t$,
\beq\label{chainrule2}
	\frc{d}{dt} f(\Sigma_t) =
	\nabla_{{\cal D}\sigma(t)} f(\Sigma_t).
\eeq
\end{lemma}
From this, we find that if ${\cal N}$ is a piecewise smoothly connected open set and $f$ is differentiable on ${\cal N}$, then $f$ is a constant if and only if its derivative is 0 everywhere on ${\cal N}$.

\subsubsection{Multiple differentiability and smoothness}

Let $A\supset\gamma(\Sigma)$ be a domain. The concept of $k$ times $A$-differentiability of $f$ at $\Sigma$, for $k\in \N$, is defined recursively as follows. Let us denote by $\N_k=\{1,2,\ldots,k\}$ the set of the first $k$ integers, and by $t:=(t_1,\ldots,t_k)$ points in $\R^k$. For every subset $S=\{j_1,j_2,\ldots,j_{|S|}\} \subseteq \N_k$ (with by convention $j_1<j_2<\cdots<j_{|S|}$), let us use the notation $g_S= g_{t}^{(j_1)}\circ \cdots \circ
 g_{t}^{(j_{|S|})}$.
\begin{defi}\label{defiFkA} We denote by $\spag^{k}(A)$ the set of functions $t\mapsto (g_{t}^{(1)},\ldots,g_{t}^{(k)})$ for $t\in\R^{k}$ in a neighbourhood of $(0,\ldots,0)$, where $g_{t}^{(j)}$ are conformal maps, with two conditions:
\begin{enumerate}
\item $g_{t}^{(j)}\to \id$ as $t\to(0,\ldots,0)$ in the $A$-topology for every $j$;
\item defining $h^{(j)}_t$ by (in local coordinates) $g_t^{(j)}(z) =: z + t_jh_t^{(j)}(z)$, we have that $\lim_{t\to(0,\ldots,0)} h_t^{(j)} = h^{(j)}$ exists uniformly on every compact subset of $A$  for every $j$; these requirements are coordinates independent.
\end{enumerate}
\end{defi}
Note that for every $\{t\mapsto (g_t^{(1)},\ldots,g_t^{(k)})\}\in\spag^{k}(A)$, we have that:
\begin{itemize}
\item[(a)] $g_S\cdot\Sigma\in{\cal N}$ and for all $t$ near enough to $(0,\ldots,0)$ by Definition \ref{defiopen} of open sets;
\item[(b)] $h^{(j)}$, $j=1,\ldots,k$ are holomorphic vector fields on $A$.
\end{itemize}
Note also that by Lemma \ref{lemcont}, if $g_t^{(j)}\cdot\Sigma \to\Sigma$ as $t\to(0,\ldots,0)$ for all $j\in\N_k$, then there exists a $A$ as in Point 1 of the definition of $\spag^{(k)}(A)$. Note finally that in the case $k=1$, the function $\sigma:t\mapsto g_t^{(1)}\cdot\Sigma$ is differentiable at $t=0$, with $\sigma(0)=\Sigma$ and ${\cal D}\sigma(0)=h^{(1)}$. Hence, the definition below specializes to that of single differentiability at $k=1$.
 
\begin{defi}\label{defikAdiff}
Let ${\cal N}\subseteq\Omega$ be an open subset, $\Sigma\in{\cal N}$, and $A\supseteq\gamma(\Sigma)$ be a domain. The function $f$ on ${\cal N}$ is $k$ times $A$-differentiable at $\Sigma$, for $k\in\N$, if:
\begin{enumerate}
\item There exists a continuous $\R$-linear functional $\nabla^{A} f(\Sigma)$ on $\lt(\spa^>(A)\rt)^n_{\otimes_\R}:=\spa^>(A)\otimes_\R \cdots\otimes_\R \spa^>(A)$ ($n$ times), the $k^{\rm th}$ differential of $f$ at $\Sigma$, such that for every $\{t\mapsto(g_t^{(1)},\ldots,g_t^{(k)})\}\in \spag^k(A)$, the following limit exists and gives
\beq\label{multder}
	\lim_{t\to (0,\ldots,0)}
	\frc1{\prod_{j=1}^k t_j}
	\sum_{S\subset \N_k} (-1)^{k-|S|} f(g_S\cdot\Sigma)
	= \nabla^{A} f(\Sigma)\, (h^{(k)},\ldots,h^{(1)}),
\eeq
where $h^{(j)}$ is the derivative of $g^{(j)}_t$ with respect to $t_j$ at $t=0$.
\item \label{exder} If $k>1$, then there exists a neighbourhood ${\cal M}$ of $\Sigma$, with the property that $A\supset\gamma(\Sigma')$ for all $\Sigma'\in{\cal M}$ (for instance, thanks to Lemma \ref{lemtopo}, a small enough uniform-convergence ball centered at $\Sigma$), such that $f$ is $k-1$ times $A$-differentiable on ${\cal M}$.
\end{enumerate}
\end{defi}
We will also use the notation
\[
	\nabla_{(h^{(k)},\ldots,h^{(1)})} f(\Sigma):=
	\nabla^{A} f(\Sigma)\, (h^{(k)},\ldots,h^{(1)}).
\]

The existence of the limit in (\ref{multder}) implies that all $(k-1)^{\rm th}$ derivatives (which exist thanks to the requirement \ref{exder}) are continuous and differentiable; hence in particular that the $(k-1)^{\rm th}$ differential is continuous at $\Sigma$ in the weak* topology.

Equation (\ref{multder}) in fact gives stronger conditions than what can be inferred merely from this topology on the differential. For instance, let us choose, in global coordinates, $h_t^{(j)}=h^{(j)}_{t_k}\in\spa^>(A)$ for $j=1,\ldots,k-1$, and $h_t^{(k)} = h^{(k)}$. Clearly, then, $dg_t^{(j)}/dt_j = h^{(j)}_{t_k}$ for $j=1,\ldots,k-1$, whence $h^{(j)}_{t_k}$ are holomorphic vector fields on $A$. By Point 2 of Definition \ref{defiFkA}, $\lim\limits_{t_k\to0} h^{(j)}_{t_k}= h^{(j)}$ in the topology on $\spa^>(A)$. In (\ref{multder}), let us take the limit $\lim\limits_{t_k\to0}\;\lim\limits_{(t_1,\ldots,t_{k-1})\to(0,\ldots,0)}$, in that order. Renaming $t_k$ by $t$, denoting $g_t(z)=z+th^{(k)}$ (in global coordinates), and denoting $v_t:=(h^{(k-1)}_t,\ldots,h^{(1)}_t)\in (\spa^>(A))^{k-1}_{\otimes_\R}$, $v:=v_0$ and $h:=h^{(k)}$, we find
\beq\label{strong2}
	\lim_{t\to0} \frc{
	\nabla_{v_t} f(g_{t}\cdot\Sigma) -
	\nabla_{v_t} f(\Sigma)}{t}
	= \nabla_{(h,v)} f(\Sigma).
\eeq
This holds for any $t\mapsto v_t$ that is continuous at $t=0$ in  $(\spa^>(A))^{k-1}_{\otimes_\R}$ and any $\{t\mapsto g_t\}\in\spag^{(1)}(A)$. Since, further, $\lim\limits_{t\to0} \nabla_{v_t}f(\Sigma)= \nabla_vf(\Sigma)$ by continuity of the differential, we also find
\beq\label{strong1}
	\lim_{t\to0}
	\nabla_{v_t} f(g_{t}\cdot\Sigma)
	= \nabla_{v} f(\Sigma).
\eeq
Equations (\ref{strong2}) and (\ref{strong1}), and related equations that can be derived in similar ways, are strong consequences of Definition \ref{defikAdiff}, which will be used in Lemmas \ref{lemintegral} and \ref{lemsimder}.

\begin{defi}\label{defikdiff}
A function $f$ on an open subset ${\cal N}\subseteq\Omega$ is $k$ times differentiable at $\Sigma\in{\cal N}$ if it is $k$ times $A$-differentiable for every domain $A\supset \gamma(\Sigma)$. The function $f$ is smooth if it is $k$ times differentiable for all $k\in\N$.
\end{defi}
Naturally, we could define $\alpha$- and $\beta$- variants as well. The ring of smooth functions on an open set ${\cal N}\subset\Omega$ will be denoted by ${\cal F}({\cal N})$. Note that it will sometimes be sufficient to consider, for instance, the weaker condition of $\C_0$-smoothness: the existence of all $\C_0$-derivatives.

Relation (\ref{strong1}) along with second-differentiability allows us to obtain a result similar to a fundamental theorem of calculus.
\begin{lemma}\label{lemintegral}
Let ${\cal N}\subseteq\Omega$ be a connected open set, $\sigma:t\to\Sigma_t,\;t\in[0,1]$ a smooth path lying in $\cal N$, and $f$ a twice-differentiable function on ${\cal N}$. Then
\beq\label{eqintegral}
	\int_0^1 dt\,\nabla_{{\cal D}\sigma(t)}f(\Sigma_t) = f(\Sigma_1)-f(\Sigma_0).
\eeq
\end{lemma}
\proof By Lemma \ref{lemchainrule} and thanks to relation (\ref{strong1}) for $k=2$, and by the definition of smooth paths which implies that ${\cal D}\sigma(t)$ is continuous at every $t$, we find that the derivative $df(\Sigma_t)/dt$ is continuous at every $t\in [0,1]$. Hence, thanks to the first fundamental theorem of calculus, its Riemann integral gives (\ref{eqintegral}).
\eproof

From the existence of the limit in (\ref{multder}) and choosing $h_t^{(j)}=h_{t_j}^{(j)}$ for all $j$, we find that if $f$ is $k$ times differentiable, then
\beq
	\nabla_{h^{(k)},\ldots,h^{(1)}} f(\Sigma)
	= \nabla_{h^{(k)}\,|\,g^{(k)}}\lt(
	\nabla_{h^{(k-1)}\,|\,g^{(k-1)}}\lt(\cdots
	\lt(\nabla_{h^{(1)}\,|\,g^{(1)}}
	f(g_{\N_k}\cdot\Sigma)\rt)\rt)\rt).
\eeq
The right-hand side is simply the multiple application of single derivatives, in the order shown:
\[
	\nabla_{h^{(k)},\ldots,h^{(1)}} f(\Sigma)=
	\nabla_{h^{(k)}}\cdots\nabla_{h^{(1)}}
	f(\Sigma).
\]
The single conformal derivative operators involved in this expression do not commute with each other, but rather form a representation of the algebra of vector fields $\spa^>(A)$.
\begin{lemma}\label{lemcomm}
Let $f$ be twice differentiable at $\Sigma\in{\cal N}$. Then for every domain $A\supset\gamma(\Sigma)$ and every $h,h'\in\spa^>(A)$, we have
\beq\label{commnabla}
	\nabla_{h} \nabla_{h'}f(\Sigma) - \nabla_{h'}\nabla_hf(\Sigma)
	=\nabla_{h\p h' - h'\p h}f(\Sigma).
\eeq
\end{lemma}
\proof Let $g_t(z)=G_{t;a}^{h}(z)$ and $g_t'(z) = G_{t;a}^{h'}(z)$ be defined as in (\ref{pathinfty}) for $a\in\hC\setminus A$ finite or infinite. Then, using second $A$-differentiability and (\ref{multder}), where for convenience we choose the path where all $t$s are equal to each other,
\[
	\nabla_h \nabla_{h'}f(\Sigma)
	-\nabla_{h'} \nabla_{h}f(\Sigma)
	= \lim_{t\to 0}
	\frc{f((g_{t}'\circ g_t)\cdot \Sigma) - 
	f((g_t\circ g_{t}')\cdot \Sigma)}{t^2}.
\]
Since both $g_t$ and $g_{t}'$ converge to $\id$ in the $A$-topology as $t\to0$, then also do both $g_t\circ g_{t}'$ and $g_{t}'\circ g_t$. Further, both $g_t$ and $g_t'$ have a compactly convergent local-coordinates Taylor series expansion in $t$, whence likewise their compositions do. Using the chain rule, and in local coordinates, we have
\[
	g_t\circ g_t' = \id + t(h + h') + t^2 h' \p h + O(t^3).
\]
Hence, in these coordinates, $g_t'\circ g_t \circ (g_t')^{-1}\circ g_t^{-1} = \id + t^2 (h\p h' - h' \p h) + O(t^3)$, whence
\[
	\lt.\frc{d}{d(t^2)}
	\lt(g_t'\circ g_t \circ (g_t')^{-1}\circ g_t^{-1}\rt)\rt|_{t^2=0} =
	h\p h' - h' \p h.
\]
Since $h\p h' - h' \p h$ transforms like a holomorphic vector field, this equations holds in a coordinate independent way. Hence, with $\t{g}_t:=g_t'\circ g_t \circ (g_t')^{-1}\circ g_t^{-1}$, we have $\{t^2\mapsto \t{g}_t\}\in\spag^{(1)}(A)$, where the derivative of $\t{g}_t$ with respect to $t^2$ at $t^2=0$ (which exists compactly on $A$) is $h\p h' - h' \p h$. Clearly,
\[
	\lim_{t\to 0}
	\frc{f((g_{t}'\circ g_t)\cdot \Sigma) - 
	f((g_t\circ g_{t}')\cdot \Sigma)}{t^2}
	=
	\lim_{t\to 0}
	\frc{f((\t{g}_t\circ g_{t}\circ g_t')\cdot \Sigma) - 
	f((g_t\circ g_{t}')\cdot \Sigma)}{t^2}.
\]
Since $f$ is twice differentiable, then certainly the following limit
\[
	\lim_{t\to 0}
	\frc{f((\t{g}_t\circ g_{t}\circ g_t')\cdot \Sigma) - 
	f((g_t\circ g_{t}')\cdot \Sigma)-
	f(\t{g}_t\cdot\Sigma) + f(\Sigma)}{t^3}
\]
exists. Hence in particular,
\[
	\lim_{t\to 0}\lt(
	\frc{f((\t{g}_t\circ g_{t}\circ g_t')\cdot \Sigma) - 
	f((g_t\circ g_{t}')\cdot \Sigma)}{t^2} -
	\frc{f(\t{g}_t\cdot\Sigma) - f(\Sigma)}{t^2}\rt)=0,
\]
which completes the proof. 
\eproof

Finally, let us display simple examples of functions that are smooth on $\Omega$. Let $z_0\in E$ and consider the functions $f_r:\Omega\to \C$ for $r=0,1,2,3,\ldots$ given by
\beqa
	f_r \;:\; \Omega &\to& \C \n
	\Sigma &\to& f_r(\Sigma):=\p ^r\Sigma(z_0). \label{simplef}
\eeqa
These functions are smooth on $\Omega$, and so are functions, perhaps on smaller open subset ${\cal N}\subseteq \Omega$, obtained by composition with smooth complex functions $F$ on domains of the complex plane, $F\circ f_r$. These statements immediately follow from the following lemma.
\begin{lemma}\label{lemsimplef}
Let $w\in\C$. For every domain $A\supset w$ and every $r=0,1,2,3,\ldots$, the function $f^{(r)}:g\mapsto \p^r g(w)$ is defined on a $A$-neighbourhood of $\id$ and is infinitely many times $A$-differentiable at $\id$. Further, the $k^{\rm th}$ $A$-derivative is
\beq\label{sfder}
	\nabla_{(h^{(k)},\ldots,h^{(1)})} f^{(r)}
	= \sum_{(m_1,\ldots,m_k)\in{\cal R}_{n,k}}
	\frc{r! C_{k}^{(m_1,\ldots,m_k)}}{\prod_{j=1}^k m_j!}
	\prod_{j=1}^k \p^{m_j}h^{(j)}(w).
\eeq
where the region of summation ${\cal R}_{n,k}$ is ${\cal R}_{n,1}:m_1=n$ and for $k\geq 2$:
\beq\label{eqqregion}
	{\cal R}_{n,k} :  m_1\geq1;\;m_1+m_2\geq2;\;\ldots;\;
	\sum_{j=1}^{k-1} m_j\geq k;\;
	\sum_{j=1}^k m_j = n+k-1,
\eeq
and where the coefficients are
\beq\label{eqqcoeff}
	C_{k}^{(m_1,\ldots,m_k)} :=
	m_1(m_1+m_2-1)\cdots \lt(\sum_{j=1}^{k-1} m_j-k+2\rt).
\eeq

\end{lemma}
\proof In order to evaluate the limit (\ref{multder}) we may choose without loss of generality a neighbourhood $A$ of $w$ that lies in $\C$ and we may use global coordinates, in order to write $g_t^{(j)}(z) = z + t_j h_t^{(j)}(z)$ for all $z\in A$. Then, we need to evaluate, for every $k\in\N$,
\beq\label{eqq0}
	\lim_{t\to (0,\ldots,0)}
	\frc1{\prod_{j=1}^k t_j}
	\sum_{S\subset \N_k} (-1)^{k-|S|} \p^r g_S(w).
\eeq

Let $u_j,\;j\in\N$ be holomorphic functions on $A$, and define $g_j := \id + u_j,\; j\in\N$. For every $z\in A$ and $k\in\N$, let $R_k(z):= \sum_{S\subset \N_k} (-1)^{k-|S|} (g_{j_1}\circ \cdots\circ g_{j_{|S|}})(z)$. Clearly, for every $k\in\N$,
\beq\label{eqq1}
	R_{k+1}(z) = R_{k}(z+u_{k+1}(z)) - R_{k}(z),
\eeq
and $R_1(z) = u_1(z)$. If for every $j\in\N$, the function $u_j$ is near enough to 0 compactly on $A$, then $R_k(z)$ has a Taylor series around $w$ of nonzero radius for every $k$, and this series can be obtained by composing the Taylor series of $u_j,\;j=1,\ldots,k$ around $w$. Let
\[
	u_j(z) =: \sum_{m=0}^\infty a_{m,j}(z-w)^m.
\]
Then we find
\beq\label{eqq2}
	R_k(z) = \sum_{n=0}^\infty A_{n,k}(z-w)^n
\eeq
where each $A_{n,k}$ is expressed as a converging power series in the coefficients $a_{m,j}:m\in\N,\,j=1,\ldots,k$.

We will show that for all $k\in\N$,
\[
	A_{n,k}=A_{n,k}^{(0)}\;\Big( 1 + O(a_{m,j}:m\in\N,\,j=1,\ldots,k)\Big)
\]
where
\beq\label{eqq3}
	A_{n,k}^{(0)}
	:= \sum_{(m_1,\ldots,m_k)\in{\cal R}_{n,k}}
	C_{k}^{(m_1,\ldots,m_k)} a_{m_1,1}\cdots a_{m_k,k},
\eeq
with (\ref{eqqregion}) and (\ref{eqqcoeff}). This is clearly true for $A_{n,1}^{(0)}=a_{n,1}$. Let us assume that it is true for $A_{n,k}$. Then, from (\ref{eqq1}) and (\ref{eqq2}),
\beqa
	R_{k+1}(z) &=&
	\sum_{n=0}^{\infty} A_{n,k}\lt(
	(z+u_{k+1}(z)-w)^n - (z-w)^n\rt) \n
	&=& 
	\sum_{n=1}^{\infty} A_{n,k}\lt[
	\sum_{j=0}^{n-1}\lt(n\atop j\rt)
	(z-w)^j \lt(\sum_{m=0}^{\infty}
	a_{m,k+1}(z-w)^m\rt)^{n-j}\rt].
\eeqa
Inside the square bracket on the second line, the term, in the series expansion in powers of $(z-w)$ and of $a_{m,k+1}:m\in\N$, that contains $(z-w)^\ell$ and that has the lowest order in $a_{m,k+1}:m\in\N$, is obtained at the maximal value of $j$, i.e.~$j=n-1$. This term is $n(z-w)^\ell a_{\ell+1-n,k+1}$, and is present for all $\ell\geq n-1$. Given $n$, for lower values of $\ell$, only terms with second or higher orders in $a_{m,k+1}:m\in\N$ are present, but given any $\ell\geq0$, there is a value of $n$ for which this term is present. Hence, we find the recursion relation
\[
	A_{\ell,k+1}^{(0)} = \sum_{n=1}^{\ell+1} n A_{n,k}^{(0)}
	a_{\ell+1-n,k+1},
\]
which shows (\ref{eqq3}).

In order to evaluate the limit (\ref{eqq0}), we need to set $u_j=t_jh_t^{(j)},\;j=1,\ldots,k$, and to evaluate
\[
	\lim_{t\to0} \frc{\p^r R_k(w)}{\prod_{j=1}^k t_j}.
\]
Clearly, $\p^rR_k(w) = r! A_{r,k}$ and $a_{m,j} = t_j \frc{\p^m h_t^{(j)}(w)}{m!}$. Hence, (\ref{eqq3}) implies $\lim_{t\to0} \frc{\p^r R_k(w)}{\prod_{j=1}^k t_j}$ exists and is exactly given by the right-hand side of (\ref{sfder}). This shows that the limit (\ref{multder}) exists, and provides the explicit result of the limit, showing that it is a continuous linear functional as on the right-hand side of (\ref{multder}). Since this holds for every $A\ni w$, every $k$ and every $r$, this shows the lemma.
\eproof

\subsubsection{Holomorphicity}

Although we have used real-linearity only, the spaces $\spa^>(A)$ of holomorphic vector fields can be seen naturally as complex linear spaces. Given a holomorphic vector field $h$ on some domain $A$, let
\beq\label{holomder}
	\Delta[h] := \frac1{2} \sum_{\eta\in\{1,i\}} \, \eta^{-1}
	\nabla_{\eta h},\quad
	\b\Delta[\b{h}] := \frc1{2} \sum_{\eta\in\{1,i\}} \, \eta
	\nabla_{\eta h}
\eeq
where $\b{h}$ is the holomorphic function $\b{z}\mapsto \overline{h(z)}:\b{z}\in\hC$. Note that
\[
	\nabla_h f(\Sigma) = \Delta[h]f(\Sigma) +
	\b\Delta[\b{h}]f(\Sigma).
\]

Clearly, thanks to real linearity of conformal derivatives, given a differentiable function $f$ at $\Sigma\in\Omega$, both functionals $\Delta[\cdot] f(\Sigma)$ and $\b\Delta[\cdot] f(\Sigma)$ are continuous, complex-linear on $\spa^>(A)$, and a similar statement holds for multiple holomorphic derivatives if $f$ is multiply differentiable.

Formula (\ref{commnabla}) implies a similar formula for these differential operators, for instance:
\beq\label{commdelta}
	\Delta[h] \Delta[h'] - \Delta[h']\Delta[h]
	=\Delta[h\p h' - h'\p h],
\eeq
which holds again on twice differentiable functions. Interestingly, the holomorphic and anti-holomorphic differential operators commute with each other:
\beqa
	\Delta[h]\b\Delta[\b{h}']-
	\b\Delta[\b{h}']\Delta[h]
	&=& \frc14 \sum_{\eta,\eta'\in\{1,i\}} \frc{\eta'}\eta
	\nabla_{\eta\eta' (h\p h' - h'\p h)} \n
	&=& 0\label{commute}
\eeqa
where the terms with $\eta=1,\eta'=1$ and $\eta=i,\eta'=i$ cancel each other, and the terms with $\eta=1,\eta'=i$ and $\eta=i,\eta'=1$ cancel each other, by real linearity of the differential $\nabla$.

The holomorphic conformal derivative is naturally obtained if in $t\mapsto g_t$ we replace $t$ by a complex number $\tau$, and we assume complex differentiability.
\begin{lemma}\label{lemtau}
Let $f$ be a function on an open subset ${\cal N}\subseteq\Omega$, that is differentiable at $\Sigma\in\cal N$, and let $A\supset\gamma(\Sigma)$ be a domain. Let $\tau\mapsto g_\tau$ be a function (valued in conformal maps) on a neighbourhood of 0 in $\C$ such that $g_\tau\to 0$ as $\tau \to 0$ in the $A$-topology, and such that its complex derivative exists at $\tau=0$ compactly on $A$. Then (i) for every continuous path $t\mapsto z_t$ in $\C$ that is differentiable at $t=0$ and such that $z_0=0$, we have $\{t \mapsto g_{z_t}\}\in\spag^{(1)}(A)$; (ii) $f(g_\tau\cdot\Sigma)$ is complex differentiable at $\tau=0$; and (iii) the complex derivative of $f(g_\tau\cdot\Sigma)$ is
\beq
	\lt.\frc{d}{d\tau} f(g_\tau\cdot\Sigma)\rt|_{\tau=0}
	= \Delta[h] f(\Sigma)
\eeq
where $h=dg_\tau/d\tau|_{\tau=0}$ (compactly on $A$).
\end{lemma}
\proof The first statement is immediate. The second follows from the fact that if $F$ is a complex function on a neighbourhood of 0 in $\C$ and $F(z_t)$ is differentiable at $t=0$ for every $z_t$ as in the lemma, then $F$ is complex differentiable at $0$. The third is a consequence of $d f(g_\tau\cdot\Sigma)/d\tau = (1/2)(df(g_t\cdot\Sigma)/dt - idf(g_{it}\cdot\Sigma)/dt)$ as well as $dg_{it}/dt|_{t=0} = i dg_t/dt|_{t=0} = ih$ by complex differentiability. \eproof

It is convenient to have a notion of  holomorphicity: we say that a complex function $f$ on an open subset ${\cal N}\subseteq\Omega$ is holomorphic on ${\cal N}$ if, for all $\Sigma\in{\cal N}$, we have $\b\Delta[\b{h}] f(\Sigma) = 0$  for all $h\in\spa^>(A)$, $A\supset \gamma(\Sigma)$. Note that if $f$ is real and holomorphic, then it has zero conformal derivatives. We also define $\alpha$-holomorphicity (resp.~$\beta$-holomorphicity) by restricting to $A\supset \alpha(\Sigma)$ (resp.~$A\supset \beta(\Sigma)$).

The notion of holomorphicity is of course close to that of the usual holomorphicity of functions on the complex plane. For instance, if $F(z_1,z_2,\ldots,z_p)$ is complex differentiable as a function of $p$ variables on some domain of $\C^p$, then the function $f(\Sigma):=F(\Sigma(z_1),\p\Sigma(z_2),\ldots,\p^p\Sigma(z_p))$ for fixed $z_1,\ldots,z_p\in E$, as a function on some open subset ${\cal N}\subseteq\Omega$, is holomorphic.

For clarity, we will sometimes use the notation
\[
	\Delta[h\,|\,g] f(g\cdot\Sigma) = \Delta[h] f(\Sigma) 
\]
paralleling the notation $\nabla_{h\,|\,g}$ introduced above; $\Delta[h\,|\,g]$ should be interpreted as a holomorphic derivative with respect to $g$ evaluated at $g=\id$. For instance, if $f(\Sigma) := \p\Sigma(z)$ for some fixed $z\in E$ (this is a smooth function), then $\Delta[h]f(\Sigma) = \Delta[h\,|\,g] \lt(\;\p g(\Sigma(z))\p \Sigma(z)\;\rt) = \p h(\Sigma(z)) f(\Sigma)$.

\begin{rema}\label{remaint}{\em 
By the general theory of continuous duals (see \cite{Rudin}), given a function $f$ on an open subset ${\cal N}\subseteq\Omega$ that is differentiable at a given $\Sigma\in{\cal N}$, there exists a unique function $z\mapsto K^\alpha(z)$ holomorphic on $\hC\setminus \alpha(\Sigma)$ such that $K^\alpha(z)\sim z^{-1}$ as $z\to\infty$, and such that
\[
	\Delta[h] f(\Sigma) = \frc1{2\pi i} \oint dz\,K^\alpha(z) h(z)
\]
for all $h\in\spa^>(A)$, $A\supset\alpha(\Sigma)$ a simply connected domain, where the contour surrounds $\alpha(\Sigma)$ once counterclockwise and lies in $A\setminus\alpha(\Sigma)$.  Likewise, there exists a unique function $z\mapsto K^\beta(z)$ holomorphic on $\hC\setminus \beta(\Sigma)$ except for pole of order 3 at $0$, such that
\[
	\Delta[h] f(\Sigma) = \frc1{2\pi i} \oint dz\,K^\beta(z) h(z)
\]
for all $h\in\spa^>(B)$, $B\supset\beta(\Sigma)$ a simply connected domain, where the contour surrounds $\beta(\Sigma)$ once counterclockwise (with respect to $\beta(\Sigma)$) and lies in $B\setminus\beta(\Sigma)$. In general, the kernel representing a given continuous linear functional can be chosen to be a singular quadratic differential with a pole of order 3, and once the position of the pole is chosen, the kernel is unique; here, for $K^\alpha$ the pole is chosen to be at $\infty$, and for $K^\beta$, at 0.}
\end{rema}

\begin{rema}\label{remahol}{\em 
We could have defined complex conformal differentiability by requiring the existence of a complex-linear, rather than real-linear, functional $\nabla^Af(\Sigma)$ in (\ref{defder}). It is then equivalent to require, in addition, the complex derivative at $t=0$ to exist rather than the real derivative in (\ref{defder}). A function that is complex conformal differentiable is then immediately holomorphic according to our definition above of holomorphicity.}
\end{rema}

\subsubsection{Particular vector fields}

For every $w\in\C$, let us introduce the holomorphic vector fields
\beq\label{defh}
	h_{n;w}(z) := (w-z)^{n+1},\quad n\in \Z
\eeq
in global coordinates on $\hC$. We will also use the notation
\beq
	h_n(z):=h_{n;0}(z).
\eeq
For $n\leq 1$, the vector fields in (\ref{defh}) are holomorphic vector fields on every domain that do not contain $w$, and for $n\geq -1$, they are holomorphic vector fields on every domain that do not contain $\infty$ (in particular, for $n=-1,0,1$, they are the holomorphic vector fields associated to M\"obius transformations). Given any function $f$ on an open subset ${\cal N}\subseteq \Omega$ that is twice differentiable at $\Sigma\in\cal N$, Equation (\ref{commdelta}) implies that
\beq\label{commh}
	[\Delta[h_{m;w}], \Delta[h_{n;w}]]f(\Sigma)
	= (m-n)\Delta[h_{m+n;w}]f(\Sigma)
\eeq
for every $w\in \C\setminus \gamma(\Sigma)$. Hence, in particular, on the space of smooth functions on $\Omega$, the operators $\Delta[h_{n}],\;n\in\Z$ form a representation of the Witt algebra.

In later sections, it will be sufficient to restrict $w$ to $\hC\setminus \beta(\Sigma)$, but the following lemmas consider the general case $w\in\C\setminus\gamma(\Sigma)$ for completeness. Below, we use the general notation
\[
	\prod_{i=k}^j \Delta[h^{(i)}]
	= \Delta[h^{(k)}] \Delta[h^{(k-1)}]\cdots \Delta[h^{(j)}],\quad
	\prod_{i=k}^{k+1} \Delta[h^{(i)}]=1
\]
for every integers $k\ge j$.
\begin{lemma}\label{lemder}
Let $\t{k},k\in\N$ with $\t{k}\geq k$, and let $f$ be a function on an open subset ${\cal N}\subseteq\Omega$ that is $\t{k}$ times differentiable at $\Sigma\in\cal N$. Let $\t\nabla$ be a linear combination of multiple conformal derivative operators in fixed directions (or be the number 1), of order at most $\t{k}-k$. Then, for every $n_1,\ldots,n_k\in\Z$, the function
\beq\label{fctmd}
	w\mapsto \t\nabla\prod_{i=k}^1 \Delta[h_{n_i;w}] f(\Sigma)
\eeq
is holomorphic on $\C\setminus \gamma(\Sigma)$. Further, its derivative with respect to $w$ is
\beq\label{der}
	\frc{\p}{\p w} \t\nabla\prod_{i=k}^1 \Delta[h_{n_i;w}]
	f(\Sigma)
	= \sum_{j=1}^k (n_j+1)
	\t\nabla\prod_{i=k}^{j+1} \Delta[h_{n_i;w}]
	\Delta[h_{n_j-1;w}]
	\prod_{i=j-1}^1 \Delta[h_{n_i;w}] f(\Sigma),
\eeq
which is obtained by using Leibnitz's rule on the formal relation
\beq\label{derform}
	\frc{\p}{\p w} \Delta[h_{n;w}] = (n+1) \Delta[h_{n-1;w}].
\eeq
\end{lemma}
\proof Let $w\in\hC\setminus\gamma(\Sigma)$. Note that $\t\nabla\prod_{i=k}^1\Delta[\cdot^{(i)}]f(\Sigma)$ is a continuous $\C$-linear functional on $\lt(\spa^>(A)\rt)^k_{\otimes_\C} $ for every $A\supset\gamma(\Sigma)$. Note also that
\beq\label{derhn}
	\frc{\p}{\p w} h_{n;w}(z) = (n+1)h_{n-1;w}(z).
\eeq

We prove by induction. The lemma obviously holds at $k=0$. Let us fix $k>0$ and assume that the lemma holds for $k$ replaced by $k-1$. Using this assumption and Equation (\ref{commh}), we may assume that $n_k\leq n_{k-1}\leq \cdots\leq n_1$. If $n_1\geq-1$, then by linearity $\Delta[h_{n_1;w}]f(\Sigma)$ is a polynomial in $w$, where the coefficients are conformal derivatives of the form $\Delta[h_n]f(\Sigma)$ for various $n\geq -1$; using (\ref{derhn}) we see that this polynomial satisfies (\ref{derform}). Hence, from the assumption of the induction and the fact that the single derivatives of $f$ are $\t{k}-1$ times differentiable, the lemma holds in this case. 

Hence let us assume that $n_j\leq -2$ for all $j$. By the fact that there exists a simple polygon separating $w$ from $\hC\setminus\gamma(\Sigma)$ \cite[Thm 3.3]{N61}, all $h_{n_j;w}$ are holomorphic vector fields on a simply connected domain $A\supset\gamma(\Sigma)$ such that $w\in\hC\setminus \cl{A}$. In particular, there is a disk $D$ centered at $w$ of (plane geometry) radius $r>0$ that lies in $\hC\setminus \cl{A}$. For all $\ep\in \C$ with $|\ep|<r$, the local-coordinates Taylor series of $h_{n_j;w+\ep}$ in powers of $\ep\in\C$ is uniformly convergent on $\hC\setminus D$. Hence, the Taylor series of the vector $(h_{n_k;w+\ep},\ldots,h_{n_1;w+\ep})$ in powers of $\ep$ is convergent in the topology of $\lt( \spa^>(A)\rt)^k_{\otimes_\C}$. Let us denote by $Q$ the function (\ref{fctmd}). Then $\C$-linearity of the functional $\t\nabla\prod_{i=k}^1 \Delta[\cdot^{(i)}]f(\Sigma)$ implies that replacing every $h_{n_j;w+\ep}(z)$ by its partial series up to power $\ep^\ell$ gives $Q(w+\ep)$ as a power series in $\ep$ up to the power $\ep^{k\ell}$. Further, continuity implies that the limit where $\ell\to\infty$ exists, whence that the resulting infinite series in nonnegative integer powers of $\ep$ equates $Q(w+\ep)$. Hence, $Q(w+\ep)$ has a Taylor series in powers of $\ep$ that holds for $|\ep|<r$. This shows holomorphicity, and the formula for the derivative is a consequence of the explicit form of the first coefficient of this Taylor series, which readily follows from (\ref{derhn}).
\eproof

An important result for the subsequent sections is that the derivatives with respect to $w$ and to $\Sigma$ in, for instance, $\Delta[h_{n;w}]f(\Sigma)$ ``interact well'' with each other: derivatives with respect to small simultaneous variations of $w$ and $\Sigma$ are obtained by Leibniz's rule.
\begin{lemma}\label{lemsimder}
Let $\t{k},k\in\N$ with $\t{k}\geq k$, and let $f$ be a function on an open subset ${\cal N}\subseteq\Omega$ that is $\t{k}+1$ times differentiable at $\Sigma\in\cal N$. Let $n_1,\ldots,n_k\in\Z$, let $w_1,\ldots,w_k\in \C\setminus\gamma(\Sigma)$, and let $A\supset \gamma(\Sigma)\cup\{w_1,\ldots,w_k\}$ be a domain. Further, let $\t\nabla$ be a linear combination of multiple conformal $A$-derivative operators in fixed directions (or be the number 1), of order at most $\t{k}-k$. Then, for every $\{t\mapsto g_t^{(j)}\}$ ($j=1,\ldots,k$), $\{t\mapsto g_t'\}\in\spag^{(1)}(A)$, the function
\[
	t\mapsto \t\nabla\prod_{i=k}^1
	\Delta[h_{n_i;g_t^{(i)}(w_i)}]
	f(g_t'\cdot \Sigma)
\]
is differentiable with respect to $t$ at $t=0$, and its derivative can be evaluated using Leibnitz's rule:
\beq\label{simder}
	\frc{d}{dt}\lt. \t\nabla\prod_{i=k}^1 
	\Delta[h_{n_i;g_t^{(i)}(w_i)}]
	f(g_t'\cdot \Sigma)\rt|_{t=0}
	= \lt(\sum_{j=1}^k
	h^{(j)}(w_j) \frc{\p}{\p w_j}
	+ \nabla_{h'}\rt)
	\t\nabla
	\prod_{i=k}^{1} \Delta[h_{n_i;w_i}]\,f(\Sigma)
\eeq
where $h^{(j)}=dg_t^{(j)}/dt|_{t=0}$, $j=1,\ldots,k$ and $h'=dg_t'/dt|_{t=0}$ (compactly on $A$). In particular, the function
\beq\label{gsimder}
	F:g\mapsto \t\nabla\prod_{i=k}^1 
	\Delta[h_{n_i;g(w_i)}]
	f(g\cdot \Sigma)
\eeq
defined on a $A$-neighbourhood of the identity, is $A$-differentiable at the identity, and its derivative can be evaluated using Leibnitz's rule.

Further, if $f$ is smooth, then the function (\ref{gsimder}) is $A$-smooth at the identity, and all its derivatives can be evaluated using Leibnitz's rule,
\beq\label{leib}
	\cdots \Delta[h]\cdots\; F\;=\;
	\cdots \lt(\sum_{j=1}^k
	h(w_j) \frc{\p}{\p w_j}
	+ \Delta[h]\rt)\cdots \;
	\t\nabla\prod_{i=k}^1 
	\Delta[h_{n_i;g(w_i)}]f.
\eeq
\end{lemma}
\proof Note that the functional
\beq
	\Upsilon:(h^{(1)},\ldots,h^{(k)},h')\mapsto
	 \lt(\sum_{j=1}^k
	h^{(j)}(w_j)\frc{\p}{\p w_j} +
	\nabla_{h'}\rt)
	\t\nabla \prod_{i=k}^{1} \Delta[h_{n_i;w_i}]\,f(\Sigma)
\eeq
where the $w_j$ derivatives are explicitly evaluated using Lemma \ref{lemder}, is real-linear and continuous on $\lt(\spa^>(A)\rt)^{k}_{\otimes_\R}\otimes_\R \spa^>(A)$. This shows that if (\ref{simder}) holds, then indeed the function (\ref{gsimder}) is $A$-differentiable. We will show that for every $\{t\mapsto g_t^{(j)}\},\,\{t\mapsto g_t'\}\in\spag(A)$, $j=1,\ldots,k$ we have
\beq\label{sd0}
	\lim_{t\to0} \frc{\t\nabla\prod_{i=k}^1\Delta[h_{n_i;g_t^{(i)}(w_i)}]
	f(g_t'\cdot\Sigma)
	- \t\nabla\prod_{i=k}^1\Delta[h_{n_i;w_i}]f(\Sigma)}{t}
	=\Upsilon(h^{(1)},\ldots,h^{(k)},h')
\eeq
where $h^{(j)}$, $j=1,\ldots,k$ and $h'$ are as in the lemma, which will show (\ref{simder}).

In order to show (\ref{sd0}), we write the right-hand side of (\ref{sd0}), by linearity, asÄ
\beqa\label{sd1}
	&&
	\lim_{t\to0} \sum_{j=1}^k
	\t\nabla\prod_{i=k}^{j+1}\Delta[h_{n_i;w_i}]
	\Delta\lt[\frc{h_{n_j;g_t^{(j)}(w_j)}-h_{n_j;w_j}}{t}\rt]
	\prod_{i=j-1}^1\Delta[h_{n_{i};g_t^{(i)}(w_i)}]
	f(g_t'\cdot\Sigma) \\
	&&
		+\;\lim_{t\to0} \frc{
	\t\nabla\prod_{i=k}^1\Delta[h_{n_i;w_i}]f(g_t'\cdot\Sigma)
	-
	\t\nabla\prod_{i=k}^1\Delta[h_{n_i;w_i}]f(\Sigma)}{t}
	\no
\eeqa
We have
\[
	\lim_{t\to0} \frc{h_{n_j;g_t^{(j)}(w_j)}-h_{n_j;w_j}}{t} =
	h^{(j)}(w_j) (n_j+1)h_{n_j-1;w_j},\quad
	\lim_{t\to0}
	h_{n_i;g_t^{(i)}(w_i)} = h_{n_i;w_i}
\]
compactly on $A$ for all $i,j$. Hence,
\beqa
	\lim_{t\to0}
	\lefteqn{(h_{n_k;w_k},\ldots,h_{n_{j+1};w_{j+1}},
	\frc{h_{n_j;g_t^{(j)}(w_j)}-h_{n_j;w_j}}{t}
	,
	h_{n_{j-1};g_t^{(j-1)}(w_{j-1})},\ldots,h_{n_1;g_t^{(1)}(w_1)})}
	&& \n
	&=&
	h^{(j)}(w_j)(n_j+1)\;
	(h_{n_k;w_k},\ldots,h_{n_{j+1};w_{j+1}},
	h_{n_j-1;w_j}
	,
	h_{n_{j-1};w_{j-1}},\ldots,h_{n_1;w_1}) \no
\eeqa
in the topology on $\lt(\spa^>(A)\rt)^k_{\otimes_\C}$ and using $\C$-linearity. Then, re-interpreting the latter equation on the real-linear space $\lt(\spa^>(A)\rt)^k_{\otimes_\R}$, Equation (\ref{strong1}) along with $(\t{k}+1)^{\rm th}$-differentiability of $f$ at $\Sigma$ imply
\beqa\label{sd4}
	\lim_{t\to0}\sum_{j=1}^k
	\t\nabla\prod_{i=k}^{j+1}\Delta[h_{n_i;w_i}]
	\Delta\lt[\frc{h_{n_j;g_t^{(j)}(w_j)}-h_{n_j;w_j}}{t}\rt]
	\lefteqn{\prod_{i=j-1}^1\Delta[h_{n_{i};g_t^{(i)}(w_i)}]
	f(g_t'\cdot\Sigma)} &&\\
	&=& \sum_{j=1}^k
	h^{(j)}(w_j) \frc{\p}{\p w_j}\t\nabla
	\prod_{i=k}^{1} \Delta[h_{n_i;w_i}]
	f(\Sigma). \no
\eeqa
Therefore, (\ref{sd4}) and (\ref{sd1}) give
\beqa
	\lefteqn{
	\lim_{t\to0} \frc{\t\nabla\prod_{i=k}^1\Delta[h_{n_i;g_t^{(i)}(w_i)}]
	f(g_t'\cdot\Sigma)
	- \t\nabla\prod_{i=k}^1\Delta[h_{n_i;w_i}]f(\Sigma)}{t}}&&\n
	&=&
	\sum_{j=1}^k
	h^{(j)}(w_j) \frc{\p}{\p w_j}\t\nabla
	\prod_{i=k}^{1} \Delta[h_{n_i;w_i}]
	f(\Sigma)
	+\lim_{t\to0} \frc{
	\t\nabla\prod_{i=k}^1\Delta[h_{n_i;w_i}]f(g_t'\cdot\Sigma)
	-
	\t\nabla\prod_{i=k}^1\Delta[h_{n_i;w_i}]f(\Sigma)}{t} \n
	&=& \Upsilon(h^{(1)},\ldots,h^{(k)},h') \no
\eeqa
where in the last step we used $(\t{k}+1)^{\rm th}$ differentiability of $f$ at $\Sigma$. This shows (\ref{sd0}), hence the first part of the lemma.

For the second part, concerning smooth functions $f$, we only provide the basic line of reasoning, as the proof, rather technical, simply extends the arguments given above for differentiability. Note that the functional of $(\ldots,h,\ldots)\in\lt(\spa^>(A)\rt)^q_{\otimes_\R}$ (for some $q\geq 1$) obtained from the right-hand side of (\ref{leib}) and from its anti-holomorphic counterpart is linear and continuous. Hence, we only need to show that the limit on the left-hand side of (\ref{multder}), with the function $F$ in place of $f$, gives rise to the right-hand side of (\ref{leib}). For this purpose, we make iterative use of the passage from (\ref{sd0}) to (\ref{sd1}) (with $g_t^{(j)}=g_t'=g_t$). Then, (\ref{strong2}), (\ref{strong1}) and related higher-order equations that are also consequences of Definition \ref{defikAdiff}, give rise to (\ref{leib}).
\eproof

It is natural to expect that the functions $h_{n;w}$ form a basis using which we may obtain expansions of derivatives in arbitrary direction. In general, however, there may be problems of convergence. There is a special case, which is of interest for our purposes, where these convergence problems disappear.
\begin{lemma}\label{lemexph}
Let $f$ be a function on an open subset ${\cal N}\subseteq\Omega$ differentiable at $\Sigma\in\cal N$, let $h$ be a holomorphic vector field on a simply connected domain $A\supset \alpha(\Sigma)$, and let $w\in \alpha(\Sigma)$. If $\Delta[h_{n;w}]f(\Sigma)$ is zero for all but a finite number of $n\geq -1$, then
\beq\label{exph}
	\Delta[h] f(\Sigma) = \sum_{n= -1}^\infty \frc{(-1)^{n+1}}{(n+1)!}
	\p^{n+1} h(w) \Delta[h_{n;w}]f(\Sigma).
\eeq
\end{lemma}
\proof By Remark \ref{remaint}, given $f$ and $\Sigma$, there exists a unique function $z\mapsto K(z)$ holomorphic on $\hC\setminus \alpha(\Sigma)$ such that $K(z)\sim z^{-1}$ as $z\to\infty$, and such that
\[
	\Delta[h] f(\Sigma) = \frc1{2\pi i} \oint_{\cal C} dz\,K(z) h(z)
\]
for all $h\in\spa^>(A)$, where the contour $\cal C$ winds around $\alpha(\Sigma)$ once counterclockwise and lies in $A\setminus\alpha(\Sigma)$. The function $K(z)$ has a Taylor series in powers of $(z-w)^{-1}$, with zero constant term, that converges for $z$ in a neighbourhood $N$ of $\infty$. By contour deformation, for $n\geq -1$,
\[
	\Delta[h_{n;w}] f(\Sigma) =
	\frc1{2\pi i} \oint_\infty dz\,K(z) h_{n;w}(z)
\]
where the contour surrounds $\infty$ in the neighbourhood $N$. Hence, the Taylor series coefficients of $K(z)$ are given by the derivatives $\Delta[h_{n;w}] f(\Sigma)$:
\[
	K(z) = \sum_{n=-1}^\infty (w-z)^{-n-2}
	\Delta[h_{n;w}] f(\Sigma).
\]
Since $f$ has a finite number of $\Delta[h_{n;w}]$ derivatives, we find that $K(z)$ is holomorphic on $\hC\setminus \{w\}$. Hence, for $h\in\spa^>(A)$,
\[
	\Delta[h] f(\Sigma) = \frc1{2\pi i} \oint_{w} dz\,K(z) h(z)
\]
where the contour surrounds $w$ in a small neighbourhood of this point. This neighbourhood can be chosen small enough so that the Taylor series
\[
	h(z) = \sum_{n=-1}^\infty \frc{(-1)^{n+1}}{(n+1)!}
	\p^{n+1} h(w) h_{n;w}(z)
\]
converges uniformly on the contour, whence the lemma follows.\eproof

\sect{Conformal Gaussians}

In this section and the rest of the paper, we concentrate on holomorphic conformal derivatives. Here, we define conformal Gaussians by the requirements that certain second-order holomorphic conformal derivatives be constant. We show that these constants can only take a very specific form, characterized by four complex parameters. One of these parameters, which we denote by $c$, is related to the central charge of the Virasoro algebra, and we define normal conformal Gaussians by requiring that the three other parameters be 0. We then show that on subsets ${\cal N}\subseteq\Omega$ that are piecewise smoothly connected, the condition that $Z$ be a normal conformal Gaussian is equivalent to the fact that certain of its first holomorphic conformal derivatives transform in a particular fashion under conformal transport. This transformation property is exactly that of the stress-energy tensor in conformal field theory, with central charge $c$.

\subsection{Conformal differential equations}

\begin{defi}\label{deficg}
Let $Z$ be a smooth function on an open set ${\cal N}\subseteq\Omega$. We say that $Z$ is a {\em conformal Gaussian} on ${\cal N}$ if there exists a $v \in \cap_{\Sigma\in {\cal N}} (\hC\setminus \beta(\Sigma))$ such that for every integers $n\ge -1,\;m\leq -2$, the quantity
\beq\label{defpf}\ba{rl}
	\Delta[h_{n;v}] \Delta[h_{m;v}] \log(Z)(\Sigma) & (\mbox{if}\ n+m\ge -1)\\
	\mbox{or}\quad \Delta[h_{m;v}]\Delta[h_{n;v}] \log(Z)(\Sigma) & (\mbox{if}\ n+m\le -2)\ea
\eeq
is independent of $\Sigma$ on ${\cal N}$.
\end{defi}
In essence, we are asking that $Z$ be the exponential of a ``quadratic expression'', an expresion whose second derivatives are constants. Note that if ${\cal N}=\Omega$, we can only have $v=0$. Here (and similarly below), for every $\Sigma$ where $Z(\Sigma)=0$, the resulting equation will be interpreted by formally multiplying through by $Z(\Sigma)^2$.

The next theorem shows that the conditions of constant second derivatives above hold as well when $v$ is replaced by $w$ for any $w \in \beta(\Sigma)$ (hence a conformal Gaussian is not characterized by the number $v$), that there is independence upon $w$, and further that these second derivatives take values of a very particular form, determined in part by the form of the central extension of the Witt algebra.
\begin{theorem}\label{theopf}
Let $Z$ be a conformal Gaussian on an open set ${\cal N}\subseteq \Omega$. Then there exist four complex numbers $c$, $d_1$, $d_2$ and $d_3$ such that, for all $\Sigma\in {\cal N}$, $w\in \hC\setminus \beta(\Sigma)$, $n\ge-1$ and $m\leq -2$,
\beq\label{relpf}\lt.\ba{rcll}\displaystyle
	\Delta[h_{n;w}] \Delta[h_{m;w}] \log(Z)(\Sigma)& (n+m\ge -1)\\
	\Delta[h_{m;w}]\Delta[h_{n;w}] \log(Z)(\Sigma) & 
	(n+m\le -2)
\ea\rt\} = (n-m) \,k_{n+m}(w) + 
	\frc{c}{12}\,(n^3-n) \,\delta_{n+m,0} 
\eeq
where
\beq
	k_{\ell}(w) = \lt\{\ba{ll} d_1w^2 + d_2w + d_3 & (\ell=-2)\\
	-2d_1w - d_2 & (\ell=-3) \\
	d_1  & (\ell=-4) \\
	0 & (\mbox{otherwise}).
	\ea\rt.
\eeq
\end{theorem}
\proof The proof is divided into four lemmas. Let $D\in\hC$ be a closed disk (greater than a point) centered at $v$, small enough so that the subset ${\cal N}':=\Omega(D)_\beta\cap {\cal N}=\{\Sigma\in{\cal N}:D\subset \hC\setminus\beta(\Sigma)\}$ of ${\cal N}$ be nonempty. For every ${\cal N}$, such a $D$ exists because for every $\Sigma\in{\cal N}$, $\beta(\Sigma)$ is closed and its complement contains $v$. By Lemma \ref{lemOmegar}, ${\cal N}'$ is open.

\begin{lemma}\label{lemmapf1}
For every $w\in D,\;n\ge -1,\;m\leq -2$, the quantity
\beq\label{defpfw}\ba{rl}
	\Delta[h_{n;w}] \Delta[h_{m;w}] \log(Z)(\Sigma) & (\mbox{if}\ n+m\ge -1)\\
	\mbox{or}\quad \Delta[h_{m;w}]\Delta[h_{n;w}] \log(Z)(\Sigma) & (\mbox{if}\ n+m\le -2)\ea
\eeq
is independent of $\Sigma$ on ${\cal N}'$.
\end{lemma}

Let us fix $n\geq-1$ and $m\leq -2$, as well as $w\in D$. Compactly on $\hC\setminus D$ for the first relation, we have
\beq\label{exphn}
	h_{m;w}= \sum_{j\leq m} a_j (w-v)^{m-j} h_{j;v} ,\quad
	h_{n;w} = \sum_{k=-1}^n b_k (w-v)^{n-k}h_{k;v}
\eeq
with appropriate constants $a_j$ and $b_k$ (whose explicit form we will not need). By continuity of the second derivatives, $\Delta[h_{n;w}]\Delta[h_{m;w}]\log(Z)$ and $\Delta[h_{m;w}]\Delta[h_{n;w}]\log(Z)$ can be expanded, at every point on ${\cal N}'$, in converging Taylor series:
\beqa
	\Delta[h_{n;w}]\Delta[h_{m;w}]\log(Z) &=& \sum_{j\leq m}\sum_{k=-1}^{n} a_jb_k \,
	(w-v)^{m+n-j-k}
	\Delta[h_{k;v}]\Delta[h_{j;v}]\log(Z),\n
	\Delta[h_{m;w}]\Delta[h_{n;w}]\log(Z) &=& \sum_{j\leq m}\sum_{k=-1}^{n} a_jb_k \,
	(w-v)^{m+n-j-k}
	\Delta[h_{j;v}]\Delta[h_{k;v}]\log(Z). \no
\eeqa
Let us denote by the symbol ``$\equiv$'' equality up to constants on ${\cal N}'$ (this indeed defines an equivalence relation). Let us denote in general by $(f(x))_{K}$ the part of the Taylor series of $f(x)$ in powers of $x$ where only terms whose powers are in the set $K$ are kept; here we have $x=w-v$. The conditions expressed in Definition \ref{deficg}, along with the above Taylor series, lead to
\beqa
	\Delta[h_{n;w}]\Delta[h_{m;w}]\log(Z)
	&\equiv&
	(\Delta[h_{n;w}]\Delta[h_{m;w}]\log(Z))_{\{\ell:\ell\geq m+n+2\}},\n
	\Delta[h_{m;w}]\Delta[h_{n;w}]\log(Z)
	&\equiv&
	(\Delta[h_{m;w}]\Delta[h_{n;w}]\log(Z))_{\{\ell:0\leq \ell\leq m+n+1\}}.
	\label{relexp}
\eeqa
Let us first assume $m+n\geq -1$. Then by the commutation relations (\ref{commh}), we have
\[
	\Delta[h_{n;w}]\Delta[h_{m;w}]\log(Z) 
	\equiv \lt(\Delta[h_{m;w}]\Delta[h_{n;w}]\log(Z)+
	(n-m)\Delta[h_{n+m;w}]\log(Z)\rt)_{\{
	\ell:\ell\ge n+m+2\}}.
\]
Since $\Delta[h_{n+m;w}]$ is a polynomial of order $n+m+1$ in $w$, the second term on the right-hand side is equivalent to 0, and by the second line of (\ref{relexp}), the first term is also equivalent to zero:
\beq\label{1stres}
	\Delta[h_{n;w}]\Delta[h_{m;w}]\log(Z) \equiv 0 \quad (m+n\geq -1).
\eeq
Let us next assume $m+n\leq -1$. Then, on the right-hand side of the second line of (\ref{relexp}) there are no terms left in the partial Taylor series, so that
\beq\label{2ndres}
	\Delta[h_{m;w}]\Delta[h_{n;w}]\log(Z) \equiv 0 \quad (m+n\leq -2).
\eeq
The combination of (\ref{1stres}) and (\ref{2ndres}) shows the lemma.

\begin{lemma}\label{lemmapf2}
There exist holomorphic functions $k_{\ell},\,\ell\in\Z$, with $k_{\ell}(w)=0$ for all $\ell\geq -1$ and for all $\ell\leq -5$, and $c$, on $D$, such that, for all $n\ge-1$, $m\leq -2$, $w\in D$ and $\Sigma\in{\cal N}'$,
\beq\label{relpf2}\lt.\ba{rl}
	\Delta[h_{n;w}] \Delta[h_{m;w}] \log(Z)(\Sigma)& (n+m\ge -1)\\
	\Delta[h_{m;w}]\Delta[h_{n;w}] \log(Z)(\Sigma) & 
	(n+m\le -2)
\ea\rt\} = (n-m)k_{n+m}(w) +
	\frc{c(w)}{12}(n^3-n) \,\delta_{n+m,0}.
\eeq
\end{lemma}
Consider the operators
\beq\label{defLn}
	L(n;w) := \lt\{\ba{ll} Z^{-1} \Delta[h_{n;w}] Z & (n\leq -2) \\
	\Delta[h_{n;w}] & (n\geq -1), \ea\rt.
\eeq
acting on the space of functions smooth on ${\cal N}'$. We use (\ref{commh}) for the commutators. Since (\ref{defLn}) gives $L(n;w)$ as an $n$-independent conjugation of $\Delta[h_{n;w}]$ for all $n\leq -2$, and since these operators are closed under commutations, if $m\leq -2$ and $n\leq -2$ then we have $[L(n;w),L(m;w)] = (n-m)L(n+m;w)$. A similar argument holds in the cases $n\ge 0$ and $m\ge -1$, or $n\ge -1$ and $m\ge 0$, giving the same bracket relations. For $n=m=-1$, the commutator is 0, which agrees again with the same bracket relation. Hence we have
\beq\label{brack}
	[L(n;w),L(m;w)] = (n-m)L(n+m;w) \quad (n\leq -2,\;m\leq -2,\;\mbox{or}\; n\geq-1,\;m\geq-1).
\eeq
Consider $[L(n;w),L(m;w)]$ in the cases $n\geq -1$ and $m\leq -2$ (the opposite is obtained by a sign change, hence it is sufficient to consider these cases). We have (where large parentheses $\Big(\Big)$ contain functions and not differential operators)
\beqa
	[L(n;w),L(m;w)] &=& \lt[\Delta[h_{n;w}],\Big(\Delta[h_{m;w}] \log(Z) \Big)+\Delta[h_{m;w}]\rt] \n
	&=& (n-m)\Delta[h_{n+m;w}] + \Big(\Delta[h_{n;w}] \Delta[h_{m;w}] \log(Z)\Big) \n
	&=& (n-m)L(n+m;w) + \Big(\Delta[h_{n;w}] \Delta[h_{m;w}] \log(Z)\Big) + \n && +
	\lt\{\ba{ll} 0 & (n+m\geq -1)\z (m-n)\Big(\Delta[h_{n+m;w}]\log(Z)\Big)&
	(n+m\leq -2)  \ea\rt. \n
	&=& (n-m)L(n+m;w) + 
	\lt\{\ba{ll} \Big(\Delta[h_{n;w}] \Delta[h_{m;w}] \log(Z)\Big)  & (n+m\geq -1)\z
	\Big(\Delta[h_{m;w}] \Delta[h_{n;w}] \log(Z)\Big) &
	(n+m\leq -2).  \ea\rt.\n \label{virder}
\eeqa
Since $Z$ is a conformal Gaussian, then by Lemma \ref{lemmapf1} we see that the $L(n;w)$s form a central extension of the Witt algebra on the space of smooth functions on ${\cal N}'$. The central extension is unique up to isomorphisms. With the standard choice of basis that we have taken, this imposes the central term in $[L(n;w),L(m;w)]$ to be of the form $(n-m)k_{n+m}(w) + (c(w)/12) (n^3-n) \delta_{n+m,0}$ (see e.g.~\cite{FLM88}). Further, from (\ref{brack}), the cases $n=-1,m\ge 0$ imply $k_\ell=0$ for all $\ell\geq -1$, and the cases $n=-2,m\leq -3$ imply the same for all $\ell\leq -5$. With Lemma \ref{lemder}, this shows the present lemma.

\begin{lemma} \label{lemmapf3}
In Lemma \ref{lemmapf2}, the functions $k_\ell,\,\ell\in\Z$ and $c$ satisfy the following differential equations on $\hC\setminus B$:
\beq\label{eqpf}
	\frc{d}{dw} k_\ell(w) = (\ell+1)k_{\ell-1}(w)\quad (\ell\in\Z), \qquad
	\frc{d}{dw} c(w) = 0,
\eeq
whence there exists complex numbers $d_1$, $d_2$, $d_3$ and $c$  such that $c(w)=c$ and
\beq
	k_{\ell}(w) = \lt\{\ba{ll} d_1w^2 + d_2w + d_3 & (\ell=-2)\\
	-2d_1w - d_2 & (\ell=-3) \\
	d_1  & (\ell=-4) \\
	0 & (\mbox{otherwise}).
	\ea\rt.
\eeq
\end{lemma}
We use (\ref{der}). We apply the complex derivative $\p/\p w$ on the left-hand side of (\ref{relpf2}), and interpret the result in terms of the right-hand side. For $n+m\geq 0$ or $n+m\leq -2$ (note that the case $n=-1$ needs special care), we find
\beqa
	\lefteqn{(n-m)\p k_{n+m} + \frc{\p c}{12} (n^3-n)\,\delta_{n+m,0}} &&\n
	&=& \Big((n+1)(n-m-1) + (m+1)(n-m+1)\Big)k_{n+m-1} +\n &&
	+ \frc{c}{12}\Big((n+1)((n-1)^3 - (n-1)) + (m+1)(n^3-n)\Big)\delta_{n+m,0} \n
	&=& (n-m)(n+m+1)k_{n+m-1}.
\eeqa
Choosing $n=-m$ for $m\leq -2$ we find $\p k_0 - k_{-1} = -\frc{\p c}{24 n} (n^3-n)$ which, since it holds for all $n\geq 2$, implies that $\p c=0$. Hence, we get all equations in (\ref{eqpf}) but for the case $\ell=-1$. This case is obtained from a similar derivation with $n+m=-1$. The difference is that we need to exchange the conformal differential operators after the $w$-derivative using (\ref{commh}). This gives:
\beqa
	\lefteqn{(n-m)\p k_{n+m} } &&\n
	&=& \Big((n+1)(n-m-1) + (m+1)(n-m+1)\Big)(k_{n+m-1} +
		\Delta[h_{n+m-1;w}]\log(Z))\n
	&=& (n-m)(n+m+1)(k_{n+m-1}+\Delta[h_{n+m-1;w}]\log(Z)).
\eeqa
This is indeed $0$ for $n+m=-1$. This completes the proof of the lemma.

\begin{lemma}\label{lemmapf4}
For all $\Sigma\in {\cal N}$, $w\in \hC\setminus \beta(\Sigma)$, $n\ge-1$ and $m\leq -2$,
\beq\lt.\ba{rcll}\displaystyle
	\Delta[h_{n;w}] \Delta[h_{m;w}] \log(Z)(\Sigma)& (n+m\ge -1)\\
	\Delta[h_{m;w}]\Delta[h_{n;w}] \log(Z)(\Sigma) & 
	(n+m\le -2)
\ea\rt\} = (n-m) \,k_{n+m}(w) + 
	\frc{c}{12}\,(n^3-n) \,\delta_{n+m,0} 
\eeq
\end{lemma}
Clearly, as the radius of $D$ tends to 0, then ${\cal N}'$ is an increasing sequence of sets with limit ${\cal N}$. Hence, for every $\Sigma\in{\cal N}$, we can find a disk $D$ such that (\ref{relpf2}) holds for all $w\in D$. Since $k_\ell(w)$ are polynomial (and $c(w)=c$ is constant), and since the conformal derivatives are holomorphic functions of $w$ in $\hC\setminus \beta(\Sigma)$ (Lemma \ref{lemder}), it follows that (\ref{relpf2}) holds for all $w\in\hC\setminus\beta(\Sigma)$.
\eproof

Note that the fact that the form (\ref{relpf}) is optimal is not immediately obvious - {\em a priori} it may be that $d_j=0$ in general. However, we believe that there are examples of conformal Gaussian with $d_j\neq0$ (although this is beyond the scope of this paper).

The number $c$ will be referred to as the {\em exponent} of the conformal Gaussian. A conformal Gaussian such that $d_1=d_2=d_3=0$ in (\ref{relpf}) will be said to be {\em normal}, and we will refer to the triplet $(d_1,d_2,d_3)$ as the {\em anomaly}. Normal conformal Gaussians play an important role in representing the Virasoro algebra on the space of smooth functions on $\cal N$, where the exponent becomes the central charge.

In certain natural cases the anomaly must be zero: when there is translation and scaling invariance. More precisely:
\begin{propo}\label{propostat}
Let $Z$ be a conformal Gaussian on an open set ${\cal N}$. If $Z$ is translation stationary, that is if $\Delta[h_{-1}]Z=0$ on ${\cal N}$, then it has an anomaly of the form $(0,0,d_3)$. If further $Z$ is scaling stationary, that is if $\Delta[h_{0;w}]Z=0$ on ${\cal N}$ for some $w\in\C$ (whence for all $w\in\C$), then it is normal.
\end{propo}
\proof
If $Z$ is translation stationary, then $\Delta[h_{-2;w}]\Delta[h_{-1;w}]\log(Z)=0$, hence from (\ref{relpf}) we must have $k_{-3}(w)=0$, whence $d_1=d_2=0$. If $Z$ is further scaling stationary, then $\Delta[h_{-2;w}]\Delta[h_{0;w}]\log(Z) = 0,$ hence from (\ref{relpf}) we must have $k_{-3}(w)=0$, whence $d_1=d_2=d_3=0$.
\eproof

Further, from a given conformal Gaussian, it is possible to obtain different conformal Gaussians, with exponent 0 and different anomalies, by differentiation:
\begin{propo}\label{propoexponent}
Let $Z$ be a conformal Gaussian on ${\cal N}$ with exponent $c$ and anomaly $(d_1,d_2,d_3)$. Then, on ${\cal N}$:
\begin{enumerate}
\item $\Delta[h_0]\log (Z)$ is a conformal Gaussian with exponent 0 and anomaly $(-4d_1,-3d_2,-2d_3)$;
\item $\Delta[h_{-1}]\log(Z)$ is a conformal Gaussian with exponent 0 and anomaly $(0,2d_1,d_2)$.
\item If $\Delta[h_{-1}]\log(Z)=0$, then $\Delta[h_1]\log(Z)$ is a conformal Gaussian with exponent 0 and anomaly $(0,4d_3,0)$
\end{enumerate}
\end{propo}
\proof For lightness of notation, let $\Delta_{m,n}:=\Delta[h_n]\Delta[h_m]$ if $n+m\ge -1$ and $\Delta_{m,n}:=\Delta[h_m]\Delta[h_n]$ if $n+m\leq -2$. Then, $[\Delta[h_0],\Delta_{m,n}] = -(m+n)\Delta_{m,n}$. Since further $\Delta[h_0]\Delta_{m,n}\log(Z) = 0$ on ${\cal N}$, we find
\[
	\Delta_{m,n}\Delta[h_0]\log(Z) = (m+n)\Delta_{m,n}\log(Z).
\]
Using (\ref{relpf}) at $w=0$ for $\Delta_{m,n}\log(Z)$, the fact that $\Delta[h_0]\log(Z)$ is smooth on ${\cal N}$, and Theorem \ref{theopf}, this shows Point 1.

Let us further use the notation $\Delta_{m,n;w}$ in the obvious way. Then, using (\ref{der}), we have $[\Delta[h_{-1}],\Delta_{m,n;w}] = - \p/\p w \Delta_{m,n;w}$ as operators on smooth functions. Since also $\Delta[h_{-1}]\Delta_{m,n}\log(Z) = 0$ on ${\cal N}$, we find
\[
	\Delta_{m,n;w}\Delta[h_{-1}]\log(Z) = \frc{\p}{\p w}
	\Delta_{m,n;w}\log(Z).
\]
Again using (\ref{relpf}) for $\Delta_{m,n;w}\log(Z)$, the fact that $\Delta[h_0]\log(Z)$ is smooth on ${\cal N}$, and Theorem \ref{theopf}, this shows Point 2.

Finally, we compute
\beq\label{eqpropoexp}
	[\Delta[h_1],\Delta_{m,n}] =
	\lt\{\ba{ll} (1-n)\Delta[h_{n+1}]\Delta[h_{m}]
	+ (1-m)\Delta[h_n]\Delta[h_{m+1}] & (n+m\geq -1)\\
	(1-m)\Delta[h_{m+1}]\Delta[h_{n}]
	+ (1-n)\Delta[h_m]\Delta[h_{n+1}] & (n+m\leq -2).
	\ea\rt.
\eeq
Let us assume that $m\leq -3$ and that $n+m\geq -1$ or $m+n\leq -3$. Then the right-hand side of (\ref{eqpropoexp}) is
\[
	(1-n)\Delta_{m,n+1} + (1-m)\Delta_{m+1,n}.
\]
Since $\Delta[h_1]\Delta_{m,n}\log(Z) = 0$ on ${\cal N}$, we can apply both sides of (\ref{eqpropoexp}) to $\log(Z)$ and use Theorem \ref{theopf} to find
\beq
	\Delta_{m,n}\Delta[h_1]\log(Z)=
	(n-m)(m+n-1)k_{n+m+1}(0).
\eeq
Hence, if $d_1=0$, we obtain the equation for a conformal Gaussian with exponent 0 and anomaly $(5d_2,4d_3,0)$. We now verify that the same hold as well for other required values of $m,n$ (that is, $m\leq-3$ and $m+n=-2$; $m=-2$ and $m+n\geq -1$; $m=-2$ and $m+n\leq -2$) if and only if $\Delta[h_{-1}]\log(Z)=0$, in which case we indeed have $d_1=0$ and also $d_2=0$ by Proposition \ref{propostat}.

Let us assume that $m\leq -3$ and that $m+n=-2$. Then in order to bring the right-hand side of (\ref{eqpropoexp}) to the correct form, we need to exchange the order of the derivatives. We find
\[
	\Delta_{m,n}\Delta[h_1]\log(Z)=
	(n-m)(m+n-1)(k_{n+m+1}(0)-\Delta[h_{m+n+1}]\log(Z)).
\]
Setting $m+n=-2$, and recalling that the anomaly must be of the form $(5d_2,4d_3,0)$ by the previous calculation, we deduce that it is necessary and sufficient to have $\Delta[h_{-1}]\log(Z)=0$ from the analysis of this case (whence $d_2=0$).

Let us assume that $m=-2$ and that $n+m\geq -1$. Applying the right-hand side of (\ref{eqpropoexp}) to $\log(Z)$, the second term on the first line is 0 by the condition $\Delta[h_{-1}]\log(Z)=0$. The first term can be evaluated using Theorem \ref{theopf}, where the only possible nonzero term in (\ref{relpf}) is the term proportional to $c$. This term, here, involves the delta-function $\delta_{m+n+1,0}$, which fixes $n=1$. But for $n=1$, the first term on the first line of the right-hand side of (\ref{eqpropoexp}) is zero, in agreement with an exponent of 0 for $\Delta[h_1]\log(Z)$.

Let us assume that $m=-2$ and that $n+m\leq -2$. There are two possibilities: $n=-1$ or $n=0$. If $n=-1$ then on the right-hand side of (\ref{eqpropoexp}), applied to $\log(Z)$, the first term on the second line is zero thanks to the condition $\Delta[h_{-1}]\log(Z)=0$. The second term can be evaluated directly using (\ref{relpf}) because $m+n+1=-2$. This gives $(1-n)(n-m+1)k_{-2}(0) = 4d_3$, in agreement with the form $(0,4d_3,0)$ of the anomaly. If $n=0$, then the first term on the second line of (\ref{eqpropoexp}) applied to $\log(Z)$ is again zero, because $\Delta[h_{-1}]\Delta[h_0]  = \Delta[h_0]\Delta[h_{-1}] - \Delta[h_{-1}]$ and thanks to the condition $\Delta[h_{-1}]\log(Z)=0$. The second term is
\[
	\Delta[h_{-2}]\Delta[h_1]\log(Z) =
	\Delta[h_1] \Delta[h_{-2}]\log(Z) -3 \Delta[h_{-1}]\log(Z)
	=0
\]
where the last equation holds thanks to (\ref{relpf}) and again the condition $\Delta[h_{-1}]\log(Z)=0$. This is again in agreement with an anomaly $(0,4d_3,0)$.
\eproof

Note that this implies that for any $w\in \cap_{\Sigma\in{\cal N}}(\hC\setminus \beta(\Sigma))$, the function $\Delta[h_{0;w}]\log(Z)$ is a conformal Gaussian on ${\cal N}$ with exponent $0$ and anomaly $(-4d_1,2d_1 w-3d_2,d_2w-2d_3)$.

\subsection{Transformation properties}

There is an alternative description of normal conformal Gaussians, where a part of the local characterization is ``globalized''. That is, normal conformal Gaussians can be characterized by specifying a certain transformation property of a single derivative, instead of specifying the form of second derivatives.

In order to obtain this, we need that certain derivatives of $\log(Z)$ be functions that are holomorphic. We refer to this as follows. For a function $f$ on an open subset ${\cal N}\subseteq\Omega$, that is twice differentiable at $\Sigma\in{\cal N}$, we define:
\begin{proper} \label{propholom} $\Delta[h_{-2;w}]f$ is $\alpha$-holomorphic at $\Sigma$ for every $w\in\hC\setminus \beta(\Sigma)$ and every $\Sigma\in\cal N$.
\end{proper}

Then we have the following theorem.
\begin{theorem}\label{theoZ}
Let $Z$ be a smooth function on an open subset ${\cal N}\subseteq \Omega$. Consider the following two statements:
\begin{enumerate}
\item The relations
\beq\label{transfZ}
	\Delta[h_{-2;w}] \log(Z)(\Sigma) =
	(\p g(w))^2 \Delta[h_{-2;g(w)}] \log(Z)(g\cdot \Sigma) +
	\frc{c}{12} \{g,w\}
\eeq
hold for all $\Sigma\in{\cal N}$, for all $w\in \hC\setminus \beta(\Sigma)$ and for all conformal maps $g$ in a set $Q(\Sigma)$ that have actions on $\Sigma$, where $\{g,w\}$ is the Schwarzian derivative of $g$ at $w$.
\item $Z$ is a normal conformal Gaussian on ${\cal N}$ with exponent $c$ and $\log(Z)$ has Property \ref{propholom} on ${\cal N}$.
\end{enumerate}
If Statement 1 holds for $Q(\Sigma)$ a $\C$-neighbourhood of the identity, then Statement 2 holds. If ${\cal N}$ is piecewise $\alpha$-smoothly connected and if Statement 2 holds, then Statement 1 holds for $Q(\Sigma)$ the set of all $\alpha$-maps at $\Sigma$ such that $g\cdot\Sigma\in{\cal N}$.
\end{theorem}
Note that thanks to Lemma \ref{lemconnect}, we may take ${\cal N} = \Omega$ or more generally ${\cal N}=\Omega(K)$ for certain closed subsets $K\subset \hC$ (as described in the lemma) for the part of theorem that goes from Statement 2 to Statement 1.

\noindent\proof  1. From Statement 1 to Statement 2.

Let us re-write the second relation of (\ref{relpf}), using (\ref{commdelta}), as
\beq\label{2nd}
	\Delta[h_n]\Delta[h_m] \log(Z) + (m-n)\Delta[h_{n+m}]\log(Z)=0\qquad(n+m\leq -2).
\eeq
Let $n\geq -1$ and $w'\in\C$. Then the functions $h_{n;w'}$ are holomorphic on $\C$, hence the derivative $\Delta[h_{n;w'}]$ is a $\C$-derivative. Then, we evaluate the following second-order derivatives using (\ref{transfZ}) and Lemma \ref{lemder}, for $n\geq -1$:
\beqa
	\lefteqn{\Delta[h_{n;w'}]\Delta[h_{-2;w}]\log(Z)(\Sigma)}&& \n &=&
	\Delta[h_{n;w'}\,|\,g]\Big(\Delta[h_{-2;w}]\log(Z)(g\cdot \Sigma) \Big) \n &=&
	\Delta[h_{n;w'}\,|\,g]\Big(
	(\p g^{-1}(w))^2 \lt(\Delta[h_{-2;g^{-1}(w)}]\log(Z)(\Sigma)
	- \frc{c}{12} \{g,g^{-1}(w)\}\rt)\Big)\n
	&=&
	-2\p h_{n;w'}(w) \Delta[h_{-2;w}]\log(Z)(\Sigma)
	- h_{n;w'}(w) \frc{\p}{\p w} \Delta[h_{-2;w}]\log(Z)(\Sigma)
	- \frc{c}{12} \p^3 h_{n;w'}(w)\n &=&
	2(n+1)(w'-w)^n\, \Delta[h_{-2;w}] \log(Z)(\Sigma) + (w'-w)^{n+1}\Delta[h_{-3;w}]\log(Z)(\Sigma)+\n && +\,
	\frc{c}{12} (n^3-n) (w'-w)^{n-2} .
	\label{exprdd}
\eeqa

We may evaluate the necessary second-order derivative from this for all $n\geq-1$ and $m\leq -2$ using (\ref{der}), that is
\beq
	\Delta[h_{n;w}]\Delta[h_{m;w}]\log(Z) = 
		\lt(\frc{(-1)^{-2-m}}{(-2-m)!}\frc{\p^{-2-m}}{\p w^{-2-m}}
		\Delta[h_{n;w'}]  \Delta[h_{-2;w}]\log(Z)\rt)_{w'=w}.
\eeq
Consider the case $n+m\geq -1$. This means that $-2-m\leq n-1$, hence the derivative with respect to $w$ at $w'=w$ will only be nonzero on the $c$-dependent term in (\ref{exprdd}), and this only for $m+n=0$, giving the first relation of (\ref{relpf}) with $c_n = (c/12)(n^3-n)$. On the other hand, if $n+m\leq -2$, then $-2-m\geq n$ so that the $c$-dependent term gives zero under the $w$ derivatives. The first two terms in (\ref{exprdd}) give nonzero contributions; the first is evaluated by applying exactly $n$ derivatives on $(w'-w)^n$ and $-2-m-n$ on the other factor, and the second, by applying $n+1$ derivatives on $(w'-w)^{n+1}$ and $-2-m-n-1$ on the other factor. Using (\ref{der}), the first term becomes $2(n+1)\Delta[h_{n+m}]\log(Z)$, and the second, $(-2-m-n)\Delta[h_{m+n}]\log(Z)$. Hence, the sum is $(n-m)\Delta[h_{n+m}]\log(Z)$ and we recover (\ref{2nd}), which is equivalent to the second relation of (\ref{relpf}).

Holomorphicity of $\Delta[h_{-2;w}]\log(Z)$ is shown simply by
\beqa
	\b\Delta[\b{h}_{n}]\Delta[h_{-2;w}]\log(Z)&=&
	\b\Delta[\b{h}_{n}\,|\,\b{g}]\Big(
	(\p g^{-1}(w))^2 \lt(\Delta[h_{-2;g^{-1}(w)}]\log(Z)
	- \frc{c}{12} \{g,g^{-1}(w)\}\rt)\Big)\n
	&=&0
	\no
\eeqa
for all $n\ge -1$.

\noindent 2. From Statement 2 to Statement 1.

Let $\sigma: [0,1]\to {\cal N} : t\mapsto \Sigma_t:=g_t\cdot\Sigma$ be a smooth $\alpha$-path starting at $\Sigma$. Then ${\cal D}\sigma(t)\in\spa^>(A)$ for a neighbourhood $A$ of $\alpha(\Sigma)$ for all $t\in[0,1]$. Consider the function
\beq\label{Ft}
	F(t):= (\p g_t(w))^2 \Delta[h_{-2;g_t(w)}] \log(Z)(g_t\cdot\Sigma)
	+ \frc{c}{12}\{g_t,w\}
\eeq
for $t\in [0,1]$. Clearly, $F(0) = \Delta[h_{-2;w}]\log(Z)(\Sigma)$. Also, thanks to Lemma \ref{lemsimder} (first part) and to the fact that $\p g_t(w)$ and $\{g_t,w\}$ are continuous and differentiable functions of $t$, we see that $t\mapsto F(t)$ is continuous and differentiable on $[0,1]$. From Lemma \ref{lemchainrule}, Equation (\ref{simder}), and the assumption that $\Delta[h_{-2;w}]\log(Z)$ is $\alpha$-holomorphic, we may evaluate the following derivative for any $t\in (0,1)$
\beqa
	\frc{d}{dt} F(t) &=&
	2\,(\p g_t(w))^2 \,\p {\cal D}\sigma(t)(g_t(w))\,
	\Delta[h_{-2;g_t(w)}]\log(Z)(g_t\cdot\Sigma) - \n&&
	-\; (\p g_t(w))^2\, {\cal D}\sigma(t)(g_t(w))\,
	\Delta[h_{-3;g_t(w)}]\log(Z)(g_t\cdot\Sigma) + \n &&
	+\;(\p g_t(w))^2\,
	\Delta[{\cal D}\sigma(t)]
	\Delta[h_{-2;g_t(w)}]\log(Z)(g_t\cdot\Sigma) +\n &&
	+\;\frc{c}{12} \p^3 {\cal D}\sigma(t)(g_t(w)) \,(\p g_t(w))^2
	\label{dFt}
\eeqa
where we used the formula
\[
	\frc{d}{dt}\{g_t,w\} = \p^3 {\cal D}\sigma(t)(g_t(w)) \, (\p g_t(w))^2
\]
which holds thanks to (\ref{gtsigma}). Thanks to the fact that $Z$ is a normal conformal Gaussian and that ${\cal D}\sigma(t)$ is holomorphic on $A$ for all $t\in[0,1]$, we can use Lemma \ref{lemexph} for evaluating the derivative $\Delta[{\cal D}\sigma(t)]$ on the function $\Delta[h_{-2;g_t(w)}]\log(Z)$. Hence, using formula (\ref{exph}) with $w$ replaced by $g_t(w)$ and $h$ replaced by ${\cal D}\sigma(t)$, we obtain
\beqa
	\lefteqn{\Delta[{\cal D}\sigma(t)]
	\Delta[h_{-2;g_t(w)}]\log(Z)}
	&& \n
	&=&  {\cal D}\sigma(t)(g_t(w))\,\Delta[h_{-3;g_t(w)}]\log(Z)
	- 2\, \p {\cal D}\sigma(t)(g_t(w))\,
	\Delta[h_{-2;g_t(w)}]\log(Z) -\n &&
	-\;\frc{c}{12} \p^3{\cal D}\sigma(t)(g_t(w)).\no
\eeqa
Putting this into (\ref{dFt}), we find that $dF(t)/dt=0$ for all $t\in [0,1]$, from which we deduce that $F(1)=F(0)$. By Definition \ref{deficonnect} of piecewise $\alpha$-smoothly connected open sets, for every $\Sigma\in{\cal N}$ and every $\alpha$-map $g$ at $\Sigma$ such that $g\cdot\Sigma\in{\cal N}$, $\Sigma$ and $g\cdot\Sigma$ are connected by a piecewise smooth $\alpha$-path, hence (\ref{transfZ}) follows for all $\alpha$-maps $g$ such that $g\cdot\Sigma\in{\cal N}$.
\eproof

The transformation property of Theorem \ref{theoZ} immediately gives us, by differentiating with respect to $w$, transformation properties for derivatives $\Delta[h_{m;w}]\log(Z)$ for all $m\leq -2$. Further, it allows us to give a much more general transformation property, for general multiple derivatives. This is written using a recursion relation, as expressed in the following proposition.

Let us introduce some notation. As we did in the previous section, given $k\in\N$, let us denote by $\N_k=\{1,2,\ldots,k\}$ the set of the first $k$ integers. Here we will also include the case $k=0$, with $\N_0=\emptyset$. For every subset $S=\{j_1,j_2,\ldots,j_{|S|}\} \subseteq \N_k$ (with by convention $j_1<j_2<\cdots<j_{|S|}$), let us use the notation $w_S=\{w_{j_1},w_{j_2},\ldots,w_{j_{|S|}}\}$ as well as
\beq
	\Delta_{-2;w_S} =
	\Delta[h_{-2;w_{j_{|S|}}}]\Delta[h_{-2;w_{j_{|S|-1}}}]
	\cdots \Delta[h_{-2;w_{j_1}}],
\eeq
with by convention $\Delta_{-2;w_S}=1$ if $S=\emptyset$. Given any such $S$ and $k$, we will use the symbol $P_S^{(k)}(g|w_{\N_k})$ to represent a differential operator of finite order in $\p/\p w_{j_l},\;l=1,\ldots,|S|$, with coefficients that are rational functions of the differences $w_{j}-w_{j'}$ for $j,j'\in\N_k$ and of the derivatives $\p^r g(w_{j})$ for $j\in\N_k,\;r\geq 0$, such that these coefficients are holomorphic functions of $w_{j}$ for every $j\in\N_k$ and in every domain of $\C$ where $g$ is defined and conformal. 
\begin{propo} \label{proporecursion}
Let $Z$ be a smooth function on an open subset ${\cal N}\subseteq \Omega$. Assume that for some $k_0\ge 0$, there exist $P_S^{(k_0)}(g|w_{\N_{k_0}})$ as described above such that, for all $\Sigma\in{\cal N}$, for all $w_1,w_2,\ldots,w_{k_0}\in\hC\setminus \beta(\Sigma)$ and for all $\alpha$-maps $g$ at $\Sigma$,
\beq\label{transfZgen0}
	\Delta_{-2;w_{\N_{k_0}}}
	Z(\Sigma) =
	\sum_{S\subseteq\N_{k_0}} P_S^{(k_0)}(g|w_{\N_{k_0}})
	\Delta_{-2;g(w_S)} Z(g\cdot\Sigma).
\eeq
Then, there exist $P_S^{(k)}(g|w_{\N_k})$ such that, for all $\Sigma\in{\cal N}$, for all $k\in\N,\;k\geq k_0$, for all $w_1,w_2,\ldots,w_k\in\hC\setminus \beta(\Sigma)$ and for all $\alpha$-maps $g$ at $\Sigma$,
\beq\label{transfZgen}
	\Delta_{-2;w_{\N_k}}
	Z(\Sigma) =
	\sum_{S\subseteq\N_k} P_S^{(k)}(g|w_{\N_k})
	\Delta_{-2;g(w_S)} Z(g\cdot\Sigma).
\eeq
Further, the following recursion relation holds:
\beqa
	P_S^{(k+1)}(g|w_{\N_{k+1}}) &\stackrel{k+1\not\in S}=&
	\frc{d}{d\tau} \lt.P_S^{(k)}(g+\tau u|w_{\N_k})\rt|_{\eta=0}
	+
	P_S^{(k)}(g|w_{\N_k})
	\sum_{l=1}^{|S|}
	\frc{u(w_{j_l})}{\p g(w_{j_l})} \frc{\p}{\p w_{j_l}}
	\n
	&\stackrel{k+1\in S}=&
	P_{S\setminus\{k+1\}}^{(k)}(g|w_{\N_k})
	(\p g(w_{k+1}))^2\,
\label{recursion}
\eeqa
where
\beq\label{fonctionu}
	u(z) = \frc{(\p g(w_{k+1}))^2}{g(w_{k+1})-g(z)}-
	\frc{\p g(z)}{w_{k+1}-z}.
\eeq
\end{propo}
\proof First, a straightforward analysis shows that if $P_S^{(k)}(g|w_{\N_k})$ has the properties stated in the paragraph above the proposition for all subsets $S\subseteq \N_k$, then so does $P_S^{(k+1)}(g|w_{\N_{k+1}})$ as defined by the recursion relation (\ref{recursion}) for all subsets $S\subseteq \N_{k+1}$. In particular, the function (\ref{fonctionu}) has no pole at $z=w_{k+1}$.

We proceed by induction; hence let us assume that (\ref{transfZgen}) holds for some $k\in \N$. Then, for $\Sigma\in{\cal N}$, for $g$ an $\alpha$-map at $\Sigma$ such that $g\cdot\Sigma\in{\cal N}$, and for $w_1,w_2,\ldots,w_{k},w\in\hC\setminus \beta(g\cdot\Sigma)$, we have
\beq\label{eq1tg}
	\Delta[h_{-2;w}] \Delta_{-2;w_{\N_k}}
	Z(g\cdot \Sigma)
	=
	\Delta[h_{-2;w}\,|\,g']\Delta_{-2;w_{\N_k}}
	Z((g'\circ g)\cdot\Sigma).
\eeq
Let $g_\tau' = \id + \tau h_{-2;w}$ for $\tau$ in a neighbourhood of 0 in $\C$. For all $|\tau|$ small enough, these maps are in a $B$-neighbourhood of the identity, for some simply connected domain $B\supset \beta(g\cdot \Sigma)$, because $w\in\hC\setminus \beta(g\cdot \Sigma)$. Let $A\supset \alpha(\Sigma)$ be a simply connected domain where $g$ is conformal.
A Taylor expansion in $\tau$ shows that, compactly on $g(A)\cap B$, the following limit exists:
\beq\label{taylinv}
	\lim_{\tau\to0}
	\frc{\t{g}_\tau\circ \t{g}_\tau' \circ (g_\tau' \circ g)^{-1}-\id}{\tau^2},
\eeq
where
\beq\label{formgeta}
	\t{g}_\tau(z) = g(z) + \tau\lt(\frc1{w-g(z)}
	- \frc{(\p g^{-1}(w))^2\,\p g(z)}{g^{-1}(w)-z}\rt),\quad
	\t{g}_\tau'(z) = z + \tau \frc{(\p g^{-1}(w))^2}{g^{-1}(w)-z}.
\eeq
Hence,
\beqa
	\lefteqn{\Delta[h_{-2;w}] \Delta_{-2;w_{\N_k}}
	Z(g\cdot \Sigma)} &&\n
	&=&
	\frc{d}{d\tau} \lt(\Delta_{-2;w_{\N_k}}
	Z((g_\tau'\circ g)\cdot\Sigma)\rt)_{\tau=0} \n
	&=&
	\lim_{\tau\to0}\tau^{-1}
	\lt(\Delta_{-2;w_{\N_k}}
	Z((g_\tau'\circ g)\cdot\Sigma)
	-
	\Delta_{-2;w_{\N_k}}
	Z((\t{g}_\tau\circ \t{g}_\tau' \circ (g_\tau' \circ g)^{-1})\cdot
	(g_\tau'\circ g)\cdot\Sigma)\,+
	\rt. \n && \qquad\lt.
	+\,
	\Delta_{-2;w_{\N_k}}
	Z(\t{g}_\tau\circ \t{g}_\tau' \cdot\Sigma)
	-
	\Delta_{-2;w_{\N_k}}
	Z(\Sigma)
	\rt)
	\n
	&=&
	\frc{d}{d\tau} \lt(\Delta_{-2;w_{\N_k}}
	Z((\t{g}_\tau\circ \t{g}'_\tau)\cdot\Sigma)\rt)_{\tau=0}
	\label{exprdertt}
\eeqa
where in the first equality we used (\ref{eq1tg}) and Lemma \ref{lemtau}, and in the third we used smoothness of $Z$ on ${\cal N}$ (in particular, second-differentiability of $\Delta_{-2;w_{\N_k}}Z$) and (\ref{taylinv}). From (\ref{formgeta}), we obtain
\beq\label{abab}
	\t{g}_\tau^{-1}(z) = g^{-1}(z)-\tau\lt(
	\frc{\p g^{-1}(z)}{w-z} - \frc{(\p g^{-1}(w))^2}{g^{-1}(w)-g^{-1}(z)}\rt)
	+O(\tau^2)
\eeq
compactly on $g(A)$. Since for every $\tau$ small enough, $\t{g}_\tau$ is an $\alpha$-map at $\t{g}'_\tau\cdot \Sigma$, then we can use the assumption that (\ref{transfZgen}) holds. Hence, from (\ref{abab}) and (\ref{exprdertt}), we obtain
\beqa
	\lefteqn{\Delta[h_{-2;w}] \Delta_{-2;w_{\N_k}}
	Z(g\cdot \Sigma)} && \n
	&=&
	\frc{d}{d\tau}\lt(
	\sum_{S\subseteq\N_k} P_S^{(k)}(\t{g}^{-1}_\tau|w_{\N_k})
	\Delta_{-2;\t{g}^{-1}_\tau(w_S)} Z(\t{g}'_\tau\cdot \Sigma)\rt)_{
	\tau=0} \n
	&=& \sum_{S\subseteq\N_{k+1}}
	P_S^{(k+1)}(g^{-1}|w_{\N_{k+1}})
	\Delta_{-2;g^{-1}(w_S)} Z(\Sigma)
\eeqa
where $P_S^{(k+1)}(g^{-1}|w_{\N_{k+1}})$ is defined by (\ref{recursion}) with the identification $w_{k+1}= w$, and where we used Lemma \ref{lemsimder} (first part) and Equation (\ref{simder}).
\eproof

\begin{rema}{\em 
Using the recursion relation (\ref{recursion}), it is simple to see that the differential operators $P_S^{(k)}(g|w_{\N_k})$ in fact have the following simplifying property: if $S$ contains all integers between $j$ and $k$ for some $j$ such that $k_0+1\leq j\leq k$, then
\[
	P_S^{(k)}(g|w_{\N_k})
	= \lt(\prod_{i=j}^{k} (\p g(w_j))^2\rt)
	P^{(j-1)}_{S\setminus \{j,\ldots,k\}}
	(g|w_{\N_{j-1}}).
\]}
\end{rema}

\begin{theorem}\label{theoZgen}
Let $Z$ be a normal conformal Gaussian with exponent $c$ on an open subset ${\cal N}\subseteq \Omega$ that is piecewise $\alpha$-smoothly connected, and assume that $\log(Z)$ has Property \ref{propholom} on ${\cal N}$. Then, there exist $P_S^{(k)}(g|w_{\N_k})$ as described above such that, for all $\Sigma\in{\cal N}$, for all $k\in\N$, for all $w_1,w_2,\ldots,w_k\in\hC\setminus \beta(\Sigma)$ and for all $\alpha$-maps $g$ at $\Sigma$, the transformation property (\ref{transfZgen}) holds, with the recursion relation (\ref{recursion}) as well as the initial conditions
\beq\label{initcond}
	P_{\{1\}}^{(1)}(g|\{w\}) = (\p g(w))^2,\quad
	P_{\emptyset}^{(1)}(g|\{w\}) = \frc{c}{12}\{g,w\}.
\eeq
\end{theorem}
\proof Thanks to Theorem \ref{theoZ}, Equation (\ref{transfZ}) holds for $Q(\Sigma)$ the set of all $\alpha$-maps at $\Sigma$. Hence, the assumptions of Proposition \ref{proporecursion} hold, with $k_0=1$ and (\ref{initcond}). In particular, for every $S$, $P_S^{(1)}(g|\{w\})$ is a rational function of the derivatives of $g$ at $w$, and is holomorphic in subdomains of $\C$ where $g$ is conformal. The theorem follows.
\eproof

Although we have not provided a proof of existence of solutions to the equations defining conformal Gaussians, there are conjectured examples, albeit somewhat implicit, of normal conformal Gaussians: the relative partition function of CFT constructed in \cite{DCFT}, and that of CLE constructed in \cite{DCLE}. Both are functions, defined in different context, that are argued to satisfy the transformation property (\ref{transfZ}) (and conjecturally, they are the same function under the identification of CLE with CFT).

These relative partition functions display the additional property of M\"obius invariance. The concept of normal conformal Gaussian is indeed closely related to M\"obius invariance, although it does not immediately imply it. The fact that it is closely related can be seen from the quadratic-differential transformation property for $\Delta[h_{-2;w}]\log (Z)(\Sigma)$ under M\"obius maps (the specialisation of (\ref{transfZ}) to $g$ being a M\"obius map). This transformation property for the derivative $\Delta[h_{-2;w}]f(\Sigma)$ is indeed a consequence of M\"obius invariance of $f$ in general \cite[Thm 3.2, Eq 3.66]{DCFT}. Note that for conformal Gaussians, we have a slightly stronger theorem: translation and scale invariance are sufficient in order to guarantee that $Z$ be a normal conformal Gaussian, hence that it transforms as (\ref{transfZ}), see Proposition \ref{propostat}. It is worth remarking that in the context of CFT, field-theoretic considerations involving the principle of locality lead to the conjecture that a model of CFT that is translation and scale invariant is necessarily M\"obius invariant (and, further, local-conformally invariant, in an appropriate sense). The present discussion gives something near to that, as if the concept of conformal Gaussian encompassed, in a sense, the idea of locality of CFT.

\sect{Virasoro vertex operator algebra structure}

In this section, we will show that the set of smooth functions on an open set ${\cal N}$ on which a normal conformal Gaussian exists forms a module for the Virasoro algebra, and that certain subsets form highest-weight modules. In this construction, Virasoro operators $L_n$ are represented via conformal derivatives $\Delta[h_n]$, which are conjugated by a normal conformal Gaussian $Z$ for $n\leq -2$. The standard vertex operator algebra construction then provides a vertex operator algebra structure as well as modules. Further, we will show if the transformation law (\ref{transfZgen}) holds on ${\cal N}$ (e.g.~if ${\cal N}$ is piecewise $\alpha$-smoothly connected), the vertex operator map has a simple expression in terms of covariant conformal derivatives. A covariant conformal derivative is defined by ``parallel transport'' of (in general) a product of conjugated conformal derivatives $Z^{-1}\Delta[h_{m;w}]Z$ for $m\leq -2$ and fixed $w$. It has the property that a product of covariant conformal derivatives at different values of $w$ transforms like the tensor product of the underlying conformal derivatives. This transformation property makes a connection with the multi-linear structure of correlation functions of conformal field theory; hence our construction encompasses both the structures of multi-linear correlation functions, and of vertex operator algebra.

\subsection{Virasoro algebra modules}

Recall that the Virasoro algebra is the vector space spanned by the basis elements $\{L_n:n\in\Z\}\cup\{C\}$ over $\C$, with the bracket relations
\beq\label{viralg}
	[L_m,L_n] = (m-n)L_{m+n} +
	\frc{C}{12}(m^3-m)\,\delta_{m+n,0},\quad
	m,n\in\Z
\eeq
and where $C$ is central. We start by showing how to construct the Virasoro algebra using conformal derivatives -- this was in fact essentially done in proving Theorem \ref{theopf}.

\begin{theorem}
Let $Z$ be a smooth function on an open subset ${\cal N}\subseteq\Omega$, and define
\beq\label{defLn0}
	L(n) := \lt\{\ba{ll} (-1)^nZ^{-1} \Delta[h_{n;w}] Z & (n\leq -2) \\
	(-1)^n \Delta[h_{n;w}] & (n\geq -1) \ea\rt.
\eeq
as operators on the space of smooth functions on ${\cal N}$. Then the $L(n)$s satisfy the bracket relations of the Virasoro algebra,
\beq\label{vir}
	[L(m),L(n)] = (m-n)L(m+n) + \frc{c}{12}(m^3-m)\,\delta_{m+n,0},\quad
	m,n\in\Z,
\eeq
if and only if $Z$ is a normal conformal Gaussian on ${\cal N}$ with exponent $c$.
\end{theorem}
\proof Note that (\ref{defLn0}) agrees with (\ref{defLn}), up to a factor $(-1)^n$, at $w=0$. Assume that (\ref{vir}) holds. Then from the derivation (\ref{virder}), which only assumes that $Z$ is smooth, we have
\[
	[L(n),L(m)] = (n-m)L(n+m)+(-1)^{n+m}\lt\{\ba{ll}
	 \Big(\Delta[h_{n}] \Delta[h_{m}] \log(Z)\Big) 
	 & (n+m\geq -1)\z
	\Big(\Delta[h_{m}] \Delta[h_{n}] \log(Z)\Big) &
	(n+m\leq -2). \ea\rt.
\]
Hence, we find (\ref{relpf}) with $k_\ell(w)=0$, so that $Z$ is a normal conformal Gaussian with exponent $c$. On the other hand, explicitly performing all possible commutators as in the proof of Lemma \ref{lemmapf2}, we find that if $Z$ is a normal conformal Gaussian with exponent $c$, then (\ref{vir}) holds on the space of smooth functions on ${\cal N}$.
\eproof

\begin{corol}
The space ${\cal F}({\cal N})$ of smooth functions on a subset ${\cal N}\subseteq \Omega$ on which a normal conformal Gaussian $Z$ of exponent $c$ exists forms a module for the Virasoro algebra (\ref{viralg}), with action $L_n\cdot f = L(n)f$ and $C\cdot f = cf$ for all $f\in{\cal F}({\cal N})$ and $n\in\Z$.
\end{corol}

The most interesting modules in the theory of infinite-dimensional Lie algebras are those with certain highest-weight properties. For the Virasoro algebra, a highest-weight module $V$ is one where there exists a vector $v$ such that $V$ is generated by $v$, that $L_n\cdot v=0$ for all $n\geq 1$ and that $L_0\cdot v = \nu v$ for some number $\nu\in\C$, the weight of the vector $v$. In particular, a highest-weight module with $\nu=0$ and with $L_{-1}\cdot v =0$ is an identity module; this is a module which acquires the structure of the Virasoro vertex operator algebra in the standard construction (see e.g.~\cite{LL04}).

It is a simple matter to construct both the identity module, and highest-weight modules with any weight $\nu\in\C$, on spaces of smooth functions. Given an open subset ${\cal N}$ on which a normal conformal Gaussian exists, any function $f_{\bf 1}$ that is invariant under $\alpha$-maps ($\alpha$-invariant function) generates, under application of the operators $L(m):m\leq -2$, an identity module. Indeed, in this case it is clear that $\Delta[h_n]f_{\bf 1} = 0$ for all $n\geq -1$. A simple example, which does give rise to a nontrivial identity module, is the identity function $f_{\bf 1}=1$. Another explicit example $\t{f}_{\bf 1}$ of such an $\alpha$-invariant function is as follows. Let $\p A\subset \gamma(\id)$ be a Jordan curve bounding a domain $A\ni 0$, and let $x,y,z,q\in \p A$. Then, given $\Sigma\in{\cal N}$, let $g_\Sigma$ be the unique map (by the Riemann mapping theorem) that is conformal on the simply connected domain $A'$ bounded by $\Sigma(\p A)$ and containing 0, extended by continuity to a map on the closure of $A'$ (by Carath\'eodory's theorem), such that $g_\Sigma(A') = A$ and that $g_\Sigma(\Sigma(x)) = x$, $g_\Sigma(\Sigma(y)) = y$ and $g_\Sigma(\Sigma(z)) = z$. Then, we may define $\t{f}_{\bf 1}(\Sigma):=g_\Sigma(\Sigma(q))$. Given an $\alpha$-map $g$ at $\Sigma$, by the uniqueness of $g_{g\cdot \Sigma}$, we have $g_{g\cdot \Sigma} = g_{\Sigma}\circ g^{-1}$, whence $\t{f}_{\bf 1}(g\cdot \Sigma)=(g_\Sigma\circ g^{-1})(g(\Sigma(q))) =\t{f}_{\bf 1}(\Sigma)$.

A highest-weight module of weight $\nu$ can be similarly constructed as follows. Let ${\cal N}$, $\Sigma$, $A$ and $A'$ be as above, and let $\t{g}_\Sigma$ be the unique conformal map on $A'$ such that $\t{g}_\Sigma(A')=A$, that $\t{g}_\Sigma(0)=0$ and that $\p \t{g}_\Sigma(0)>0$. Then the number $r_\Sigma:=\p \t{g}_\Sigma(0)$ is a version of the conformal radius of $A'$ with respect to the point 0 (in the usual definition of the conformal radius, $A'$ must be mapped to the unit disk). It has the property that if $g$ is an $\alpha$-map at $\Sigma$ with $g(0)=0$, then $r_{g\cdot\Sigma} = \p (\t{g}_\Sigma\circ g^{-1})(0) = (\p g(0))^{-1}\,r_\Sigma$.  Let us defined $f_\nu(\Sigma):= (r_\Sigma)^\nu$. Since $h_n(0)=0$ for all $n\geq 0$, then we have
\beqa
	\Delta[h_0]f_\nu(\Sigma) &=& 
	\lt.\frc{d}{d\tau}
	f_\nu((\id + \tau h_0)\cdot \Sigma)\rt|_{\tau=0}
	= \lt.\frc{d}{d\tau} (1-\tau)^{-\nu} \rt|_{\tau=0}
	f_\nu(\Sigma)
	= \nu f_\nu(\Sigma)\n
	\Delta[h_n]f_\nu(\Sigma)&=&
	\lt.\frc{d}{d\tau}
	f_\nu((\id + \tau h_n)\cdot \Sigma)\rt|_{\tau=0}
	= \lt.\frc{d}{d\tau} (1)^\nu \rt|_{\tau=0}
	f_\nu(\Sigma)
	= 0 \quad (n\geq 1)
\eeqa
where we used $\p h_n(0)=0$ for $n\geq 1$, and Lemma \ref{lemtau}. The module is then generated by application of conformal derivatives on $f_\nu$.

In a similar fashion, it is also a simple matter to construct logarithmic modules: define $f_{\nu,\rho}(\Sigma):=\rho\,(r_\Sigma)^\nu\,\log(r_\Sigma)$. Then $\Delta[h_0]f_{\nu,\rho}(\Sigma) = \nu f_{\nu,\rho}(\Sigma) + \rho f_{\nu}(\Sigma)$. Hence in this case, as is usual in logarithmic modules, the action of $L(0)$ on the two-dimensional subspace $\C f_{\nu,\rho} \oplus \C f_{\nu}$ of smooth functions is not diagonalisable.

Note that $f_0=1=f_{\bf 1}$ generates the identity module. Note also that all functions $f_{\bf 1}$, $f_\nu,\;\nu\in\C$ and $f_{\nu,\rho},\;\nu,\rho\in\C$ constructed above are smooth on $\Omega$. We will not provide the proof of this, but below we only need $f_{\bf 1}$, which  is obviously smooth.

\subsection{Vertex operator algebra structure}

An identity module for the Virasoro algebra has the structure of a vertex operator algebra \cite[Thm 6.1.5]{LL04}. Further, highest-weight modules give rise to vertex operator algebra modules \cite[Thms 6.1.9, 6.1.2]{LL04}. The construction is standard and we refer the reader to \cite{LL04}. We recall that a vertex operator algebra structure on a quasi-finite vector space $V$ is a structure whereby an infinity of products are defined, $V\otimes V\to V: (v,w)\mapsto v_n w$ for every $n\in\Z$ (that is, for every $v\in V$ and $n\in\Z$, we associate an endomorphism $v_n\in{\rm End}(V)$), with commutativity and associativity properties. We recall in Appendix \ref{appvoa} for the full definition.

Our main result here is the relation between the vertex operator map and conformal derivatives (which we will express in the context of the identity module only).

\subsubsection{The spaces $\Omega_j$}

Recall that the space $\Omega$ is the space of maps conformal on neighbourhoods of a given continuum $E=\gamma(\id)$ whose complement has exactly two components, such that $0$ and $\infty$ lie in different components of $\hC\setminus E$ (the full requirements on the conformal maps in $\Omega$ are expressed in Section \ref{sectOmega}). Recall also the open subsets $\Omega(K)$ of Lemma \ref{lemOmegar}, which, for appropriate closed subsets $K$ as in Lemma \ref{lemconnect} (in particular, for $\Omega(\{0\})=\Omega$), have the property that they are piecewise $\alpha$-smoothly connected. Let us consider a sequence of increasing continua $E_0,\,E_1,\,E_2,\,\ldots$ with $E_0=E$, all of which have the two properties stated above for $E$, such that $\cl{\hC\setminus E_{j+1}}\subset \hC\setminus E_j$ for all $j=1,2,\ldots$. As from $E$ we construct $\Omega(K)$, let us construct the topological spaces $\Omega_0,\,\Omega_1,\,\Omega_2,\,\ldots$ from $E_0,\,E_1,\,E_2,\,\ldots$ (with $\Omega_0=\Omega(K)$), for some $K$ as in Lemma \ref{lemconnect}. It is a simple matter to see that, under the identification of maps with their analytic continuation from $E_j$ to $E_{j+1}$ ($j=0,1,2,\ldots$), these topological spaces form a sequence of decreasing subsets, $\Omega_0\supset \Omega_1\supset\Omega_2\supset\cdots$. It will be convenient to implicitly perform these analytic continuations, and understand $\Omega_j$ in both ways: as topological spaces like $\Omega$ on which the theorems of the previous sections apply, and as subsets. Since when seen as subsets the information about the domain $E_j$ in $\Sigma$ is lost, we will use the notation $\alpha_j$, $\beta_j$ and $\gamma_j$ for the $j$-dependent functions that recover the conformally mapped continua in the spaces $\Omega_j$; for instance, $\gamma_j(\Sigma) = \Sigma(E_j)$. We will also continue to use $\alpha=\alpha_0$, $\beta=\beta_0$ and $\gamma=\gamma_0$ for the space $\Omega_0$. Note that an $\alpha_{j+1}$-map is an $\alpha_{j}$-map for every $j=0,1,2,\ldots$ (likewise for $\beta_j$- and $\gamma_j$-maps).

Note that $\Omega_{j+1}$ is not an {\em open} subset of $\Omega_{j}$ for any $j=0,1,2,\ldots$. Yet, of course, all results of the previous sections apply to all spaces $\Omega_j$, because these results are independent of the choice of the annular continuum $E$. The subtlety lies in the use of the functions $\alpha_j$, $\beta_j$ and $\gamma_j$ when applying these results to the space $\Omega_j$; it is these functions that encode the information of the underlying annular continuum $E_j$, whence of the topology.

\subsubsection{Covariant conformal derivatives}

If $f$ is a smooth function on $\Omega_0$ and is $\alpha$-invariant, then the transformation property of its $\Delta[h_{-2;w}]$ derivative is obtained by replacing $\log(Z)$ by $f$ (or by $\log(f)$ -- this is equivalent) in (\ref{transfZ}) with $c=0$ for all $\alpha$-maps $g$. This is an immediate consequence of \cite[Cor 3.11]{DCFT}. It is also a simple consequence of Proposition \ref{proporecursion}, where we take $k_0=0$ and $P_{\emptyset}^{(0)}(g|\emptyset)=1$. Hence, by Theorem \ref{theoZ}, $f$ is a normal conformal Gaussian with exponent 0, and both $f$ and $\log(f)$ have Property \ref{propholom} on $\Omega_0$.

For the rest of this section, let us assume that a normal conformal Gaussian $Z$ on $\Omega_0$, with exponent $c$ and such that $\log(Z)$ has Property \ref{propholom}, exists. Then clearly, for $f$ as above, $Zf$ is also a normal conformal Gaussian with exponent $c$, and $\log(Zf)$ has Property \ref{propholom}.

For any given $k\in\N$ and $w\in\C$, let us use the notation ${\bf m} = (m_k,m_{k-1},\ldots,m_1)$ for an ordered set of integers with the property that $-2\geq m_1\geq m_2\geq\cdots\geq m_k$, as well as the notation
\beq
	\Delta_{{\bf m};w} := \Delta[h_{m_k;w}] \Delta[h_{m_{k-1};w}]\cdots
	\Delta[h_{m_1;w}].
\eeq
By Theorems \ref{theoZ} and \ref{theoZgen}, the transformation equations (\ref{transfZ}) and (\ref{transfZgen}) hold for the normal conformal Gaussians $Z$ and $Zf$ on $\Omega_0$ (for $f$ as above). By taking derivatives with respect to $w_1,\,w_2,\,\ldots$ in these equations using Lemma \ref{lemder}, and then setting $w_1=w_2=\ldots$, we obtain transformation equations for $\Delta_{{\bf m};w}(Zf)$. It is a simple matter to see that these have the form
\beq\label{trans}
	\Delta_{{\bf m};w}(Zf)(\Sigma)=
	\sum_{{\bf m}'} P_{{\bf m},{\bf m}'}(g|w)
	\Delta_{{\bf m}';g(w)} (Zf)(g\cdot\Sigma)
\eeq
for all $\Sigma\in\Omega_0$ and all $\alpha$-maps $g$ at $\Sigma$, such that $g\cdot\Sigma\in\Omega_0$, where the sum is finite, and where $P_{{\bf m},{\bf m}'}(g|w)$ are rational functions of $g$ and its derivatives at $w$ and are holomorphic, as functions of $w$, on $\hC\setminus \beta(\Sigma)$. In particular, in order to establish that there is no explicit dependence on $w$, we used the fact that the coefficients of the differential operator $P_S^{(k)}(g|w_{\N_k})$ are rational functions of the differences $w_{j}-w_{j'}$ for $j,j'\in\N_k$. The fact that we can set $w_1=w_2=\ldots=w_k$ follows from the fact that these coefficients are holomorphic functions of $w_{j}$ for every $j\in\N_k$ in every domain of $\C$ where $g$ is conformal. Of course, we have
\[
	P_{{\bf m},{\bf m}'}(\id|w) = \delta_{k,k'}\prod_{j=1}^k \delta_{m_j,m_j'}
\]
if ${\bf m}$ has $k$ elements and ${\bf m}'$ has $k'$ elements, and
\[
	P_{(-2),{\bf m}'}(g|w) =\lt\{\ba{ll}
	\frc{c}{12} \{g;w\} & ({\bf m}' = ())\\
	(\p g(w))^2 & ({\bf m}' = (-2))\\
	0 & (\mbox{otherwise}).
	\ea\rt.
\]

Let us remark that in (\ref{trans}), if $\Sigma\in\Omega_{1}$ and $w\in \beta_{1}(\Sigma)\setminus \beta_0(\Sigma)$, then the right-hand side is defined for any map $g$ (not necessarily an $\alpha$-maps) that has an action in $\Omega_{1}$. This remark allows us to do the following construction. For every ${\bf m}$, $j\in\N$, $\Sigma\in\Omega_{j}$, $w\in \beta_{j}(\Sigma)\setminus \beta_{j-1}(\Sigma)$, we define the linear operators
\beq\label{defopgamma}
	\ba{rcl} \Gamma_{{\bf m};w;\Sigma}\ :\ 
	{\cal F}(\Omega_{j-1})&\to&{\cal F}(\Omega_{j}) \\
	f &\mapsto& \Gamma_{{\bf m};w;\Sigma}f \ea
\eeq
by
\beqa
	\Gamma_{{\bf m};w;\Sigma}f(\Sigma)
	&=& Z(\Sigma)^{-1}\Delta_{{\bf m};w}(Zf)(\Sigma) \n
	\Gamma_{{\bf m};w;\Sigma}f(g\cdot \Sigma) &=&
	Z(g\cdot\Sigma)^{-1}\sum_{{\bf m}'} P_{{\bf m},{\bf m}'}(g|w)
	\Delta_{{\bf m}';g(w)}(Zf)(g\cdot\Sigma) \label{defgamma}
\eeqa
for all $g$ that have an action on $\Sigma$ in $\Omega_{j}$. Note that the set of $g\cdot\Sigma$ for all such $g$ is $\Omega_{j}$, hence this indeed gives a definition of the mapped function $\Gamma_{{\bf m};w;\Sigma}f$. Further, thanks to Lemma \ref{lemsimder} (last part), the right-hand side of the second line of (\ref{defgamma}) is smooth in $g$, hence this indeed defines an operator (\ref{defopgamma}).

The important point to notice is that these linear operators preserve the $\alpha$-invariance property of smooth functions: thanks to (\ref{trans}), if $f$ is an $\alpha_{j-1}$-invariant smooth function, then $\Gamma_{{\bf m};w;\Sigma}f$ is an $\alpha_{j}$-invariant smooth function. The operators $\Gamma_{{\bf m};w;\Sigma}$ will be referred to as covariant conformal derivatives (their covariance property is discussed below). In a sense, these operators give conformal derivatives obtained under ``parallel transport'' from $\Sigma$, where the function $\Delta_{{\bf m};w}\log(Z)$ plays the role of a ``connection''.

Let us now consider the multiple covariant conformal derivatives of a smooth function $f$ on $\Omega_0$: for $\Sigma\in\Omega_i$, $w_i\in\hC\setminus\beta_i(\Sigma)$ and $w_j\in \beta_j(\Sigma)\setminus\beta_{j-1}(\Sigma)$ for all $j=1,2,\ldots,i-1$, this can be expressed as
\beqa
	\lefteqn{\Gamma_{{\bf m}_i;w_i;\Sigma}
	\Gamma_{{\bf m}_{i-1};w_{i-1};\Sigma}\cdots
	\Gamma_{{\bf m}_1;w_1;\Sigma} f(\Sigma)}&&
	\label{multider} \\
	&=& Z(\Sigma)^{-1} \Delta_{{\bf m}_i;w_i\,|\,g_{i-1}}
	\sum_{{\bf m}_{i-1}'}
	P_{{\bf m}_{i-1},{\bf m}_{i-1}'}(g_{i-1}|w_{i-1})
	\Delta_{{\bf m}_{i-1}';g_{i-1}(w_{i-1})\,|\,g_{i-2}} \cdot \n
&&	\sum_{{\bf m}_{i-2}'}
	P_{{\bf m}_{i-2},{\bf m}_{i-2}'}(g_{i-2}\circ g_{i-1}|w_{i-2})
	\Delta_{{\bf m}_{i-2}';
		(g_{i-2}\circ g_{i-1})(w_{i-2})\,|\,g_{i-3}} \cdot \n
&&	\cdots
	\sum_{{\bf m}_{1}'}
	P_{{\bf m}_{1},{\bf m}_{1}'}(g_1\circ\cdots\circ g_{i-1}|w_1)
	\Delta_{{\bf m}_{1}';(g_1\circ\cdots \circ g_{i-1})(w_1)}
	(Zf)((g_1\circ\cdots \circ g_{i-1})\cdot\Sigma).\no
\eeqa
Again, if $f$ is $\alpha$-invariant, then the result is a function that is $\alpha_i$-invariant. Clearly, the right-hand side does not depend on the choice of the sets $E_j,\;j=1,\ldots,i$. Since the $P_{{\bf m},{\bf m}'}(g|w)$s are rational functions of $g$ and its derivatives at $w$, and thanks to Lemma \ref{lemder}, a straightforward analysis shows that the expression above is a holomorphic function of the variable $w_j$ on $\hC\setminus (\beta(\Sigma)\cup \{w_l:l\neq j\})$, for every $j=1,\ldots,i$. It is also important to note, from Equation (\ref{multider}), that the function
\[
	\Sigma\mapsto 
	\Gamma_{{\bf m}_i;w_i;\Sigma}
	\Gamma_{{\bf m}_{i-1};w_{i-1};\Sigma}\cdots
	\Gamma_{{\bf m}_1;w_1;\Sigma} f(\Sigma)
\]
is smooth on $\Omega_0$ whenever $f$ is.

A clear understanding of the operators $\Gamma_{{\bf m};w;\Sigma}$ is obtained by observing that they transform in a covariant manner, as their name suggest. First note that the validity of (\ref{trans}) for every $\alpha$-invariant function $f$ implies
\beq\label{gpp}
	P_{{\bf m},{\bf m}'}(g_1\circ g_2|w)
	=\sum_{{\bf m}''} P_{{\bf m},{\bf m}''} (g_2|w)
	P_{{\bf m}'',{\bf m}'}(g_1|g_2(w))
\eeq
(this can be verified explicitly from the construction of $P_{{\bf m},{\bf m}'}(g|w)$). Equations (\ref{multider}) and (\ref{gpp}) then give the following transformation property for the multiple covariant derivatives, for any smooth function $f$ on $\Omega_0$ that is $\alpha$-invariant:
\beqa\label{covtrans}
	\lefteqn{\Gamma_{{\bf m}_i;w_i;\Sigma}
	\Gamma_{{\bf m}_{i-1};w_{i-1};\Sigma}\cdots
	\Gamma_{{\bf m}_1;w_1;\Sigma} f(\Sigma)} &&\\
	&=&
	\sum_{{\bf m}_{i}',{\bf m}_{i-1}',\ldots,{\bf m}_{1}'}
	P_{{\bf m}_i,{\bf m}_{i}'}(g|w_i)
	P_{{\bf m}_{i-1},{\bf m}_{i-1}'}(g|w_{i-1})
	\cdots
	P_{{\bf m}_1,{\bf m}_{1}'}(g|w_1)
	\cdot\n&&\qquad\qquad\quad\ \;
	\Gamma_{{\bf m}_i';g(w_i);g\cdot\Sigma}
	\Gamma_{{\bf m}_{i-1}';g(w_{i-1});g\cdot\Sigma}\cdots
	\Gamma_{{\bf m}_1';g(w_1);g\cdot\Sigma} f(g\cdot\Sigma)\no
\eeqa
for every $\alpha_i$-map $g$ at $\Sigma$ such that $g\cdot\Sigma\in\Omega_i$. Hence, the multiple covariant derivatives $\Gamma_{{\bf m}_i;w_i;\cdot}\,\Gamma_{{\bf m}_{i-1};w_{i-1};\cdot}\,\cdots\,\Gamma_{{\bf m}_1;w_1;\cdot}\, f(\cdot)$ transform covariantly, like the tensor product
\[
	(\;Z^{-1}\Delta_{{\bf m}_i;w_i}Z\;\otimes\;
	Z^{-1}\Delta_{{\bf m}_{i-1};w_{i-1}}Z\;\otimes\;\cdots\;\otimes\;
	Z^{-1}\Delta_{{\bf m}_1;w_1}Z\;)\;f.
\]
It is this covariant transformation property that allows us to identify multiple covariant conformal derivatives with correlation functions of products of multiple stress-energy tensor descendants in conformal field theory, and hopefully with expectation values of certain random variables in conformal loop ensembles.

\subsubsection{Vertex operator algebra}

Let $V$ be the vector subspace of smooth functions on $\Omega_0$ generated by $f_{\bf 1}=1$ under applications of finite numbers of operators $L(m):m\leq -2$. For ${\bf m}=(m_k,m_{k-1},\ldots,m_1)$ as above, let us use the notation $L({\bf m}) = L(m_k)\cdots L(m_1)$. The vector space $V$ has the structure of a Virasoro vertex operator algebra $(V,Y,\omega,{\bf 1})$. In this structure, the identity vector is ${\bf 1}=f_{\bf 1}$, the conformal vector is $\omega = L(-2)f_{\bf 1}$, and the vertex operator map $Y(\cdot,x)$ is defined by
\beq\label{defY1}
	Y(L(m) {\bf 1},x) = \frc1{(-2-m)!}\frc{d^{-m-2}}{dx^{-m-2}}
	\lt(\sum_{n\in\Z} L(n) x^{-n-2} \rt)
\eeq
for $m\leq -2$, and, recursively for $j=2,3,\ldots$,
\beq
	Y(L({\bf m}_j) {\bf 1},x)
	= \;\:Y(L(m_j) {\bf 1},x) \;Y(L({\bf m}_{j-1}){\bf 1},x)\:
	\label{defY}
\eeq
with ${\bf m}_j=(m_j,m_{j-1} \ldots,m_1)$. The normal-ordering is defined by \cite[p 95]{LL04}
\beq\label{normord}
	\:Y(L(m){\bf 1},x)\bullet\: = Y^+(L(m){\bf 1},x)\bullet\, + \,\bullet
	Y^-(L(m){\bf 1},x)
\eeq
where
\beqa
	Y^+(L(m){\bf 1},x) &=& 
	\frc1{(-2-m)!}\frc{d^{-m-2}}{dx^{-m-2}}
	\lt(\sum_{n\leq -2} L(n) x^{-n-2} \rt)\n
	Y^-(L(m){\bf 1},x) &=&
	\frc1{(-2-m)!}\frc{d^{-m-2}}{dx^{-m-2}}
	\lt(\sum_{n\geq -1} L(n) x^{-n-2} \rt)
\eeqa
are, respectively, the regular and singular parts of the vertex operator. As is usual in this context, we will also understand the normal-ordering symbol \:\: as a re-ordering symbol for the products of operators $L(n):n\in\Z$ as per the coefficients of the formal series (\ref{normord}).

We now show that products of vertex operators correspond to multiple applications of covariant conformal derivatives. It is convenient to introduce the following notation: given a function $a(w_1,\ldots,w_i)$ of complex variables $w_1,\ldots,w_i$ holomorphic in each of these variables in the region $|w_1|<|w_2|<\cdots<|w_i|<r$ for some $r>0$, and given a formal series $b(x_1,\ldots,x_i)$ with formal variables $x_1,\ldots,x_i$, we will say that 
\[
	a(w_1,\ldots,w_i)\equiv b(x_1,\ldots,x_i)
\]
under $w_j\equiv x_j,\;j=1,\ldots,i$, if the multiple-variable Laurent series expansion of the left-hand side in the region $|w_1|<|w_2|<\cdots<|w_i|<r$ is reproduced by the formal series on the right-hand side under the replacement of the complex variables $w_j$ by the formal variables $x_j$, $j=1,\ldots,i$. Note that Fubini's theorem guarantees that the Laurent series expansion of $a(w_1,\ldots,w_i)$ in the region $|w_1|<|w_2|<\cdots<|w_i|<r$ is unique\footnote{One first Laurent expands in $w_j$ for some $j$, fixing $w_{j'},\;j'\neq j$. Each coefficient is a contour integral in $w_j$ at fixed $|w_j|$. Each such coefficient is then holomorphic in $w_{j'}$ for any $j'\neq j$ in the region $|w_1|<|w_2|<\cdots<|w_i|<r$, because the integral in $w_{j'}$ on every close contour in this region gives zero by Fubini's theorem (exchanging the order of the $w_j$ and $w_{j'}$ integrals). Hence each coefficient can be Laurent expanded in $w_{j'}$. Continuing the process, one obtains a multiple-variable Laurent expansion, where each coefficient is a multiple contour integral. By Fubini's theorem again, the order of the integrals involved can be exchanged. Hence, the result does not depend on the choice of the order of variables in which the sequential Laurent expansions were made.}.
\begin{theorem}\label{theovoa}
For every $\Sigma\in\Omega_0$, there exists a $r>0$ such that
\beq
	\Gamma_{{\bf m}_i;w_i;\Sigma}\cdots
	\Gamma_{{\bf m}_1;w_1;\Sigma}f_{\bf 1}(\Sigma)
\eeq
is holomorphic in all variables $w_j,\;j=1,\ldots i$ in the region $|w_1|<|w_{2}|<\cdots<|w_i|<r$, and in this region,
\beq
	\Gamma_{{\bf m}_i;w_i;\Sigma}\cdots
	\Gamma_{{\bf m}_1;w_1;\Sigma}f_{\bf 1}(\Sigma) \equiv
	(-1)^{\sum_{j=1}^i |{\bf m}_j|}
	Y(L({\bf m}_i){\bf 1},x_i)\cdots Y(L({\bf m}_1){\bf 1},x_1)f_{\bf 1} (\Sigma)
\eeq
under $w_j\equiv x_j,\;j=1\ldots,i$. Here, we use the notation $|{\bf m}| = \sum_{l=1}^k m_k$ for any ${\bf m} = (m_k,m_{k-1},\ldots,m_1)$.
\end{theorem}
Note that the order of the absolute values $|w_j|$ above is the opposite of the ``natural'' order provided by the inclusion $E_0\subset E_1\subset E_2\subset\cdots$. However, by the remark after (\ref{multider}), the resulting function does not depend on the choice of sets $E_1,E_2,\ldots$, and can be analytically continued.

\proof We prove by induction. First, since $\beta(\Sigma)$ is closed and does not contain the point 0, there exists an open disk $D$ of radius $r>0$ centered at 0, included in $\hC\setminus \beta(\Sigma)$. Hence, by the remark after (\ref{multider}), there exists a $r>0$ as in the theorem. It will then be sufficient to show that, for every $j=1,\ldots,i$ and every $\Sigma\in\Omega$, we have
\beqa\label{toshow}
	\lefteqn{
	\Gamma_{{\bf m}_j;w_j;\Sigma} \Gamma_{{\bf m}_{j-1};w_{j-1};
	\Sigma}\cdots \Gamma_{{\bf m}_{1};w_{1};\Sigma}
	f_{\bf 1}(\Sigma)} &&\\
	&\equiv& (-1)^{|{\bf m}_j|} \;Y(L({\bf m}_j){\bf 1},x_j)_{|\,g}\;
	\Gamma_{{\bf m}_{j-1};w_{j-1};g\cdot\Sigma} \Gamma_{{\bf m}_{j-2};w_{j-2};
	g\cdot\Sigma}\cdots \Gamma_{{\bf m}_{1};w_{1};g\cdot\Sigma}
	f_{\bf 1}(g\cdot\Sigma)\no
\eeqa
under $w_j\equiv x_j$, where for $j=1$ there is no covariant conformal derivative on the right-hand side.

First, it is straightforward to verify that the formal Taylor series expansion of the operator $Z^{-1}\Delta[h_{m;w}]Z$ in $w$, for $m\leq -2$, exactly reproduces, under $w\equiv x$, the formal series $(-1)^mY^+(L(m){\bf 1},x)$. The case $j=1$ is then immediate: the function
\[
	\Gamma_{{\bf m};w;\Sigma}f_{\bf 1}(\Sigma) = Z(\Sigma)^{-1}\Delta_{{\bf m};w}(Zf_{\bf 1})(\Sigma)
\]
is holomorphic in $w$ around 0, whence, by Lemma \ref{lemder}, its expansion is obtained by expanding each conformal derivative $\Delta[h_{m_l;w}]$, $l=1,\ldots,k$ in Taylors series in $w$. Under $w\equiv x$, this expansion is equivalent, up to the factor $(-1)^{|{\bf m}|}$, to
\beq
	Y^+(L(m_k) {\bf 1},x)\cdots Y^+(L(m_1) {\bf 1},x) f_{\bf 1}
	=\;
	Y(L({\bf m}) {\bf 1},x) f_{\bf 1} .
\eeq

For the cases $j\geq 2$, we use the following important result.
\begin{lemma}\label{lemmamt}
Let $A$ be a simply connected domain containing $w_1,\ldots,w_{j-1}$ but not 0 (assuming $w_1\neq 0$), such that $\cl{A}\subset D$. Let $B\supset\gamma(\Sigma)$ be a domain not intersecting $D\cup \{\infty\}$. Let $(g,\t{g})\mapsto f(g,\t{g})$ be a function of two variable conformal maps, defined on a $A$-neighbourhood $U$ of $\id$ and $A$-smooth on the first variable, and defined on a $B$-neighbourhood $\t{U}$ of $\id$ and $B$-smooth on the second variable. Further, assume that for every $g'\in \t{U}$ and every linear combination of multiple conformal derivative operators in fixed directions $\t\nabla$ with respect to the second variable (or $\t\nabla=1$), the function $g\mapsto \t\nabla f(g,g\circ g')$: 1. is $\C_0$-smooth at $g=\id$, and 2. its derivatives can be evaluated by Leibnitz's rule. Finally, assume that for every $g'\in \t{U}$, the function $g\mapsto f(g,g\circ g')$ has zero $\C$-derivatives at $g=\id$. Then, the function $\Delta_{{\bf m};w\,|\,g}f(g,g)$ is holomorphic on $D\setminus A$. Further, on the annulus $r_0<|w|<r$, where $r_0$ is the radius of the smallest disk centered at 0 containing $A$, for every ${\bf m}$, we have
\beq\label{eqlemmamt}
	Z(\Sigma)^{-1}\Delta_{{\bf m};w\,|\,g}Z(g\cdot\Sigma)
	f(g,g)
	\equiv (-1)^{|{\bf m}|}\;Y(L({\bf m}){\bf 1},x))_{|\,g}f(\id,g)
\eeq
under $w\equiv x$.
\end{lemma}

Before showing the lemma, let us show how it implies (\ref{toshow}), hence the theorem. We write the left-hand side of (\ref{toshow}) as $Z(\Sigma)^{-1}\Delta_{{\bf m}_j;w_j\,|\,g}Z(g\cdot\Sigma) f(g,g)$ with some function $f$ of two arguments such that
\beq\label{fgg}
	f(g,g) = \Gamma_{{\bf m}_{j-1};w_{j-1};
	\Sigma}\cdots \Gamma_{{\bf m}_{1};w_{1};\Sigma}
	f_{\bf 1}(g\cdot\Sigma).
\eeq
Using (\ref{multider}) and (\ref{defgamma}), we choose $f$ as
\beqa
	\lefteqn{f(g,\t{g})}&&
	\label{multiderf} \\
	&=&
	Z(\t{g}\cdot \Sigma)^{-1} \sum_{{\bf m}_{j-1}'}
	P_{{\bf m}_{j-1},{\bf m}_{j-1}'}(g|w_{j-1})
	\Delta_{{\bf m}_{j-1}';g(w_{j-1})\,|\,g_{j-2}} \cdot \n
&&	\sum_{{\bf m}_{j-2}'}
	P_{{\bf m}_{j-2},{\bf m}_{j-2}'}(g_{j-2}\circ g|w_{j-2})
	\Delta_{{\bf m}_{j-2}';
		(g_{j-2}\circ g)(w_{j-2})\,|\,g_{j-3}} \cdot \n
&&	\cdots
	\sum_{{\bf m}_{1}'}
	P_{{\bf m}_{1},{\bf m}_{1}'}(g_1\circ\cdots\circ g_{j-2}\circ g|w_1)
	\Delta_{{\bf m}_{1}';(g_1\circ\cdots \circ g_{j-2}\circ g)(w_1)}
	(Zf_{\bf 1})((g_1\circ\cdots \circ g_{j-2}\circ \t{g})\cdot\Sigma).\no
\eeqa
We choose $U$ and $\t{U}$ small enough, such that $g(w_l)\in A$ for every $l$ and every $g\in U$, that $\t{g}(\gamma(\Sigma))\in B$ for every $\t{g}\in\t{U}$, and such that every map in $\t{U}$ has an action on $\Sigma$ (such choices exist by Lemma \ref{lemtopo}). Note that the function $f$ thus defined is a finite sum of products of multiple conjugated derivatives of the form $Z^{-1}\Delta_{{\bf m};g(w_l)}Z$ for $l=1,\ldots,j-1$, applied to $f_{\bf 1}$, and evaluated at $\t{g}\cdot\Sigma$, with coefficients that depend rationally on $g(w_l)$ for $l=1,\ldots,j-1$.

This function is $A$-smooth on the first variable by Lemma \ref{lemsimplef} because the only dependence on the first variable $g$ comes from holomorphic functions of $g(w_l)$ for $l=1,\ldots,j-1$. It is $B$-smooth on the second variable because $f_{\bf 1}$ and $Z$ are. Taking any derivative with respect to the second variable and setting $\t{g}=g\circ g'$ for some $g'\in\t{U}$, we obtain a function that is $\C_0$-smooth with respect to $g$ at $\id$ and whose derivatives may be evaluated using Leibnitz's rule, thanks to Lemma \ref{lemsimder} (last part). Further, we have
\beq\label{fggp}
	f(g,g\circ g') =
	\Gamma_{{\bf m}_{j-1};w_{j-1};g'\cdot\Sigma}\cdots
	\Gamma_{{\bf m}_1;w_1;g'\cdot\Sigma}
	f_{\bf 1}(g\cdot g'\cdot \Sigma).
\eeq
This has zero $\C$-derivatives with respect to $g$ at $\id$ because the multiple covariant conformal derivative of $f_{\bf 1}$ is $\alpha_{j-1}$-invariant. Equation (\ref{toshow}) then follows from the lemma and from (\ref{fggp}) with $g=\id$.

We now show the lemma. First, the holomorphicity in $w$ on $D\setminus A$ is an immediate consequence of the use of Leibnitz's rule and of the differentiability property of $f$ in its individual variables. For (\ref{eqlemmamt}), we proceed by induction on the number of elements of ${\bf m}$. For 0 elements, there is no differential operator on either side, and the equality is trivial.

Assume that (\ref{eqlemmamt}) holds for all ${\bf m}=(m_k,\ldots,m_1)$ with $k$ elements, for some $k\ge 1$. Let ${\bf m}' = (m,m_k,\ldots,m_1)$ for $m\leq -2$. Then we have
\beqa
	\lefteqn{Z(\Sigma)^{-1}
	\Delta_{{\bf m}';w\,|\,g} Z(g\cdot\Sigma)f(g,g)} && \label{initex} \\
	&=&
	Z(\Sigma)^{-1} \Delta[h_{m;w}\,|\,g]
	\Delta_{{\bf m};w'\,|\,g'} Z((g'\circ g)\cdot\Sigma)f(g'\circ g,
	g'\circ g)\;|_{w'=w}. \no
\eeqa
Using the assumption of the induction, the function of $w'$ involved is holomorphic on $D\setminus A$. Hence, let us keep $w'\neq w$ for now, and use (\ref{eqlemmamt}) to expand in $w'$ in the region $r_0<w'<r$:
\beqa
	\lefteqn{Z(\Sigma)^{-1} \Delta[h_{m;w}\,|\,g]
	\Delta_{{\bf m};w'\,|\,g'} Z((g'\circ g)\cdot\Sigma)f(g'\circ g,
	g'\circ g)} &&\n
	&=&
	Z(\Sigma)^{-1} \Delta[h_{m;w}\,|\,g]Z(g\cdot\Sigma)\; 
	Z(g\cdot\Sigma)^{-1}
	\Delta_{{\bf m};w'\,|\,g'} Z(g'\cdot g\cdot\Sigma)f(g'\circ g,
	g'\circ g) \n
	&\stackrel{w'\equiv x'}\equiv&
	Z(\Sigma)^{-1} \Delta[h_{m;w}\,|\,g]Z(g\cdot\Sigma)
	\;(-1)^{|{\bf m}|}
	\;Y(L({\bf m}){\bf 1},x')_{|\,g'}
	f(g,g'\circ g). \no
\eeqa
Note that if ${\bf m}$ has 0 element, then there is no formal series in $x'$ and we simply set $g'=\id$. At each power in $x'$ in the formal series formed by the last factors
\[
	(-1)^{|{\bf m}|}\;Y(L({\bf m}){\bf 1},x')_{|\,g'}f(g,g'\circ g)
	=:\t\nabla(x')_{|\,g'}f(g,g'\circ g)
\]
there is a multiple $\C_0$-derivative of $f(g,g'\circ g)$ with respect to $g'$ at $\id$; we denote by $\t\nabla(x')_{|\,g'}$ the corresponding formal series. Let us consider the function
\[
	F(x')(g,\t{g}):=\t\nabla(x')_{|\,g'}f(g,g'\circ \t{g}).
\]
Clearly this is (at each power of $x'$) $A$-smooth in the first variable and $B$-smooth in the second variable.

Hence for every $m\leq -2$ and each power of $x'$, $Z(\Sigma)^{-1}\Delta[h_{m;w}\,|\,g]Z(g\cdot\Sigma)F(x')(g,g)$ is a holomorphic function of $w$ in a neighbourhood of 0. By Leibnitz' rule, we have
\beqa
	Z(\Sigma)^{-1}\Delta[h_{m;w}\,|\,g]Z(g\cdot \Sigma)
	\lefteqn{F(x')(g,g)} &&\n  &=&
	\Delta[h_{m;w}\,|\,g]F(x')(g,\id)
	+ Z(\Sigma)^{-1}\Delta[h_{m;w}\,|\,g]Z(g\cdot\Sigma)F(x')(\id,g).
	\no
\eeqa
Since $F(x')$ is $A$-smooth on the first variable, then $\Delta[h_{m;w}\,|\,g]F(x')(g,\id)$ is holomorphic for $w\in\hC\setminus A$, and since $F(x')$ is $B$-smooth on the second variable, then $Z(\Sigma)^{-1}\Delta[h_{m;w}\,|\,g]Z(g\cdot\Sigma)F(x')(\id,g)$ is holomorphic for $w\in D$. Hence, the resulting function $Z(\Sigma)^{-1}\Delta[h_{m;w}\,|\,g]Z(g\cdot\Sigma)F(x')(g,g)$ is holomorphic on $D\setminus A$. In particular, it is holomorphic on the annulus $r_0<|w|<r$. Evaluating the formal series in $x$ corresponding to the Laurent expansion in $w$ in that region, under $w\equiv x$, and then setting $x=x'$, we obtain the formal series expansion corresponding to the Laurent expansion in $w$ of the initial expression (\ref{initex}) (i.e.~the left-hand side of (\ref{eqlemmamt})) under $w\equiv x$.

Let us assume that $r_0<|w|<r$. Then, by continuity of the conformal derivative,\\ $\Delta[h_{m;w}\,|\,g]F(x')(g,\id)$ has a convergent Taylor expansion in powers of $w^{-1}$, given by
\beq
	\Delta[h_{m;w}\,|\,g]F(x')(g,\id) =
	\frc{(-1)^{m}}{(-2-m)!}
	\frc{d^{-2-m}}{dw^{-2-m}} \sum_{n\geq -1}
	w^{-n-2} (-1)^{n+1} \Delta[h_{n}\,|\,g]F(x')(g,\id).
\eeq
We have
\[
	\Delta[h_n\,|\,g]F(x')(g,\id) =
	\Delta[h_n\,|\,g]\t\nabla(x')_{|\,g'}f(g,g')=
	\t\nabla(x')_{|\,g'}\Delta[h_n\,|\,g]f(g,g').
\]
Since $f(g,g\circ g')$ has zero $\C$-derivatives as a function of $g$ at $\id$, then, by Leibnitz' rule, $\Delta[h_{n}\,|\,g]f(g,g') + \Delta[h_{n}\,|\,g]f(\id,g\circ g')=0$ for all $n\geq-1$. Hence,
\beq\label{res1mt}
	\Delta[h_{m;w}\,|\,g]F(x')(g,\id)
	= 
	\frc{(-1)^{m}}{(-2-m)!}
	\frc{d^{-2-m}}{dw^{-2-m}} \sum_{n\geq -1}
	w^{-n-2} (-1)^{n} \t\nabla(x')_{|\,g'} \Delta[h_{n}\,|\,g]f(\id,g\circ g').
\eeq
Similarly, by continuity of the conformal derivative, $Z(\Sigma)^{-1}\Delta[h_{m;w}\,|\,g]Z(g\cdot\Sigma)F(x')(\id,g)$ has a convergent Taylor expansion in powers of $w$, given by
\beqa
	\lefteqn{
	Z(\Sigma)^{-1}\Delta[h_{m;w}\,|\,g]Z(g\cdot\Sigma)F(x')(\id,g)}
	&& \label{res2mt} \\ &=&
	\frc{(-1)^{m}}{(-2-m)!}
	\frc{d^{-2-m}}{dw^{-2-m}} \sum_{n\leq -2}
	w^{-n-2} (-1)^{n} Z(\Sigma)^{-1}\Delta[h_{n}\,|\,g]
	Z(g\cdot\Sigma)\t\nabla(x')_{|\,g'}f(\id,g'\circ g).\no
\eeqa
Combining (\ref{res1mt}) and (\ref{res2mt}) and using (\ref{defLn0}) as well as the definition of the normal ordering (\ref{normord}), we obtain
\[
	Z(\Sigma)^{-1}\Delta[h_{m;w}\,|\,g]Z(g\cdot\Sigma)F(x')(g,g)
	=\frc{(-1)^{m}}{(-2-m)!}
	\frc{d^{-2-m}}{dw^{-2-m}} \sum_{n\in\Z}
	w^{-n-2} \:L(n)\Delta(x')\:_{|\,g}f(\id,g).
\]
That is, we find the crucial result that, for $f$ as described and using (\ref{defY}),
\beqa
	\lefteqn{Z(\Sigma)^{-1} \Delta[h_{m;w}\,|\,g]Z(g\cdot\Sigma)
	\;(-1)^{|{\bf m}|}
	\;Y(L({\bf m}){\bf 1},x')_{|\,g'}
	f(g,g'\circ g)} &&\n
	&\stackrel{w\equiv x}\equiv&
	(-1)^{|{\bf m}'|} \;\:Y(L(m){\bf 1},x)Y(L({\bf m}){\bf 1},x')\:_{|\,g}f(\id,g)\n
	&\stackrel{x'=x}=& 
	(-1)^{|{\bf m}'|} \;Y(L({\bf m}'){\bf 1},x)_{|\,g}f(\id,g).
\eeqa
From (\ref{initex}) this completes the induction process, and shows (\ref{eqlemmamt}).\eproof

\subsubsection{Connection with the stress-energy tensor of CFT and its descendants}

We now mention the relation of the present construction to standard concepts in conformal field theory (CFT) \cite{BPZ,DFMS97}, in order to make the connection with the work \cite{DCFT}. In the latter work, the relative partition function $Z$ was involved, which is defined in CFT and which was argued, by CFT arguments, to be M\"obius invariant and to satisfy the transformation law (\ref{transfZ}). Then, it was argued that the differential operator $Z^{-1}\Delta[h_{-2;w}]Z$ should be identified, in the case where the functions to which it is applied are $\alpha$-invariant, with the stress-energy tensor of CFT. The argument was based on the analytical structure in terms of $w$, the transformation law, and the explicit application to correlation functions in boundary CFT (seen as certain $\alpha$-invariant functions): it was shown to reproduce the conformal Ward identities and boundary conditions. In a more or less standard symbolism (we refer to \cite{DCFT}),
\beq\label{T}
	Z(\id)^{-1}\Delta[h_{-2;w}\,|\,g]
	Z(g) \bra g\cdot \Or(g(x_1))\cdots g\cdot \Or(g(x_q))\ket_{g\cdot C}
	= \bra T(w)\Or(x_1)\cdots \Or(x_q)\ket_C.
\eeq
This identification was the basis for formulating the concept of stress-energy tensor in the context of conformal loop ensembles (CLE) \cite{DCLE}, where probability functions were seen as functionals of conformal maps in a similar way.

The present construction generalises this to all descendants of the stress-energy tensor, and further to multiple insertions of the stress-energy tensor inside correlation functions. In the CFT context, the points $x_1,\ldots,x_q$ as well as the domain boundary $\p C$ should lie in the closed set $E$ from which the space $\Omega$ is defined; in the CLE context, the support of the event and the domain boundary should lie in $E$.

In the CFT language, the stress-energy tensor $T(w)$ corresponds to the vertex operator $Y(L(-2){\bf 1},x)$ (see (\ref{defY1})) under $w\equiv x$. Theorem \ref{theovoa} shows that the normal-ordered products of $k$ stress-energy tensors $( \,T(w)\,(\, T(w)\,(\,\cdots\, (\,T(w) T(w))\cdots )))$ (here using the CFT notation of, for instance, \cite{DFMS97}), which correspond to the vertex operators
\[
	Y(L(-2)\cdots L(-2){\bf 1},x)
\]
(see (\ref{defY})), are identified with multiple conformal derivatives
\[
	(-1)^{k} Z^{-1}\Delta[h_{-2;w}]\cdots\Delta[h_{-2;w}]Z,
\]
and a similar identification holds for normal-ordered products involving derivatives of the stress-energy tensor. This identification is in the sense of (\ref{T}): the application of such derivatives to correlation functions of boundary CFT seen as $\alpha$-invariant functions, reproduce the Ward identities and boundary conditions.  This identification is a new result: there is no equivalent, to our knowledge, either in the context of CFT or of vertex operator algebras (but see the relation with the representation theory of vertex operator algebras below). Understanding what CLE objects are obtained upon application of such derivatives to probability functions or expectations, will give an identification in CLE of descendants of the stress-energy tensor (generalising \cite{DCLE}).

Theorem \ref{theovoa} also shows that multiple insertions of descendants of the stress-energy tensor are reproduced by multiple applications of covariant conformal derivatives. The use of covariant conformal derivatives is expected in the context of the CFT construction in \cite{DCFT}, in the sense that it is the natural mean of making $\alpha$-invariant functions out of correlation functions with insertions of stress-energy tensor descendants, along the lines of the formula (\ref{T}).

Formula (\ref{transfZgen}) can be seen as a generating function for transformation laws for all descendants of the stress-energy tensor, explicitly expressed as normal-ordered products. Interestingly, multiple covariant conformal derivatives transform like the tensor product of (conjugated) conformal derivatives, Equation (\ref{covtrans}). This means that from the viewpoint of the transformation laws, it is possible to identify multiple covariant conformal derivatives with multilinear functions of the stress-energy tensor descendants. This is natural in CFT, where correlation functions are multilinear. Here, however, it is here a general structural result, valid beyond its application to CFT correlation functions. In particular, this would be consistent with identifying stress-energy tensor descendants as (certain limits of) random variables in CLE expectation values, as expectation values of products of such variables are multilinear.

\subsubsection{Connection with Haisheng Li's representation theory for vertex operator algebras}

The representation theory of vertex operator algebras was first developed by Li in \cite{Li96}; see the nice exposition in Chapter 5 of the book \cite{LL04}. This representation theory gives rise to construction theorems for vertex operator algebras, and is probably the most efficient way of obtaining vertex operator algebras and modules from infinite-dimensional Lie algebras (see Chapter 6 of \cite{LL04}).

In the representation theory, the vector space which acquires a vertex operator algebra structure is that of weak vertex operators: doubly-infinite formal series with coefficients that are endomorphisms of another vector space $V$, with a certain truncation condition \cite[Def 5.1.1]{LL04}. The formal series
\[
	a(x):=Y(L(-2){\bf 1},x)
\]
defined in (\ref{defY1}) is a weak vertex operator, where in this case $V$ is the identity module for the Virasoro algebra (as above). By the representation-theory construction of the vertex operator map, the operators (\ref{defY}) are built by multiple applications of the endomorphisms of weak vertex operators associated to the weak vertex operator $a(x)$, as follows:
\[
	Y(L({\bf m}){\bf 1},x) =
	a(x)_{m_k+1}\cdots a(x)_{m_1+1}1_V.
\]
We refer to the literature cited above for the details of how the endomorphism $a(x)_n$ associated to $a(x)$ is defined (see for instance \cite[Eq 5.2.6]{LL04}).

This along with Lemma \ref{lemmamt} immediately suggests the following. Consider the Virasoro operators (\ref{defLn0}) as differential operators with respect to the variable conformal map $g_2$, and let them act on a space generated by a function $f(g_1,g_2)$ with the properties stated in Lemma \ref{lemmamt}, as well as the condition that it forms a restricted module ($L(n)$ is zero for $n$ large enough). Let $b(x)$ be any weak vertex operator on this space. Then, we may define
\[
	(-1)^m  a(x)_{m+1}b(x) :=
	\lt(
	Z^{-1} \Delta[h_{m;w}\,|\,g_1,g_2]Z\; b(x) \rt)_{w\equiv x}
\]
for $m\leq -2$. That is, the endomorphism $a(x)_n$ associated to $a(x)$, for $n\leq -1$, is obtained by mapping to a conformal derivative where the formal variable is replaced by a complex variable $w$, interpreting it as a derivative with respect to the two conformal maps $g_1$ and $g_2$ simultaneously, and expanding the result in a formal series with $w\equiv x$. One can guess how to complete the definition by giving the endomorphisms $a(x)_n$ for $n\geq 0$, as well as the endomorphisms associated to other weak vertex operators. We hope to come back to this in a future work.

\sect{Conclusion}

We have constructed a topological space $\Omega$ of conformal maps, with a structure of smooth paths and a concept of differentiable complex functions on $\Omega$. From this, we defined conformal Gaussians as smooth functions on $\Omega$ that satisfy certain differential equations: essentially, that the second derivatives of their logarithm are constants. We have shown that these constants can only take very special values, characterized by 4 complex numbers. When 3 of these are zero (we refer to the resulting functions as normal conformal Gaussians), we have shown that these differential equations are essentially equivalent to a certain conformal transformation property of a particular conformal derivative. From this and from the differential equations, we have represented the Virasoro algebra using conformal derivatives, and we have constructed from certain representations of this type the structure of the Virasoro vertex operator algebra. Our most important result is the equivalence of the product of vertex operators with the multiple application of certain covariant conformal derivatives, Theorem \ref{theovoa}. Covariant conformal derivatives are defined in terms of conformal derivatives, in such a way that the transformation property of a multiple application of covariant conformal derivatives is obtained by a tensor product of the transformation properties of the individual conformal derivatives. We have discussed how this main result and these transformation properties connect with concepts of conformal field theory, as well as with concepts of the representation theory of vertex operator algebras.

There are many extensions suggested by the present work. First, it is important to show that an explicit normal conformal Gaussian $Z$ exists on $\Omega$, and perhaps to construct it explicitly (ideas developed in of \cite{BB04} may be useful). But also, it would be very interesting to have a CFT or algebraic interpretation of conformal Gaussians with nonzero anomaly. It is further very likely that the transformation property of the conformal derivative $\Delta[h_{-2;w}]\log(Z)$ can be generalized to the anomalous case. Preliminary calculations suggest that to the right-hand side in the transformation property (\ref{transfZ}), we should add a term $K(g,w)$, a functional of the conformal map $g$ and the complex number $w$, that depends on the three parameters $d_1$, $d_2$ and $d_3$, and that satisfies the following conditions:
\beqa
	K(g_1\circ g_2,w) &=&
	(\p g_2(w))^2 K(g_1,g_2(w)) + K(g_2,w) \n
	\Delta[h_n\,|\,g]K(g,w) &=& 2(-1)^n\lt(
	(n+2)d_1 w^{n+2} + \lt(n+\frc32\rt)d_2 w^{n+1} +
	(n+1)d_3 w^{n}\rt)\quad (n\geq -1). \no
\eeqa
Note that the first condition is satisfied by the Schwarzian derivative $\{g,w\}$, and that it implies $K(\id,w)=0$. We believe that these conditions can be solved, and that the solution is unique. In particular, for $g(z)=g_a(z): = z+a$, $a\in\C$, we find $K(g_a,w)=-(2d_1w +d_1a+d_2)a$.

Second, it would be interesting to understand the relation between the present analytic construction and the null-vector equations of CFT. For instance, certain functions on $\Omega$, which are Virasoro highest-weights, should satisfy these equations; what are the conditions?

Third, a theory of integration should be developed, in order to have a Hilbert space structure whereby representations of the Virasoro algebra in terms of conformal derivatives are unitary. This, for instance, would lead to conditions for null-vector equations to hold, and would give a very interesting relation between the analysis on the infinite-dimensional space $\Omega$ and algebraic results pertaining to the Virasoro algebra and vertex operator algebra.

Fourth, the present construction may provide principles upon which one may base the abstract structure of vertex operator algebra. Generalizing it to Riemann surfaces of higher genus (instead of the Riemann sphere $\hC$, at the basis of the present construction) could lead to the notion of vertex operator algebras on higher genus.

Fifth, a natural extension would be that whereby additional degrees of freedom are considered, for instance associated to a Lie algebra. This should lead to a similar analytic understanding of Wess-Zumino-Witten models of CFT (or of vertex operator algebras associated to affine Lie algebras). Likewise, it would be interesting to develop a similar construction that would connect with the free boson model of CFT (or the Heisenberg vertex operator algebra).

Sixth, as mentioned in the introduction, it would be very interesting to apply the present construction to conformal loop ensembles, in such a way as to obtain expressions for descendants of the stress-energy tensor.

Finally, of course, a manifold structure for $\Omega$, perhaps in relation with the Teichm\"uller theory, could clarify some important aspects developed here (smooth paths, differentiability).

\appendix

\sect{Vertex operator algebras}\label{appvoa}

We recall {}from \cite{FLM88} the definition of the notion of vertex
operator algebra, a variant of Borcherds' notion \cite{Bo86} of vertex
algebra:
\begin{defi}\label{VOA}
A {\bf vertex operator algebra} $(V,Y,{\bf 1},\omega)$, or $V$ for
short, is a $\mathbb{Z}$--graded vector space
\[
    V=\coprod_{n \in \mathbb{Z}} V_{(n)};
    \ \mbox{\rm for}\ v\in V_{(n)},\;\mbox{\rm wt}\ v = n,
\]
such that
\beqa
     &&  V_{(n)} = 0 \;\; \mbox{ for }\; n \mbox{ sufficiently
    negative,} \n &&  \mbox{\rm dim }V_{(n)}<\infty\;\;\mbox{ for }\;
    n \in {\mathbb Z} , \no
\eeqa
equipped with a linear map $Y(\cdot,x)$:
\begin{eqnarray}
    Y(\cdot,x)\,: \ V&\to&(\mbox{\rm End}\; V)[[x, x^{-1}]]\nonumber \\
    v&\mapsto& Y(v, x)={\displaystyle \sum_{n\in{\mathbb Z}}}v_{n} x^{-n-1}
    \,,\;\;v_{n}\in \mbox{\rm End} \;V,
\end{eqnarray}
where $Y(v,x)$ is called the {\em vertex operator} associated with
$v$, and two particular vectors, ${\bf 1},\,\omega\in V$, called
respectively the {\em vacuum vector} and the {\em conformal
vector}, with the following properties:\\
{\em truncation condition:} For every $v,w \in V$
\beq
    v_n w=0
\eeq
for $n\in\Bbb Z$ sufficiently large;\\
{\em vacuum property:}
\beq
    Y({\bf 1},x) = 1_V \quad \mbox{($1_V$ is the identity on $V$)};
\eeq
{\em creation property:}
\beq
    Y(v,x){\bf 1} \in V[[x]] \quad{and}\quad \lim_{x\to0} Y(v,x){\bf 1} = v \;;
\eeq
{\em Virasoro algebra conditions:} Let
\begin{equation}
    L(n)=\omega _{n+1}\;\; \mbox{\rm for} \;n\in{\mathbb Z},
        \;\;{\rm i.e.},\;\;
        Y(\omega, x)=\sum_{n\in{\mathbb Z}}L(n)x^{-n-2} \;.
\end{equation}
Then
\[
    [L(m),L(n)]=
    (m-n)L(m+n)+c_V\frc{m^{3}-m}{12}\,\delta_{n+m,0}\,1_V
\]
for $m, n \in {\mathbb Z}$, where $c_V\in {\mathbb C}$ is the
central charge (also called ``rank'' of $V$),
\[
    L(0) v=({\rm wt}\ v) v
\]
for every homogeneous element $v$, and we have the {\em
$L(-1)$--derivative property:}
\beq\label{Lm1der}
    Y(L(-1)u,x)=\frac{d}{dx}Y(u,x) \;;
\eeq
{\em Jacobi identity:}
\beqa\label{jacobi}
    && x_0^{-1}\delta\lt(\frc{x_1-x_2}{x_0}\rt) Y(u,x_1) Y(v,x_2)
        - x_0^{-1}\delta\lt(\frc{x_2-x_1}{-x_0}\rt) Y(v,x_2) Y(u,x_1) \no\\
    &&  \qquad = x_2^{-1} \delta\lt(\frc{x_1-x_0}{x_2}\rt)
            Y(Y(u,x_0)v,x_2)\;.
\eeqa
\end{defi}

In this definition, the formal delta-function is defined by
\beq
	\delta(x) = \sum_{n\in\Z} x^n
\eeq
and we use the binomial expansion convention: $(x_1-x_2)^m$, for $m\in\Z$, is the formal series in two variables obtained from the Taylor expansion in nonnegative powers of $x_2$.

\end{document}